\documentclass[fleqn,usenatbib]{mnras}
\usepackage{newtxtext,newtxmath}

\usepackage[T1]{fontenc}

\DeclareRobustCommand{\VAN}[3]{#2}
\let\VANthebibliography\thebibliography
\def\thebibliography{\DeclareRobustCommand{\VAN}[3]{##3}\VANthebibliography}

\newcommand{\SchouwsPrep}{\citeauthor{Schouws22InPrep} \citeyear{Schouws22InPrep} (in prep.)\xspace}

\usepackage{graphicx}	
\usepackage{amsmath}	

\usepackage{amssymb}	

\usepackage{xspace}
\usepackage{xcolor}

\newcommand{\CII}{[\ion{C}{ii}]\xspace}
\newcommand{\CIImu}{[\ion{C}{ii}]\,158\,$\upmu$m\xspace}
\newcommand{\OIII}{[\ion{O}{iii}]\xspace}
\newcommand{\OIIImu}{[\ion{O}{iii}]\,88\,$\upmu$m\xspace}

\title[{REBELS \CII Luminosity Function at $z\sim7$}]{The ALMA REBELS Survey: The \CII Luminosity Function of UV-Selected Galaxies at $z\sim7$}

\author[T. Herard-Demanche et al.]{
Thomas Herard-Demanche,$^{1}$\thanks{E-mail: herard@strw.leidenuniv.nl}
Pascal A. Oesch,$^{2,3}$
Rychard Bouwens,$^{1}$
Jacqueline A. Hodge,$^{1}$ \and
Hiddo S. B. Algera,$^{4}$
Manuel Aravena,$^{5,6}$
Karin Cescon,$^{1}$
Elisabete da Cunha,$^{7,8}$
Pratika Dayal,$^{9,10,11}$ \and
Andrea Ferrara,$^{12}$
Rebecca Fisher,$^{13}$
Yoshinobu Fudamoto,$^{14}$
Kasper E. Heintz,$^{3}$
Hanae Inami,$^{15}$\and
Andrea Pallottini,$^{16}$ 
Lucie E. Rowland,$^{1}$
Renske Smit,$^{17}$
Laura Sommovigo,$^{18,19}$
Mauro Stefanon,$^{20,21}$ \and
Michael W. Topping,$^{22}$
Paul van der Werf$^{1}$ and
Ivana F. van Leeuwen$^{1}$
\\
$^{1}$Leiden Observatory, Leiden University, PO Box 9513, 2300 RA Leiden, The Netherlands \\
$^{2}$Department of Astronomy, University of Geneva, Chemin Pegasi 51, CH-1290 Versoix, Switzerland \\
$^{3}$Cosmic Dawn Center (DAWN), Niels Bohr Institute, University of Copenhagen, Jagtvej 128, K\o benhavn N, DK-2200, Denmark \\
$^{4}$Institute of Astronomy and Astrophysics, Academia Sinica, 11F of Astronomy-Mathematics Building, No.~1, Sec.~4, Roosevelt Rd, Taipei 106319, Taiwan, ROC \\
$^{5}$Instituto de Estudios Astrof\'isicos, Facultad de Ingenier\'ia y Ciencias, Universidad Diego Portales, Av. Ej\'ercito 441, Santiago, Chile \\
$^{6}$Millennium Nucleus for Galaxies (MINGAL), Chile \\
$^{7}$International Centre for Radio Astronomy Research (ICRAR), The University of Western Australia, M468, 35 Stirling Highway, Crawley, WA 6009, Australia \\
$^{8}$ARC Centre of Excellence for All Sky Astrophysics in 3 Dimensions (ASTRO 3D), Australia \\
$^{9}$Canadian Institute for Theoretical Astrophysics, 60 St George St, University of Toronto, Toronto, ON M5S 3H8, Canada \\
$^{10}$David A. Dunlap Department of Astronomy and Astrophysics, University of Toronto, 50 St George St, Toronto, ON M5S 3H4, Canada \\
$^{11}$Department of Physics, University of Toronto, 60 St George St, Toronto, ON M5S 3H8, Canada \\
$^{12}$Scuola Normale Superiore, Piazza dei Cavalieri 7, 56126 Pisa, Italy \\
$^{13}$Jodrell Bank Centre for Astrophysics, University of Manchester, Oxford Road, Manchester M13 9PL, UK \\
$^{14}$Center for Frontier Science, Chiba University, 1-33 Yayoi-cho, Inage-ku, Chiba 263-8522, Japan \\
$^{15}$Hiroshima Astrophysical Science Center, Hiroshima University, 1-3-1 Kagamiyama-Higashi, Hiroshima 739-8526, Japan \\
$^{16}$Dipartimento di Fisica ``Enrico Fermi'', Universit\`a di Pisa, Largo Bruno Pontecorvo 3, I-56127 Pisa, Italy \\
$^{17}$Astrophysics Research Institute, Liverpool John Moores University, 146 Brownlow Hill, Liverpool L3 5RF, UK \\
$^{18}$Department of Astronomy, Columbia University, 550 W 120th St, New York, NY 10025, USA \\
$^{19}$Center for Computational Astrophysics, Flatiron Institute, 162 5th Ave, New York, NY 10010, USA \\
$^{20}$Departament d'Astronomia i Astrof\'isica, Universitat de Val\`encia, Av. Vicent Andr\'es Estell\'es 19, E-46100 Burjassot, Val\`encia, Spain \\
$^{21}$Unidad Asociada CSIC ``Grupo de Astrof\'isica Extragal\'actica i Cosmolog\'ia'' (Instituto de F\'isica de Cantabria -- Universitat de Val\`encia), Spain \\
$^{22}$Steward Observatory, University of Arizona, 933 North Cherry Avenue, Tucson, AZ 85721, USA
}

\date{Accepted 2026 October 14. Received 2026 October 3; in original form 2026 July 15}

\pubyear{2026}

\begin{document}
\label{firstpage}
\pagerange{\pageref{firstpage}--\pageref{lastpage}}
\maketitle

\begin{abstract}
Luminosity functions (LF) are some of the most basic quantities providing insights into galaxy evolution across cosmic time and enabling a direct comparison between observations and simulations.
Here, we present the first \CIImu LF at $z\sim7$ from a sizeable UV-selected ALMA sample. Our analysis makes use of 27 galaxies with significant \CII emission line detections in the epoch of reionization from the REBELS (Reionization Era Bright Emission Line Survey) ALMA Large Program and its pilot programs.  We used these line detections and similar from ALPINE to calibrate UV luminosity-$L_\mathrm{\CII}$ and stellar mass-$L_\mathrm{\CII}$ relations at $z\sim7$.  Given the challenge in precisely defining a selection volume for the REBELS sample itself, we rely on published UV LF and stellar mass function (SMF) determinations at $z\sim7$ and three complementary approaches to derive the \CII LF. The UV LF–based method provides a lower limit on the \CII LF, due to its missing the most dust-obscured systems. Our SMF-based estimate yields a Schechter function with a characteristic luminosity of $L^*_\mathrm{\CII}=10^{9.11_{-0.27}^{+0.11}}\,L_\odot$ and a steep faint-end slope. From $z\sim7$ to $z\sim5$, our inferred \CII LF shows an evolution of $+0.3\,$dex in normalisation and $+0.1\,$dex in $L^*_\mathrm{\CII}$.  Our results indicate a general underprediction from models of the abundance of \CII emitters at high-redshift and thus potentially the number densities of massive yet UV-faint systems implied by our SMF-based method. Additionally, they suggest that 63\% of the star formation rate density occurs in massive \CII-bright systems and that $\sim 60$--$90$\% of \CII emitters at $z\sim7$ may be missed by bright UV selections. This highlights the importance of surveys focusing on especially massive, UV-faint sources at $z>6$.
\end{abstract}

\begin{keywords}
galaxies: high-redshift -- galaxies: abundances -- galaxies: ISM -- galaxies: star formation -- galaxies: evolution
\end{keywords}



\section{Introduction}

The detailed characterization of the first generation of galaxies is one of the prime goals of observational astrophysics.
Great progress has been achieved to reveal the star-formation rates (SFR) and stellar masses of galaxies in the epoch of reionization (EoR) based on large galaxy samples selected from the Hubble Space Telescope (HST) and the James Webb Space Telescope (JWST) out to the highest redshifts \citep[e.g.][]{Dayal2018,Robertson21,Matthee2025,Stark2026,Bouwens2026}. However, the interstellar medium (ISM) properties of these early sources are more difficult to study \citep[e.g.,][]{Carilli13,Hodge20}.

Evidence has emerged that the ISM conditions of $z>6$ galaxies differed significantly compared to their descendants at later cosmic epochs. These systems exhibit intense nebular emission, high ionization parameters, hard radiation fields produced by young massive stars, elevated electron densities, and rapid chemical enrichment despite their young ages \citep[e.g.,][]{Curti2023,Cameron2023,Sanders2024,Heintz2024_dla,Santon26,Faisst26,rowland26a}. Together, these observations point to compact, gas-rich galaxies with highly dynamic interstellar media that likely promoted efficient star formation and strong stellar feedback during the first few hundred million years of cosmic history.

The revolutionary sensitivity of ALMA is now providing detailed insights into the chemical and physical properties of early galaxies, in particular, with the use of the \CII emission line at 158\,$\upmu$m.

\CII is the dominant cooling line of the ISM, coming primarily from photo-dissociation regions and the cold neutral medium of molecular clouds \citep[e.g.,][]{Wolfire95,Hollenbach99,Wolfire22}.
However, the relative contributions from different ISM phases to the \CII emission could well differ at higher redshifts compared to the local Universe \citep[e.g.,][]{Vallini18,Ferrara19,Katz19,RamosPadilla22,Pallottini22}.
Nevertheless, \CII correlates surprisingly well with the
star formation rates of galaxies out to the highest redshifts 
\citep[see e.g.,][]{Stacey91,DeLooze14,Capak15,HerreraCamus15,Carniani18,
Matthee19,Schaerer20,Romano22}. Beyond SFR, \CII has been used to 
estimate molecular gas masses \citep[e.g.,][]{Zanella18,Dessauges20,Vallini25}, 
outflows \citep[e.g.,][]{Gallerani18,Ginolfi20,Pizzati23}, the extended ISM 
\citep[e.g.,][]{Fujimoto19,Fudamoto22}, and even neutral gas and metal 
content \citep{Heintz21,Heintz23,Vallini24}.

Dedicated ALMA programmes have applied these diagnostics 
to the earliest massive, star-forming galaxies. The Reionization Era Bright Emission Line Survey \citep[REBELS]{Bouwens21REBELS}, in particular, has revealed that UV-luminous $z\sim7$ galaxies host large molecular gas reservoirs ($M_\mathrm{mol}\sim10^{10.5}-10^{11}\,M_\odot$; \citealt{Aravena24,Cescon26} ), together with significant dust masses and elevated dust temperatures ($T_\mathrm{d}\sim40$--$60\,\mathrm{K}$; 
\citealt{Inami22,Sommovigo22}). Between $30$ and $90\%$ of their star 
formation is dust-obscured, depending on stellar mass 
\citep{Algera23,Bowler24}. High-resolution follow-up of the most 
\CII-luminous source, REBELS-25 at $z=7.31$, has further uncovered a 
fully-assembled, dynamically cold rotating disc with 
$V_\mathrm{rot}/\sigma\approx11$ \citep{Rowland24}. Taken together, these 
results show that massive, gas-rich, dusty, and kinematically mature 
systems are already in place within the first billion years of cosmic 
time.

JWST is now providing a complementary, rest-frame optical 
view of these same systems. Through the REBELS-IFU programme, 
NIRSpec/IFU spectroscopy has delivered gas-phase metallicities, 
demonstrating that an enriched mass--metallicity relation is already in 
place at $z\sim7$ \citep{rowland26a}, the first direct constraints on 
their dust attenuation curves \citep{Markov23, Markov25, Shivaei25,Fisher25}, and scaling relations 
between dust, gas, and metals that begin to disentangle the dominant 
dust-production pathways at these epochs \citep{Algera26}. Together, 
ALMA and JWST are therefore enabling a panchromatic characterization of 
the most massive star-forming galaxies in the EoR.

Alongside these UV-selected \CII emitters, however, 
ALMA and JWST have also revealed populations of heavily dust-obscured 
galaxies that evade UV-selection altogether \citep{barrufet23b,herard-demanche25}. These, along with serendipitously identified companions to targeted \CII systems may contribute a non-negligible fraction of the integrated stellar mass \citep{Gottumukkala24}, star formation rate density \citep{vanleeuwen24}, and must be accounted for in any complete census of star-forming systems in the EoR.

The \CII line offers several advantages for such a census: it is unaffected by dust attenuation, and falls in a favourable ALMA frequency window, when redshifted to $z=6-8$ in the heart of the EoR. It has thus been widely used both to probe the ISM of early galaxies and to secure their redshifts 
\citep[e.g.,][]{Maiolino15,Carniani18,Hashimoto19,Bakx20,LeFevre20,Bouwens21REBELS,HerreraCamus25}. 
First \CII luminosity functions (LF) have been measured up to $z\sim6$ 
from the ALPINE survey \citep{Yan20}, while at $z=6-8$ only upper limits 
exist so far \citep[e.g.,][]{Yamaguchi17,Uzgil21}. Measurements at these epochs offer particularly powerful 
constraints on theoretical models and cosmological simulations, which at present predict widely differing LFs at these redshifts, as we show in the later sections.

In this paper, we exploit the ALMA--JWST synergy 
outlined above to derive the first \CII LF at $z\sim7$, leveraging data from the REBELS ALMA Cycle-7 Large Program and its pilot programmes. To account both for the UV-luminous massive galaxies that dominate REBELS and for the dust-obscured population largely missed by UV-selected surveys, we 
complement the sample with a JWST-derived galaxy stellar mass function \citep[SMF,][]{Weibel24}. We apply multiple complementary methods, including the approach developed by \citet{Yan20} for the ALPINE sample at $z\sim4-6$ \citep{LeFevre20,Bethermin20,Faisst20}, to derive LFs in our measurement range using the UV LF at this epoch as proxy and help constrain the intrinsic \CII LF in the EoR. This paper is organized as follows: In Sect. \ref{sec:Sample}, we describe the galaxy sample. In Sect. \ref{sec:Methods}, we describe our approach to derive the \CII LF, before we compare our results with other works in Sect. \ref{sec:comp_discussion} and conclude in Sect. \ref{sec:Conclusion}.
Throughout this paper, magnitudes are given in the AB system and the adopted cosmological parameters are $\Omega_{\Lambda,0}=0.7$, $\Omega_{M,0} = 0.3$, $H_0=70$\,km\,s$^{-1}$\,Mpc$^{-1}$.

\begin{figure}
\centering
\includegraphics[width=\linewidth]{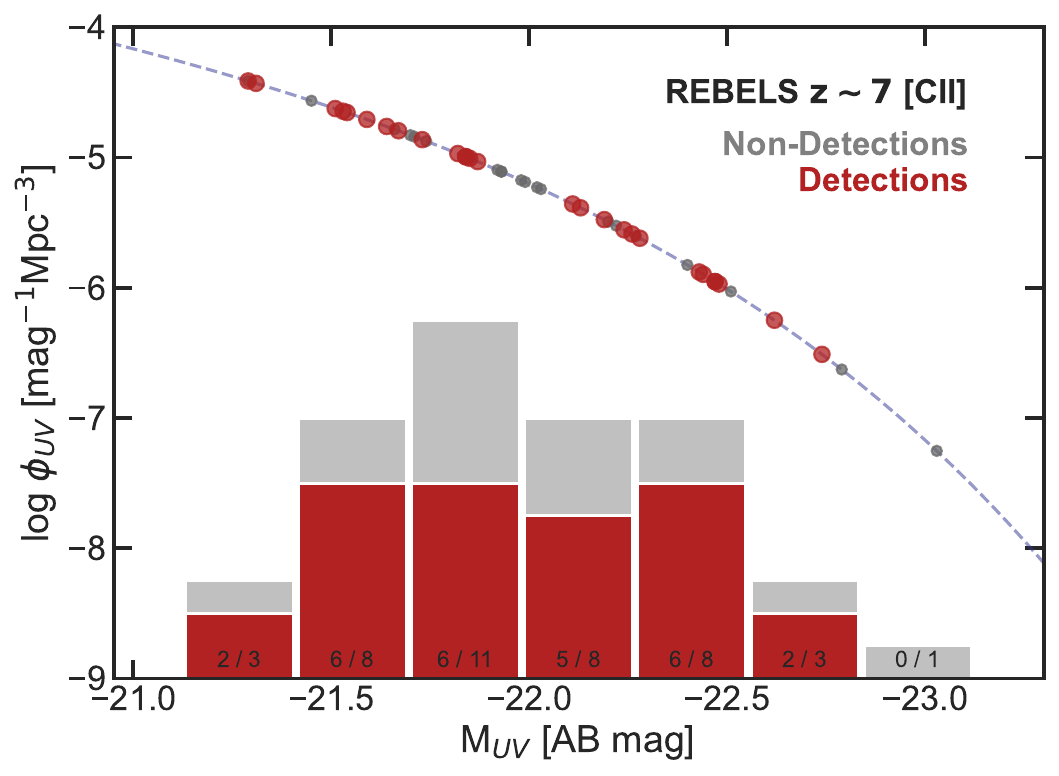}
\caption{Histogram of the REBELS+pilot \CII detected vs. non-detected galaxies as a function of the absolute magnitude M$_{UV}$ in the rest-UV. The red-shaded area shows the detections ($>4\sigma$), while gray corresponds to non-detections. The numbers presented at the base of the histogram indicate the number of \CII-detected galaxies relative to total numbers in each M$_{UV}$ bin. Generally, the detection fraction seems to be independent of UV luminosity. The dashed line further shows the UV luminosity function from \citet{Bouwens21a}.  The small red and gray dots indicate the locations of REBELS targets for detected and non-detected galaxies, respectively. Our method of estimating a step-wise \CII LF is based on the expected numbers of galaxies in M$_{UV}$ bins based on the UV LF (see Sect. \ref{sec:MethodUV} for more details). }
\label{fig:CiiDetHist}
\end{figure}

\section{Observational Data}
\label{sec:Sample}

\subsection{REBELS \texorpdfstring{$z\sim7$}{z~7} Sample}
The primary dataset for this paper is based on the ALMA Large program REBELS \citep[Reionization Era Bright Emission Line Survey;][]{Bouwens21REBELS}, as well as its pilot programs \citep[][]{Smit18,Schouws22a,Schouws22b}. These programs selected a sample of 40 (REBELS) and 9 (pilot) UV luminous galaxies with photometric redshifts $z>6.5$ for ALMA line scan follow-up. In total $\mathrm{\sim 7\,deg^{2}}$ of ground-based imaging was used to identify the best targets, with accurate photometric redshift estimates.  The presence of strong rest-frame optical line emission and a sharp Lyman-break were both of significant utility in ensuring the accuracy of our photometric redshift estimates.
The total REBELS sample spans a range of UV absolute magnitudes $M_\mathrm{UV}=-23.0$ to $-21.3$ and redshifts $z=6.4-8.6$. Their stellar masses and star-formation rates place them at the high end 
of the main-sequence of star-formation at these redshifts \citep[][]{Topping21,Bouwens21REBELS}.

For sources in the redshift range $z=6.4-7.7$, a scan is conducted for the \CIImu emission line in frequency windows spanning the photometric redshift probability distribution function of the target galaxies. 
For higher redshift sources ($z>7.7$), the spectral scans focused on the \OIIImu line \citep{vanleeuwen25}.  Given the differences in emission line strengths and sky transparency at the specific observed wavelengths, this is the most efficient approach for detecting ISM cooling lines for galaxies at $z>6$, while also probing the dust continua.
The typical \CII luminosity limits of the REBELS line scans are $\sim3\times10^{8}L_\odot$.
\citet{Bouwens21REBELS} present the spectral windows used for the line scans relative to the photometric redshift probability distribution functions. Meanwhile, the dust continuum detections are described in \citet{Inami22}.
In addition to the \CII luminosities of the REBELS galaxies, we also make use of their stellar masses and absolute UV magnitudes derived from SED fitting from \citet{Topping22}. These fits are primarily based of ground-based photometry supplemented by Spitzer/IRAC and HST/ACS and WFC3 data where available \citep{Bouwens21REBELS,Topping22}. The fits are performed at the \CII redshift for the sources considered in this work.

\subsection{\CII Emission Line Sample}
In this paper, we only focus on the galaxy sample for which we have searched for \CII emission lines, i.e., with either photometric or spectroscopic redshifts $z=6.4-7.7$. This includes 42 sources in total, 35 and 7 of which are from the REBELS and pilot programs, respectively. We refer the reader to \SchouwsPrep for details of the \CII emission line measurements. The line sensitivity limit can vary significantly from source to source, with a median $4\sigma$ line sensitivity of 280\,mJy\,km\,s$^{-1}$, corresponding to a luminosity limit of $\log(L_\mathrm{\CII}/L_\odot)=8.6$.

The spectroscopic success rate for a significant detection of a \CII line at $>4\sigma$ is 64\% with little indication that this success rate depends on UV luminosity (see Fig. \ref{fig:CiiDetHist}). Therefore, a total of 27 sources have obtained new spectroscopic redshift measurements thanks to REBELS and its pilots (22 and 5, respectively; see \SchouwsPrep). These 27 \CII detected galaxies provide the main sample for this work.
The mean redshift of these sources is $z_\mathrm{\CII} = 6.94$ and span the redshift range $z=6.50-7.68$.
Note that we do not use the remaining REBELS sources that did not reveal a detectable \CII line, since we cannot be sure whether the line was not detected due to its faintness or because we have not scanned at the proper redshifted frequency given the photometric redshift probability distribution function. There is some evidence that the former applies in many cases: the lack of continuum detections among the \CII-undetected sources suggests that most may genuinely fall below the \CII detection limit, in which case their exclusion would systematically bias $L_{\rm \CII}/L_{\rm UV}$ high. Unfortunately, this cannot presently be established with certainty for individual non-detections, while treating all of them as intrinsically \CII-faint would introduce a bias in the opposite direction.

\subsection{\texorpdfstring{$z\sim4$--6}{z~4--6} Galaxies from ALPINE}
In addition to the REBELS $z\sim7$ galaxy sample, we also consider the publicly available catalogs from the ALPINE large program\footnote{\url{https://cesam.lam.fr/a2c2s/data_release.php}} for lower redshift comparisons \citep{LeFevre20}. In particular, ALPINE targeted the \CII emission line of 118 galaxies on the main sequence with pre-existing spectroscopic redshifts in the range $z_\mathrm{spec}=4.4-5.9$ (see \citet{Bethermin20} and \citet{Faisst20} for the \CII catalogues and ancillary data analysis respectively). In the following, we make use of the UV magnitudes and stellar masses of the ALPINE galaxies derived from the \textsc{LePhare} SED fits \citep{Faisst20}.
The UV magnitude range of ALPINE includes somewhat fainter ($M_{UV}\sim-20.5$ mag) sources than REBELS does. Nevertheless, the overall success rate of ALPINE for \CII detections is almost exactly the same (64\%) as for the REBELS sample at $z\sim7$, resulting in 75 \CII ($>3\sigma$) detections and 43 non-detections, for which conservative upper limits on \CII are derived \citep[see][for details]{Bethermin20}.  

\begin{figure}
\centering
\includegraphics[width=\linewidth]{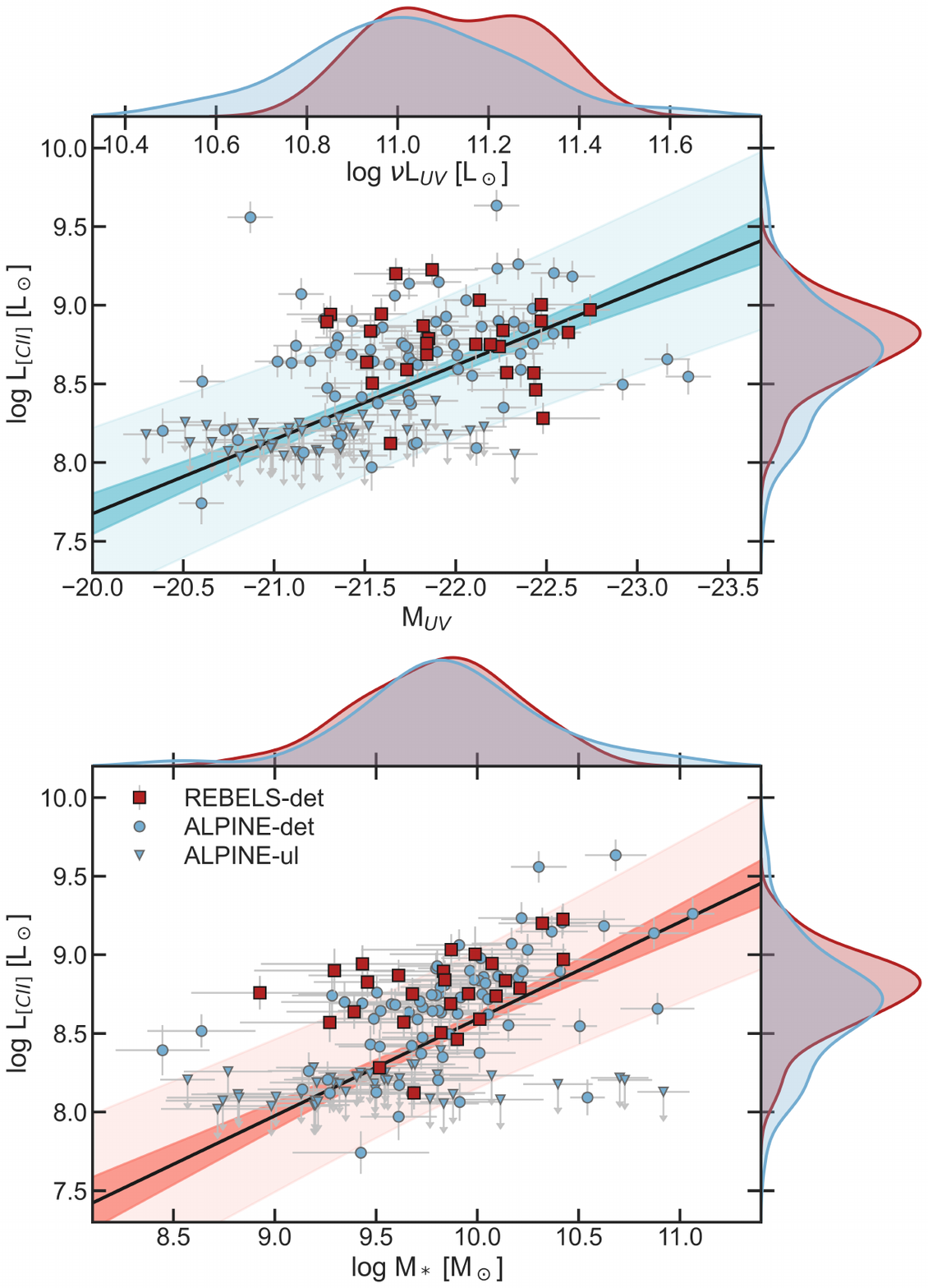}
\caption{Empirical relations between rest-frame UV and \CII luminosity (upper), and stellar mass and \CII luminosity (lower) for sources at $z=4-8$ based on the ALPINE and REBELS galaxy samples. The smoothed distribution using kernel density estimation for $L_\mathrm{\CII}$, $M_\star$ and $M_{UV}$ is shown on the right side and the top of each figure. Blue points and triangles correspond to the ALPINE $z=4-6$ \CII detections and $3\sigma$ upper limits, respectively. The red squares are the REBELS $z\sim7$ galaxies with \CII detections. The modelled linear relation between UV and \CII luminosity is shown as the black line with $1\sigma$ uncertainties as blue and salmon coloured bands for the fit to $M_{UV}$ and $M_\star$ respectively. The fainter band is the relation when including intrinsic dispersion around the mean relation, which is found to be $\sigma_\mathrm{UV}=0.42\pm 0.17$ for the upper panel, and $\sigma_\mathrm{M_\star}=0.39\pm 0.17$ dex for the lower one (see text for details).}
\label{fig:LuvLciifit}
\end{figure}

A comparison of the ALPINE and REBELS targets can be seen in Fig. \ref{fig:LuvLciifit}. At fixed UV luminosity, the two samples show consistent \CII luminosities, which span a very similar range. However, the ALPINE sample extends to lower UV luminosities, while for the REBELS sample bright \CII emitters make up a higher percentage of the total sample. Despite this, the two samples span a similar range in stellar mass. We will thus use these measurements to derive a relation between both the UV and \CII luminosities and the $M_\star$ and \CII luminosities (see Sect. \ref{sec:MethodUV} and \ref{sec:MethodSMF}).

\begin{figure}
\centering
\includegraphics[width=\linewidth]{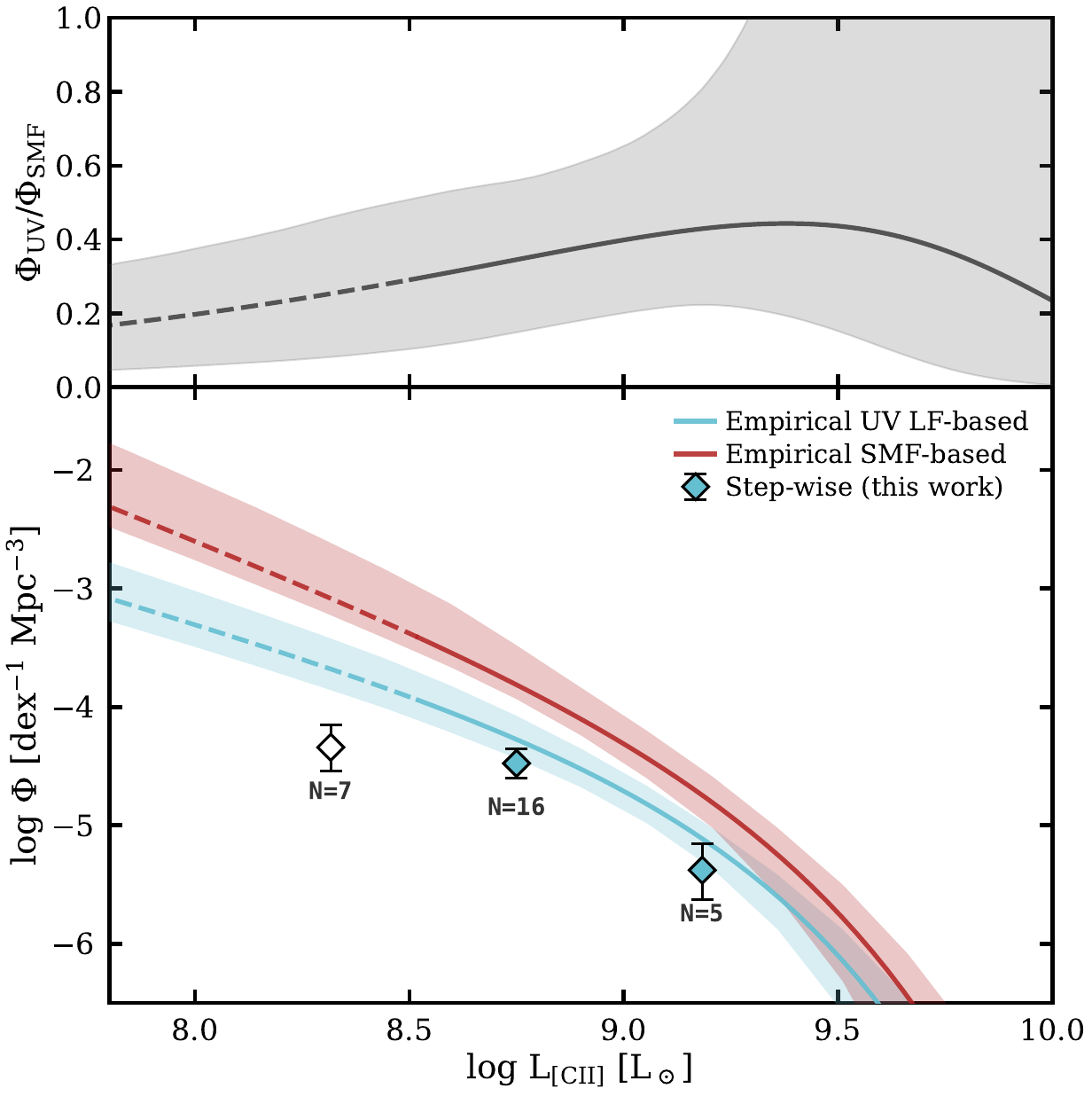}
\caption{Comparison of the different methods used in this work to estimate the \CII luminosity function at $z\sim7$. The lower panel shows both empirical modelling methods with solid lines. The Schechter fit to the method using the UV luminosity function from \citet{Bouwens21a} and the empirical $M_{UV}-L_\mathrm{\CII}$ is shown in sky blue. The method using the SMF from \citet{Weibel24} and the empirical $M_\star-L_\mathrm{\CII}$ relation is shown in dark red. We show $1\sigma$ contours for both methods as well as indicate incompleteness due to our $M_{UV}$ selection with a dashed line below $\log(L_\mathrm{\CII}/L_\odot) = 8.5$. Sky blue diamonds show the step-wise method adapted from the method developed in \citet{Yan20} with empty symbols below the completeness limit. We indicated the number of sources $N$ in each $L_\mathrm{\CII}$ bin. The ratio between the UV LF and SMF-based methods is shown as a function of $\log L_\mathrm{\CII}$ on the upper plot in black.}
\label{fig:methodsplot}
\end{figure}

\begin{figure*}
\centering
\includegraphics[width=0.495\linewidth]{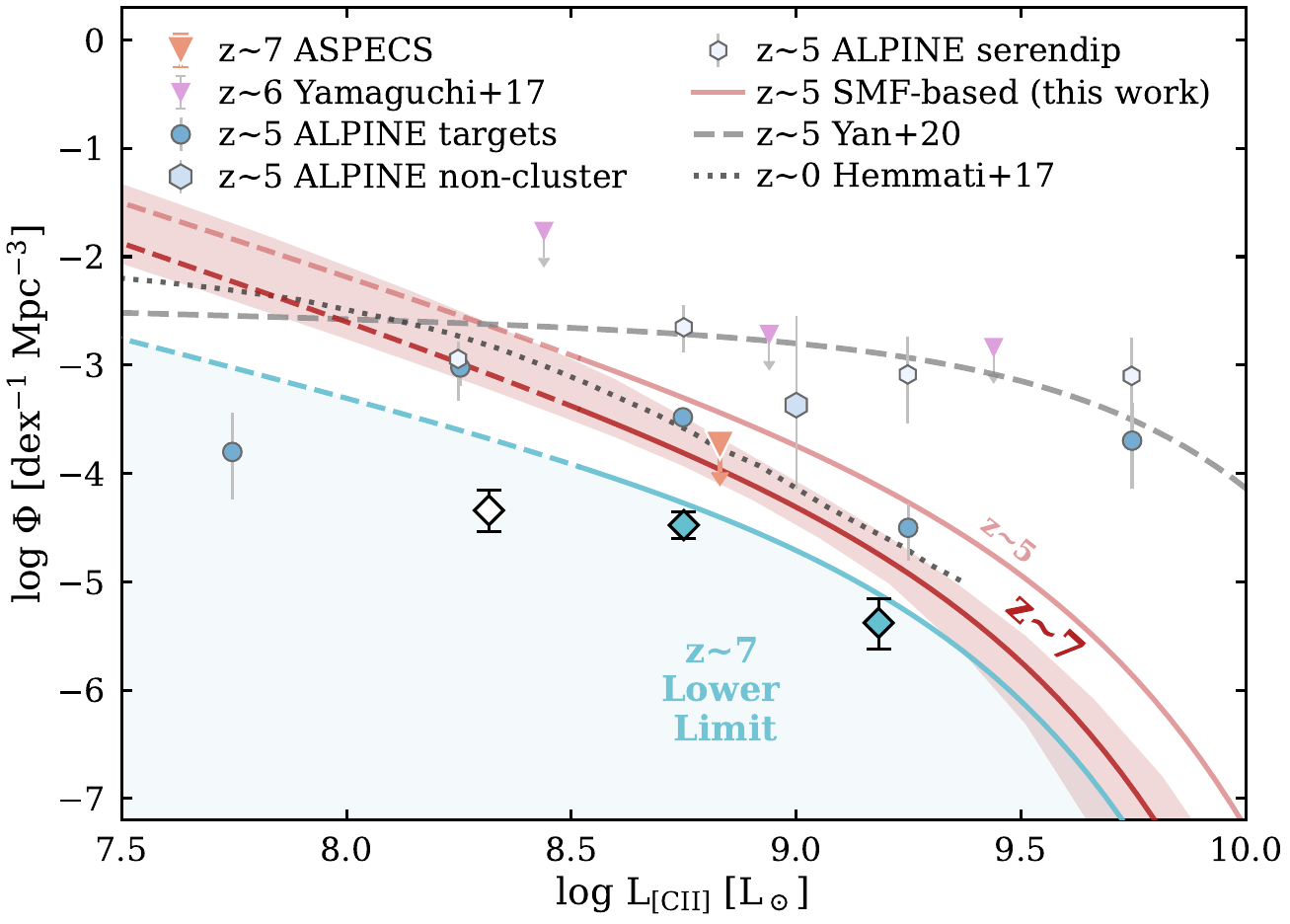}
\includegraphics[width=0.495\linewidth]{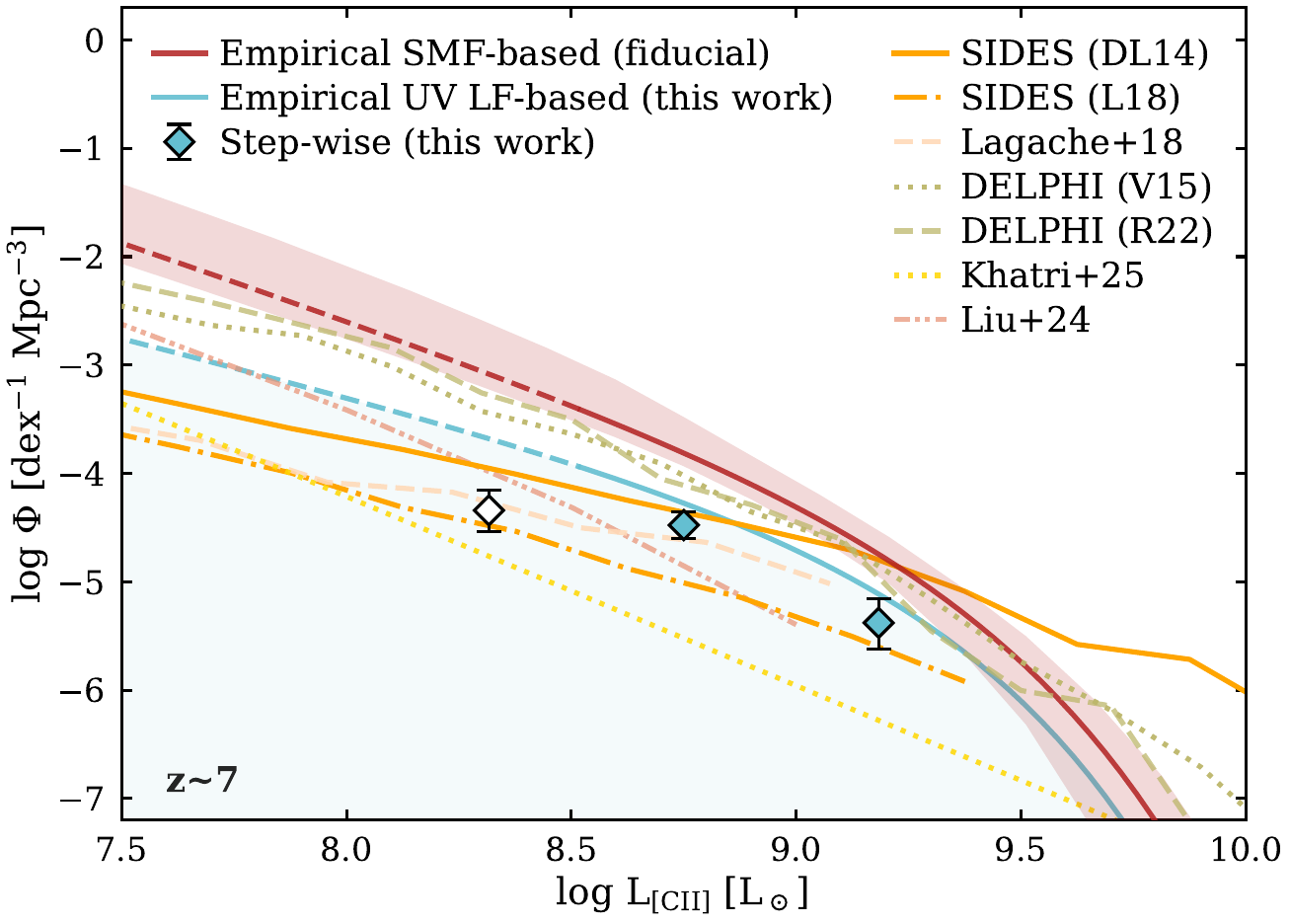}
\caption{Comparison between \CII luminosity function estimates from different observations (left) and various semi-analytical models and simulations (right). In both panels, the sky blue lines show the LFs from the UV-based method  while the dark red lines show the SMF-based one; solid above $\log(L_\mathrm{\CII}/L_\odot)=8.5$ and dashed below to indicate incompleteness due to our $M_{UV}$ selection. As the UV LF based method may miss a modest fraction of \CII emitters (see Sect. \ref{Sec:discussion}), it is shown as a lower limit on both panels. We show the step-wise estimate based on the UV LF as filled blue diamonds and empty symbols below the completeness limit. \textbf{Left} We show the measured \CII LF at $z\sim0$ \citep{Hemmati17} and limits at $z\sim7$ from ASPECS \citep{Uzgil21} and at $z\sim6$ \citep{Yamaguchi17}. The $z\sim4-6$ determinations from ALPINE are shown in blue circles for the target sample \citep{Yan20} and grey hexagons for the serendipitous sample including a large overdensity at $z=4.5$ and in the general field \citep[hexagon with dark gray outline;][]{Loiacono20}. The gray dashed line shows the Schechter function for $z\sim5$ derived in \citet{Yan20} from a fit to a combination of ALPINE results and \CII LF estimates derived from CO observations. Using the same SMF-based method but relying on the $z\sim5$ SMF from \citet{Weibel24}, we derive an independent estimate of the \CII LF at $z\sim5$ (salmon line).  Unsurprisingly, this is in agreement with the targeted sample. \textbf{Right}  We show simulated \CII LFs at $z\sim7$ from various theoretical works: the simulated LFs from the SIDES simulation in orange \citep{Bethermin22}, using two different L$_\mathrm{\CII}$-SFR relations (DL14 for \citet{DeLooze14} and L18 for \citet{Lagache18}), the \citet[][dashed peach]{Lagache18} estimate at $z\sim6.7$, \citet[][dotted gold]{kathri25} from the MARIGOLD simulation, the fiducial result from analytical bursty star formation model of \citet[][dash-dotted pink]{Liu24} with $\epsilon_{\rm ff}=0.015$ at $z\sim7.5$ and the \texttt{Delphi} semi-analytical model in a dotted and dashed khaki lines \citep{dayal2014a, Dayal22}. The \texttt{Delphi} predictions, using either the \citet{Vallini15} or \citet{Romano22} relation (V15 and R22 in the figure respectively), are the most consistent with our results.}
\label{fig:CiiLF}
\end{figure*}

\section{Estimates of the \CII Luminosity Function}
\label{sec:Methods}

In this section, we describe three methods to determine the \CII LF at $z\sim7$ based on the REBELS results. The main challenge with this endeavor is that the REBELS targets were identified from rest-UV observations over a 7$\,$deg$^2$ area and no equivalent survey for the \CII line exists over the same volume.  One way of circumventing these challenges is by tying the derivation of the \CII LF to galaxy distribution functions. The latter functions will be used to provide a representative volume for each source. Our first two approaches, both based on empirical modelling, are identical in structure but use different ingredients. In those two empirical modellings, we convert a galaxy distribution function based on a certain galaxy property,  i.e., UV luminosity or stellar mass, into a \CII LF with the use of an empirical relation on that quantity with respect to \CII luminosity. The first method uses the UV LF  and a relation between $L_\mathrm{UV}$ and $L_\mathrm{\CII}$ to build the \CII LF. Similarly, our second method uses the SMF from \citet{Weibel24} and converts it into a \CII LF using the $M_\mathrm{\star}$-$L_\mathrm{\CII}$ relation we derived. In the following sections, we will refer to these two methods as UV LF-based and SMF-based methods respectively. In the third method, we derive step-wise estimates of the \CII LF following the approach used for the ALPINE sample in \citet{Yan20}.

\subsection{Empirical Modelling Based on the \texorpdfstring{$L_\mathrm{UV}-L_\mathrm{\CII}$}{LUV-L\CII} Relation}
\label{sec:MethodUV}

The first method to derive the \CII LF relies on the UV LF from \citet{Bouwens21a} and an empirical relation between the UV and \CII luminosities. The first step is to derive the $L_\mathrm{UV}-L_\mathrm{\CII}$ relation including the intrinsic dispersion. In both cases, we use our SED fitting results to derive the UV luminosities at 1500\,\AA.
The upper panel of Fig. \ref{fig:LuvLciifit} shows the $L_\mathrm{UV}-L_\mathrm{\CII}$ relation for both the REBELS sources at $z\sim7$ and for the ALPINE sample at $z\sim5$. As can be seen, at fixed UV luminosity, the two samples show a very similar mean \CII luminosity and scatter.  This is perhaps not surprising, since the relation between SFR and $L_\mathrm{\CII}$ has been shown to show little evolution from the local Universe out to $z\sim6$, likely continuing out to $z\sim8$ \citep[see, e.g.,][Schouws et al., in prep]{Matthee19,Schaerer20}. 

In order to increase the dynamic range in our modeling of the $L_\mathrm{UV}-L_\mathrm{\CII}$ relation, we consider both the ALPINE and REBELS samples. The computation of the Spearman coefficient for the total sample shows a weak monotonicity and correlation between the parameters with a value of 0.28 and a p-value of 0.005. For the REBELS detections, this coefficient falls to $\rho=-0.04$. We can therefore not use this sample alone, mostly due to its narrow range spanned in UV ($\Delta M_{\rm UV}\sim 1.5$ mag versus $\sim$3 mag for ALPINE).
We thus include the combined sample of ALPINE and REBELS sources assuming they trace similar representative galaxy populations at high-redshift.

The best-fit relation between the \CII and UV luminosities is modeled using the python implementation\footnote{\url{https://github.com/jmeyers314/linmix}} of the \texttt{linmix} package \citep{Kelly07}. This is used to perform a Bayesian linear regression with uncertainties in both $L_\mathrm{UV}$ and $L_\mathrm{\CII}$, including upper limits. Only the ALPINE upper limits will be considered (see Sect. \ref{sec:Sample}).

We thus derive the linear relation between $L_\mathrm{UV}$ and $L_\mathrm{\CII}$ allowing for an intrinsic Gaussian dispersion ($\sigma_\mathrm{UV}$) around the mean relation
\begin{equation}
\label{eq:linear}
\log(L_\mathrm{\CII}/L_\odot) = a_{UV} + b_{UV} \times \log(L_{UV}/L_\odot) + \sigma_\mathrm{UV} 
\end{equation}
In this equation we use $L_\mathrm{UV}=\nu \, L_\nu$, evaluated at rest-frame 1500 \AA.
The posterior mean values and their $1\sigma$ uncertainties of the different parameters are $a_{UV}=-4.49\pm 2.01$, $b_{UV}=1.18\pm 0.19$, and $\sigma_\mathrm{UV}=0.42\pm 0.17$.
This is also shown in Fig. \ref{fig:LuvLciifit}. As a test we also fit the relation to the REBELS sample alone, fixing the slope to the $b_{UV}=1.18$ value and leaving the normalisation free. This leads to $+0.15$\,dex offset above the combined relation, well within $\sigma_{\rm UV}$. Additionally, while there is some uncertainty as to the nature of those sources in REBELS whose spectral scans did not yield a \CII detection, if we assume this is due to their line being too faint for detection, this only impacts a small fraction of the sources in our total fits, resulting in $<0.1$ dex lower $L_{\CII}-M_{UV}$ (or $M_*$) relation.  When compared to other issues we will discuss (e.g. UV-LF vs. SMF-based determinations), the impact of this issue is relatively small and will be ignored in the discussion which follows.

With this relation in hand, we can now derive the \CII LF, based on the known $z\sim7$ UV LF from \citet{Bouwens21a}. We adopt the Schechter parameterisation of the $z\sim7$ UV LF from \citet{Bouwens21a}. Out to $M_{\rm UV}\sim-23$, the brightest magnitudes probed by REBELS, the Schechter and double-power-law fits (from \citealt{Bouwens21a} and \citealt{Bowler2020}, respectively) agree within the uncertainties of the two parameterisations, so the choice has a negligible effect on the resulting \CII LF.

In order to compute uncertainties, we adopt a Monte Carlo approach. We sample 500,000 points from the UV LF and convert their corresponding $L_\mathrm{UV}$ values to $L_\mathrm{\CII}$, using Eq. \ref{eq:linear} and the intrinsic dispersion $\sigma_\mathrm{UV}$. Finally, we fit a Schechter function \citep{Schechter1976} to the resulting
$L_\mathrm{\CII}$ values, which in logarithmic form reads
\begin{equation}
\label{eq:schechter}
\begin{split}
\phi(\log L)\,d\log L =\;
& \ln(10)\,\phi^{*}\,
10^{(\alpha+1)(\log L - \log L^{*})} \\
& \times \exp\!\left(-10^{\,\log L - \log L^{*}}\right)d\log L,
\end{split}
\end{equation}
with the three parameters $\phi^{*}$, $L^{*}$,
and $\alpha$, representing the normalisation, the characteristic
luminosity, and the faint-end slope of the \CII LF, respectively.

This procedure is repeated 10$^4$ times, varying the input Schechter parameters of the UV LF according to the appropriate covariance matrix from \citet{Bouwens21a}. This procedure results in 10$^4$ \CII LFs, from which the mean and standard deviation for all three Schechter parameters are computed. The resulting values are listed in Tab. \ref{tab:resultsLF}.

\subsection{Empirical Modelling Based on the \texorpdfstring{$M_\mathrm{\star}-L_\mathrm{\CII}$}{Mstar-L\CII} Relation}
\label{sec:MethodSMF}
The second method is effectively the same as the first albeit with different ingredients. Here, we derive the \CII LF using the $z\sim7$ SMF from \citet{Weibel24} \footnote{We convert all stellar masses in the REBELS and ALPINE catalogues from a Chabrier IMF to a Kroupa IMF using the usual 0.05 dex correction in order to use the SMF from \citet{Weibel24}.} and an empirical relation between stellar mass and \CII luminosity. We derive the $M_\mathrm{\star}-L_\mathrm{\CII}$ along with its intrinsic dispersion using the same tool and sample as in section \ref{sec:MethodUV}.  The empirical relation is shown in the lower panel of Fig. \ref{fig:LuvLciifit}, and follows:
\begin{equation}
\label{eq:linearM}
\log(L_\mathrm{\CII}/L_\odot) = a_{M_\star} + b_{M_\star} \times \log(M_\star/M_\odot) + \sigma_\mathrm{M_\star} 
\end{equation}
The Spearman coefficient shows a stronger correlation between $M_\star$ and \CII luminosity compared to UV luminosity, with a value of 0.56 and a p-value of $2\times10^{-9}$. We also find a positive correlation using the REBELS sample alone $\rho=0.39$ but with a lower confidence $p=0.04$ and with much larger uncertainties on the slope.

The posterior mean values of the \texttt{linmix} fit and their $1\sigma$ uncertainties of the different parameters are $a_{M_\star}=2.46\pm 0.93$, $b_{M_\star}=0.61\pm 0.09$, and $\sigma_{M_\star}=0.39\pm 0.17$. As in Sect. \ref{sec:MethodUV}, we fit the relation using the REBELS sample alone at the fixed slope of the combined sample leading to a $+0.23$ dex offset again attributed to the exclusion of non-detections from our fits and uncertainties associated with that (see Sect. \ref{sec:caveats} for implications on the resulting \CII LF).
Similarly, we derive the \CII LF using that relation and the SMF at $z\sim7$ from \citet{Weibel24}. We adopt the same Monte Carlo approach and sample points from the SMF and convert their $M_\star$ values into $L_\mathrm{\CII}$ using the corresponding empirical relation. The results are listed in Tab. \ref{tab:resultsLF}.
Additionally, we repeat the same analysis at $z\sim5$. We use the SMF from \citet{Weibel24} at the $z\sim5$ in order to derive the \CII LF appropriate for a comparison with the ALPINE sample. The $z\sim5$ Schechter parameters for the resulting \CII LF are also listed in Tab. \ref{tab:resultsLF}.

\subsection{Step-Wise Estimate Based on the REBELS \CII Sample}
\label{sec:MethodPTWISE}

In our third approach, we derive step-wise \CII LFs, closely following the method developed in \citet{Yan20} for the derivation of the \CII LF from the ALPINE sample (see also \citealt{Barrufet23}, where the same approach is used to estimate the IR LF).

As in the first approach, we use the $z\sim7$ UV LF from \citet{Bouwens21a} as the reference to derive a representative volume for individual galaxies. In particular, we derive the total number of expected galaxies in a volume limited survey and compare this to the number of galaxies that are actually targeted by REBELS.
The expected number of galaxies in a given bin $\mathrm{M_{UV}\pm\Delta M_{UV}/2}$ is given by:

\begin{equation}
    N_\mathrm{exp} = \phi_\mathrm{UV}(M_\mathrm{UV})\,\Delta M_\mathrm{UV}\,V_\mathrm{tot}
    \label{eq:Nexp}
\end{equation}
where $\mathrm{ \phi_{UV}}$ is the UV LF (expressed in units of mag$^{-1}$), and $\mathrm{V_{tot}}$ is the total survey volume over which REBELS sources were selected.  While we assume an area of 7 deg$^2$ and $z=6.4-7.7$ in computing $V_{tot}$, normalisation of the \CII LF is done using the UV LF, and thus the final result will not be sensitive to the assumed survey volume.

REBELS only targeted the brightest galaxies in the rest-UV identified over a 7 deg$^2$ where tight constraints could be set on the redshift likelihood distribution. We can derive the `completeness' factor in each UV luminosity bin as
$\mathrm{f_{UV}= N_{exp}/N_{obs}}$, where $\mathrm{N_{obs}}$ is the number of targeted REBELS galaxies in the given bin.

The requirement for a $>4\sigma$ \CII line detection in our target sample introduces a reduction in the survey volume for each source, which can be accounted for using the available volume $\mathrm{V_{max}}$ factor \citep[see][]{Schmidt68}. For each source, we thus compute the maximum redshift up to which it would be detected at $>4\sigma$.

We now have all quantities to calculate the \CII LF $\mathrm{\phi_\mathrm{\CII}}$ in bins of $L_\mathrm{\CII}$. This is given by 
\begin{equation}
\phi_\mathrm{\CII} = \frac{1}{\Delta\log L}\sum_{i\in\mathrm{bin}}\frac{f_{\mathrm{UV},i}}{V_{\mathrm{max},i}} 
\end{equation}
where $\mathrm{\Delta log\,L}$ is the logarithmic width of the bins, $i$ runs over all sources in a given \CII luminosity bin, and $\mathrm{{V_{max,i}}}$ is the maximum comoving volume of each source (see Eq. 3 in \citealt{Yan20}). 
The above equation is also used to derive the appropriate Poisson uncertainties for each LF bin. The resulting \CII LF stepwise results are shown in Fig. \ref{fig:methodsplot} as well as Tab. \ref{tab:LFstepwise}.

\begin{table}
\centering
\caption{\CII LF parameters derived from Schechter function
fits at $z \sim 5$ and $z \sim 7$.}
\label{tab:resultsLF}
\begin{tabular}{lcccc}
\hline
Method & Redshift & $\log\phi^*_{\mathrm{\CII}}$ &
$\log L^*_{\mathrm{\CII}}$ & $\alpha_{\mathrm{\CII}}$ \\
\hline
UV LF & $z\sim7$ & $-4.71^{+0.40}_{-0.30}$ & $9.04^{+0.15}_{-0.16}$ & $-2.04^{+0.10}_{-0.11}$ \\
SMF   & $z\sim7$ & $-4.50^{+0.91}_{-0.23}$ & $9.11^{+0.11}_{-0.27}$ & $-2.41^{+0.14}_{-0.11}$ \\
SMF   & $z\sim5$ & $-4.20^{+0.67}_{-0.23}$ & $9.25^{+0.15}_{-0.22}$ & $-2.34^{+0.12}_{-0.10}$ \\
\hline
\end{tabular}
\par\medskip
\begin{minipage}{\columnwidth}
\small \textit{Notes.} $\phi^*_{\mathrm{\CII}}$ is in $\mathrm{Mpc^{-3}} \mathrm{dex}^{-1}$
and $L^*_{\mathrm{\CII}}$ in $L_\odot$.
\end{minipage}
\end{table}

\begin{table}
\centering
\caption{Step-wise \CII LF at $z\sim7$.}
\label{tab:LFstepwise}
\begin{tabular}{ccc}
\hline
$\log L_\mathrm{\CII}$ & $N$ & $\log\phi$ \\
\hline
$[8.10 - 8.53]$ &  7 & $-4.34^{+0.19}_{-0.20}$ \\
$[8.53 - 8.97]$ & 16 & $-4.48^{+0.12}_{-0.12}$ \\
$[8.97 - 9.40]$ &  5 & $-5.38^{+0.23}_{-0.24}$ \\
\hline
\end{tabular}
\par\medskip
\begin{minipage}{\columnwidth}
\small \textit{Notes.} $L_\mathrm{\CII}$ is in $L_\odot$ and $\phi$ in
$\mathrm{Mpc^{-3}\,dex^{-1}}$. Three equal-width bins over
$8.1 \le \log L_\mathrm{\CII}/L_\odot \le 9.4$ ($\Delta\log L = 0.433$\,dex),
corresponding to the range of $L_\mathrm{\CII}$ measurements of REBELS.
The number and width of bins were chosen to have $N>4$ sources per bin.
Uncertainties are $1\sigma$ Poisson.
\end{minipage}
\end{table}

\subsection{Comparison of Methods}
The mean Schechter functions from the empirical modelling methods as well as the step-wise estimates and their respective uncertainties are shown on the lower panel of Fig. \ref{fig:methodsplot}. To compare our methods and the implied number densities at each luminosity, we also show the ratio of both empirical methods in the top panel.

Overall, both our SMF-based method and UV LF-based empirical modelling methods agree very well with one another at the bright end but display an increasingly large offset at $\log(L_\mathrm{\CII}/L_\odot) < 9 $ reaching  $\sim0.5$ dex at  $\log(L_\mathrm{\CII}/L_\odot) = 8.5$. We will  refrain from drawing any significant conclusions based on the inferred \CII LF at $\log(L_\mathrm{\CII}/L_\odot) < 8.5$ due to the initial $M_{\mathrm{UV}}$ cut of our sample (see section \ref{sec:caveats}).
Because our sample is UV--selected, the inferred \CII LF is biased against galaxies that are massive but UV--faint. If we assume that the SMF-based \CII LF captures the totality of \CII emitters, then the ratio of the SMF-based to the UV LF-based \CII LF quantifies the fraction of \CII emitters missed by UV-based selection alone. At $\log(L_\mathrm{\CII}/L_\odot) = 9$, the UV-based method recovers only 40\% of the sources captured by the SMF-based method, decreasing to 29\% at $\log(L_\mathrm{\CII}/L_\odot) = 8.5$. We therefore interpret the difference between the two methods as a measure of UV-selection incompleteness, implying
that UV-based selections systematically underestimate the \CII luminosity function.

As for the step-wise method, it shows good agreement with the UV--LF--based approach  in the brightest two bins, increasing our confidence in the procedures used to derive those two results.
However, the determination in the lowest luminosity bin at  $\log L_\mathrm{\CII}/L_\odot < 8.5$ lies significantly below the derived Schechter function. This is most certainly due to the initial $M_\mathrm{UV}$ cut of the primary REBELS sample at $M_\mathrm{UV}<-21.3$. At those magnitudes, the expected mean $\log L_\mathrm{\CII}/L_\odot$ is 8.3, according to our relation in Eq. \ref{eq:linear}. Hence, our input sample is missing galaxies with fainter UV luminosities that can still have \CII luminosities up to $\log L_\mathrm{\CII}/L_\odot= 8.7$ including the 1$\sigma$ intrinsic dispersion (see discussion in the next section).

As shown earlier, stellar mass exhibits a tighter correlation with $L_{\mathrm{\CII}}$ than UV luminosity does. In contrast, UV luminosity is strongly affected by dust attenuation and bursty star formation, introducing additional scatter in the $L_{\mathrm{UV}}$--$L_{\mathrm{\CII}}$ relation. Moreover, UV-selected samples are intrinsically biased toward UV-bright
systems. Such selections can miss massive, \CII-luminous galaxies that are
sufficiently UV-faint to fall below the detection threshold of REBELS-like surveys, while still contributing significantly to the total \CII luminosity density. For these reasons, in the following sections and figures we interpret our UV-based determinations as lower limits on the true \CII luminosity function at $z\sim7$. In contrast, the SMF-based approach, is less affected by dust-selection biases and is therefore expected to provide a more complete census of \CII emitters--and a closer estimate of the intrinsic \CII luminosity function at this epoch.\\

\subsection{Uncertainties and Possible Systematics}
\label{sec:caveats}

The accuracy of our analysis is dependent on the quality of the multiple ingredients used for estimating the \CII LF, namely the samples used for the empirical relations as well as the validity of the UV LF and the SMF. Furthermore, these samples only provide us with a good probe of the \CII LF over the range of luminosities and masses sampled by the observations. The luminosity bins we adopt for Schechter fits follow the distribution seen in our samples, but ultimately render the estimate at the upper and lowermost bins uncertain. This is due to the scatter in the empirical relations, leading to less reliable estimates of the number densities there.  Furthermore, the faint-end slope $\alpha_{\mathrm{\CII}}$ of the Schechter fit is particularly sensitive to results in the lowest luminosity bin -- where incompleteness can be an issue. Nevertheless, using our empirical relations, both the knee of the UV LF and the SMF should lie in the regime where our sample is largely complete.  This allows us to constrain both the normalisation $\phi^*_{\CII}$ and the knee $L^*_{\mathrm{\CII}}$ of the \CII LF.

Another potential systematic comes from the normalisation of the empirical relations. In Sect. \ref{sec:MethodUV} and Sect. \ref{sec:MethodSMF}, we tried fitting the REBELS detections sample at fixed slope which placed them $+0.15$ and $+0.23$ dex above the combined $L_{\rm UV}$--$L_{\rm \CII}$ and $M_\star$--$L_{\rm \CII}$ relations respectively. We note these are still within the intrinsic scatter of both relations fitted on the combined ALPINE and REBELS sample. Adopting these normalisations at fixed $b$ and $\sigma$ would translate the per-dex LF, $\phi'(\log L)=\phi(\log L-\Delta)$, raising $\log L^*_{\rm \CII}$ by the same amount $\Delta$ with $\phi^*_{\rm \CII}$ and $\alpha_{\rm \CII}$ left unchanged. We attribute this effect to REBELS sampling the upper side of the scatter with its higher detection floor in $M_{UV}$ (and $M_\star$) and challenges associated with interpreting the \CII non-detections.

Furthermore, while the $z\sim7$ \CII LF derived with the UV-based and step-wise methods agree well down to $\log(L_\mathrm{\CII}/L_{\odot})=8.5$, both remain lower limits for the reasons presented earlier: they cannot account for sources sufficiently obscured in the UV to escape REBELS-like selection, such as extreme sub-mm galaxies already known to exist at these redshifts \citep[e.g.,][] {Riechers17,Marrone18}.

It is worth noting that even in the REBELS survey itself, a population of `normal' obscured galaxies at $z\sim7$ has been unveiled. \citet{Fudamoto21} serendipitously found two galaxies as neighbors to two REBELS targets at $z\sim7$. At face value, their number density is $\sim10^{-4}\,$Mpc$^{-3}$ with \CII luminosities of $\sim8\times10^8\,$L$_\odot$. After accounting for the clustering excess given that these two galaxies were found as neighbors to luminous targets, their expected \CII LF is in excellent agreement with our SMF-based method. Nevertheless, the detection of these sources shows that there are some UV-faint, \CII-bright galaxies that are missing from our analysis. 

Similarly, the ALPINE primary sample certainly revealed some galaxies which had exceptionally high \CII luminosities but modest UV luminosities. One example is a source which has a \CII luminosity in excess of 10$^{9.5}$ $L_{\odot}$ but is nevertheless one of the faintest galaxies in the rest-UV in ALPINE (see Fig. \ref{fig:LuvLciifit}). This source is  responsible for the extremely high \CII LF estimate in the ALPINE target at $\log(L_\mathrm{\CII}/L_{\odot})=9.75$, which lies almost one dex above the step-wise LF estimate in the next fainter bin (see Fig. \ref{fig:CiiLF}). 

Furthermore, the ALPINE serendipitous sample of \citet{Loiacono20} illustrates
how overdensities can inflate the inferred \CII LF: 11 of their 12
serendipitous \CII detections belong to a single large overdensity at $z=4.5$
in the COSMOS field. \citet{Loiacono20} address this by deriving separate
"field" and "clustered" luminosity functions. The clustered points lie well
above the field determination which indicates that number densities in such regions are not representative of the cosmic average. This highlights the sensitivity of \CII LF measurements to cosmic variance in small survey volumes. The field LF, however, rests on the single remaining galaxy outside the overdensity and is correspondingly uncertain. Nevertheless it is consistent within 1$\sigma$ with the targeted sample from \citet{Yan20} and with our derived $z\sim5$ \CII LF using the \CII-UV luminosity relation.

In summary, the \CII LFs we derived with both UV-based methods in the previous sections are most likely underestimated by some degree, possibly even in the highest luminosity bins, due to the proven existence of UV-faint galaxies with significant \CII luminosities. We therefore adopt our SMF-based approach as the fiducial result of this paper, which therefore tries to account for UV-faint and dusty massive \CII emitters. This determination is nevertheless not itself free of selections effects: it inherits the systematics of the input SMF, which carries  uncertainties from SED-fitting and model assumptions, may incompletely recover red or dust-obscured galaxies, and is subject to cosmic variance in the limited volumes from which it is derived. Mapping the intrinsic \CII LF directly will ultimately require wider blind \CII surveys at $z\sim7$.

\section{Comparisons with other observations and theoretical works}
\label{sec:comp_discussion}
\subsection{Comparison with Lower Redshift Results}

In this section, we will compare our $z\sim7$ LF estimates with other estimates from the literature. The most direct comparison can be made with the ASPECS analysis from \citet{Uzgil21}. In particular, ASPECS blindly scanned the full ALMA band 6 for line emitters over the HUDF, i.e., its scan covers the \CII line at $z=6-8$ \citep[see also][]{Decarli20}. While no serendipitous lines were found at high significance, the resulting upper limit on the \CII LF is still quite constraining. As can be seen in Fig. \ref{fig:CiiLF}, the upper limit is only $\sim0.5\,$dex above our step-wise determination. Hence, the two measurements are fully consistent. The ASPECS upper limit lies essentially at the same level as the measured $z\sim0$ \CII LF from \citet{Hemmati17} and $1\sigma$ above our fiducial \CII LF of our SMF-based method.

The comparison to $z\sim4-6$ from the ALPINE survey is done in two steps. First, we discuss the `total' Schechter function that was provided by \citet{Yan20}. This LF was derived from a fit to the serendipitous sources from \citet{Loiacono20} without any correction for clustering, as well as to LF points derived from converting CO LFs to \CII. As can be seen in Fig. \ref{fig:CiiLF}, this fit does not really pass through the majority of the ALPINE targeted sample, having an extremely shallow faint-end slope. At $L_\mathrm{\CII}=10^9\,$L$_\odot$ this Schechter function is $>1\,$dex higher than our $z\sim7$ estimate (as well as the ALPINE targeted sample). This would indicate an extremely fast evolution across the $\sim400\,$Myr from $z\sim7$ to $z\sim5$.

However, when deriving the $z\sim5$ \CII Schechter parameters in the same way as for the REBELS sample at $z\sim7$ using the SMF from \citet{Weibel24}, we obtain a much more moderate evolution of only $\sim$0.5-0.6$\,$dex across this redshift range and over the luminosities shown in Fig. \ref{fig:CiiLF}. Our $z\sim5$ Schechter function indeed passes through the step-wise estimates from the ALPINE targeted sample, except in the brightest and faintest bin, likely due to the UV-luminosity cut in the input sample (similar to REBELS). At the bright end, the step-wise estimate is significantly boosted by one particular galaxy that is very faint in UV luminosity, but \CII luminous. Its faint UV luminosity results in a very high $\mathrm{f_{UV}}$ factor, thus boosting the brightest LF point. Nevertheless, we consider our $z\sim5$ Schechter function derivation as more representative of the ALPINE targeted sample, indicating a much more gradual evolution from $z\sim7$ to $z\sim5$ and consistent with that seen in the UV LF.

\begin{figure}
\centering
\includegraphics[width=\linewidth]{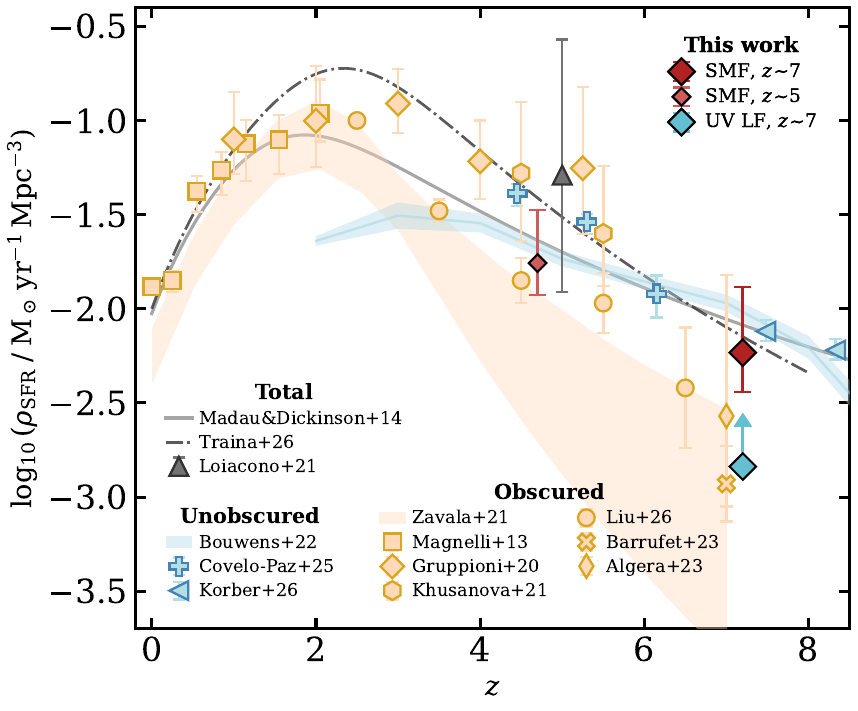}
\caption{Cosmic star formation rate density as a function of redshift. Literature determinations are grouped by the star-formation
tracer they probe --- total (grey), unobscured (blue), and dust-obscured
(tan) --- while our \CII-based results are shown as diamonds following the colour scheme of the previous figures. Our points at $z\sim7$ are plotted with a slight offset for clarity. The UV-LF-based determination at $z\sim7$
(sky blue diamond) is plotted as a lower limit (upward arrow), since UV-based determinations fail to account for the contribution from  UV-faint galaxies (see Sect.~\ref{sec:sfrd}). Our \CII-derived values are obtained by integrating the corresponding \CII LF down to $\log(L_{\rm\CII}/L_\odot)=8.1$ ($\mathrm{SFR}\approx11\,M_\odot\,
\mathrm{yr^{-1}}$ using the \citet{DeLooze14} calibration). Total determinations (grey) comprise the parametric fits of \citet{madaudickinson14} (solid) and \citet{traina26} (dash-dotted), and the \CII-LF-based estimate from the ALPINE field sample (\citealt{Loiacono20}, triangle). Unobscured
tracers (blue) comprise the rest-UV SFRD of \citet{bouwens22} (shaded band) and the nebular-line determinations from
H$\alpha$ (\citealt{covelopaz25}) and \OIII{}+H$\beta$ (\citealt{Korber26}, GLIMPSE).  Like UV-selected samples, these recover star formation only in line-detected galaxies and miss heavily dust-obscured sources. Dust-obscured, IR-based determinations (tan) comprise \citet{zavala21} (shaded band), \citet{magnelli13},
\citet{Gruppioni20}, \citet{khusanova21}, \citet{Liu26}, \citet{Barrufet23}, and \citet{Algera23}. The convergence of our
UV-LF-based lower limit with the obscured-only determinations at the bright end, together with the $\sim0.5$--$0.7$\,dex offset from our fiducial SMF-based point, illustrates the contribution UV-faint,
\CII-bright systems make to the total SFRD at $z\sim7$.}
\label{fig:sfrd}
\end{figure}

Essentially, the shape of our \CII LF is very similar to that of the $z\sim0$ \CII LF from \citet{Hemmati17} and shows only mild evolution to $z\sim5$. This is not especially surprising given that \CII is a SFR-tracer and that we observe similar values for the SFRD at $z\sim0$ and at $z=5-7$ (see Sect. \ref{sec:sfrd} and Fig. \ref{fig:sfrd}).

\subsection{Comparison with Semi-Analytical Models and Simulations}

Fig. \ref{fig:CiiLF} also shows recent \CII LF predictions from different theoretical studies at $z\sim7$. These are all in reasonable agreement with the UV-based methods but in tension with our SMF-based procedure. Since we use a sample of galaxies at $z\sim5-7$ to calibrate the  empirical relations, we also compute a $z\sim6$ \CII LF using the SMF at the same redshift from \citet{Weibel24}. A more direct comparison can be made with other theoretical works at $z\sim6$ shown in appendix \ref{app:z6comp}.

Firstly, \citet{Lagache18} use a semi-analytical model of galaxy evolution combined with the photoionisation code CLOUDY to predict the \CII luminosity of galaxies at $z\sim6.7$. The resulting \CII LF has a relatively shallow slope, but it agrees with our step-wise results. While its normalisation is below our lower limit, the bright-end seems in agreement with our determinations from both UV- and SMF-based methods.

We show the model curves from \citet{Bethermin22} estimated at $z\sim7$, based on an updated version of the SIDES simulation\footnote{\url{https://cesamsi.lam.fr/instance/sides/home}}. The simulation predicts the \CII LF based on two different conversions between star-formation rate (SFR) and \CII luminosity. In the first approach, the `standard' relation from \citet{DeLooze14} is used, which has also been shown to hold at higher redshifts \citep[e.g.][]{Schaerer20}. The second approach uses the relation from \citet{Lagache18}, which was calibrated with semi-analytical models and includes a redshift evolution in the conversion to account for lower metallicities. This predicts lower \CII emission, resulting in a lower LF, as can also be seen in Fig. \ref{fig:CiiLF}. 

This comparison with the SIDES models is particularly interesting.  
All results from the SIDES simulations either are in agreement or lie below our lower limit to the \CII LF (from the UV-based method) at the faint end. The bright end of all simulations show better agreement with our results. We note, however, that the simulation using the \citet{DeLooze14} relation has overall a more similar normalisation across $L_\mathrm{\CII}$ and redshift evolution as our results. This is consistent with the limited evolution seen in this conversion factor from the local Universe out to $z\sim7$ (\citealt{Schaerer20,Matthee19}: see also \SchouwsPrep). Regarding the evolution from the presented $z\sim6$ result in \citet{Bethermin22} to $z\sim7$, both uses of the SFR-L$_\mathrm{\CII}$ relations from \citet{DeLooze14} and \citet{Lagache18} relations show a $0.5$ dex decrease from $z\sim6$ to $z\sim7$. A similar shift is seen using our method (see App. \ref{app:z6comp}).
Interestingly, this model also predicts an excess of objects with $L_\mathrm{\CII}>10^{9.5}\,$L$_\odot$, in tension with our fiducial determination. As discussed earlier, such luminous sources may still exist among UV-fainter galaxies missed by the current selection.

Additionally, we consider the results from \citet{kathri25} obtained from post-processing galaxies from the MARIGOLD cosmological simulation at redshifts $3<z<7$. Their double power law relation sits well below our determination but nevertheless with a similar steepness on the faint-end slope.

We also consider comparisons against analytical model results of \citet{Liu24} which accounts for bursty star formation due to delayed supernova feedback. In this analytical model, $L_{\rm \CII}$ scales with the gas mass and metallicity of each halo, and star formation bursts are driven by delayed supernova feedback, whose amplitude is set by the cloud-scale star formation efficiency $\epsilon_{\rm ff}$. We show their fiducial model ($\epsilon_{\rm ff}=0.015$) at $z=7.5$, the redshift closest to our sample. Similarly to other works, this model lies below our determination, however their $z\sim6$ models are in good agreement with our results (see App. \ref{app:z6comp}). The discrepancy seen here can be attributed to redshift evolution between $z\sim7.5$ and our mean sample redshift of $z\sim7$. We also note that their model results adopting instead $\epsilon_{\rm ff}=0.005$ and $0.045$ differ from their $\epsilon_{\rm ff}=0.015$ results by only $\sim0.2$--$0.4$ dex over $8.5<\log(L_{\rm \CII}/L_\odot)<9.5$, which is less than the offset between our UV- and SMF-based determinations. Also, the effects of burstiness itself are concentrated at $\log(L_{\rm \CII}/L_\odot)\lesssim8$, which is below our completeness limit and we therefore cannot discriminate between them. Nevertheless, all three model variants they present predict a similarly steep faint end slope, consistent with our $\alpha_{\rm \CII}<-2$. Our data therefore cannot constrain the burstiness of star formation at $z\sim7$, which will require deeper blind surveys or line intensity mapping.

Finally, we compare to the \texttt{Delphi} semi-analytical model \citep{dayal2014a, Dayal22}, the latest version of which incorporates small-scale ISM physics calibrated on the \texttt{Sphinx} simulations \citep{mauerhofer25}. \texttt{Delphi} uses a binary merger-tree approach to jointly track the assembly of dark matter halos and their gas, stellar, metal, and dust content, tuned with only two free parameters to reproduce the observed UV LF at high-redshifts with JWST. \CII luminosities are then assigned to each galaxy using the SFR- and metallicity-dependent relation of \citet[Eq. 8]{Vallini15} and that of \citet{Romano22}. The \CII LFs presented here are evaluated at $z\sim6.4-7.7$. Of all the models considered here, \texttt{Delphi} most closely reproduces our fiducial (SMF-based) \CII LF, following it within the $1\sigma$ uncertainties in both shape and normalisation across the observed luminosity range.

\subsection{Implications for the SFRD}
\label{sec:sfrd}
We make use of our fiducial \CII LF to estimate the contribution from massive REBELS-like galaxies to the SFRD at $z\sim7$. The SFRD is derived by integrating our \CII luminosity function and applying the empirical \CII–SFR calibration of \citet{DeLooze14} as follows: 
\begin{equation}
\label{eq:sfrd_cii}
\log\!\left(\frac{\rho_{\rm SFR}}{\mathrm{M_\odot\,yr^{-1}\,Mpc^{-3}}}\right) =
-7.06 + \log\!\left(\frac{\rho_{\rm \CII}}{\mathrm{L_\odot\,Mpc^{-3}}}\right).
\end{equation}
Uncertainties are propagated via Monte Carlo by drawing $10^4$ samples from the MCMC posterior over $(\log\phi^{}, \log L^{}, \alpha)$, from which we report the median and 16th/84th percentile interval of $\log(\rho_{\rm SFR})$.
In our case, $\alpha < -2$ would make the integral of the Schechter function diverge as $L_{min} \to 0$ rendering the choice of $L_{min}$ crucial. Given this and our uncertainty on the faint-end slope, we integrate down to $\log L_{min}/L_\odot = 8.1$ corresponding to the faintest \CII detection in REBELS to avoid extrapolating the results below that directly observed with the probed sample. This corresponds to integrating down to a SFR of 11\,M$_\odot$\,yr$^{-1}$.

We obtain $\log(\rho_{\rm SFR}/M_\odot \mathrm{yr}^{-1} \mathrm{Mpc}^{-3}) =-2.23^{+0.35}_{-0.21}$ for the SMF-based method and $\log\rho_{\rm SFR}/M_\odot \mathrm{yr}^{-1} \mathrm{Mpc}^{-3} = -2.84^{+0.21}_{-0.17}$ for the UV-based one. A direct comparison to other works is not meaningful, as they often use different integration limits depending on their SFR tracers and completeness. Nevertheless, the values we infer can be interpreted as the contribution from REBELS-like sources to the SFRD, as a fractional contribution to the total. Relative to the \citet{madaudickinson14} fit at $z\sim7$, our SMF-based value implies that massive, \CII-bright galaxies comprise ${\sim}63\%$ of the total SFRD at this epoch, with a comparable fraction obtained against the \citet{traina26} parametrisation, which sits at a similar level at $z\sim7$. As an indication, we also integrate our \CII LF to a lower limit adopting the \citet{madaudickinson14} convention, integrating here to $0.03L_\mathrm{\CII}^{*}$, corresponding to $\log L_{min}/L_\odot =7.6$. We obtain a value of $\log\rho_{\rm SFR}/M_\odot \mathrm{yr}^{-1} \mathrm{Mpc}^{-3}=-1.87^{+0.40}_{-0.25}$, which is consistent within errors to \citet{madaudickinson14}. As the \CII LF suggests, $\rho_{\rm SFR}$ at $z\sim7$ is similar to that of the local universe at $z\sim0$ \citep{Hemmati17}.

Other works have estimated the obscured contribution to the SFRD from IR tracers \citep{magnelli13,zavala21,Gruppioni20,khusanova21, Liu26}, showing that the SFRD is dominated by obscured star formation up to $z\sim5$-7. More recently, other works based on REBELS have estimated that obscured SF accounts for 30\% of the cosmic SFRD at $z\sim7$ \citep[][]{Algera23, Barrufet23}. These works focus on the obscured SF occurring inside UV-selected sources and dark companions. In our case, the ratio between the SFRD from the SMF-based and UV-based methods does not correspond to the obscured fraction, but can instead be interpreted as the fraction of SFRD hosted by UV-faint systems missed by ALPINE and REBELS-like selections, hence does not include the SF hosted by UV-bright galaxies themselves. By construction, that contribution comes from \CII bright and therefore star-forming sources which are massive but UV-faint, hence heavily dust-obscured. Their contribution should be quantified and added to the estimation from IR measurements to completely account for the SFRD at high redshift. The tension between different IR censuses underscores a remaining systematic uncertainty, however all approaches agree that the obscured/UV-faint contribution to the SFRD is still significant at $z\sim7$.

\subsection{Possible explanation for discrepancies}
\label{Sec:discussion}
In summary, our $z\sim7$ \CII luminosity function lies in agreement with the ASPECS blind upper limits and aligns with the $z\sim0$ relation, suggesting a moderate evolution in the number density of \CII emitters with cosmic time. The UV-based result using the UV LF from \citet{Bouwens21a} and hence the REBELS sample likely underestimates the true abundance of \CII emitters, missing the contribution from UV-faint and dust-obscured galaxies. In contrast, our SMF–based determination, which traces the underlying stellar mass distribution, yields higher number densities and aligns well with the ALPINE targeted sample at $z\sim5$, indicating a smooth evolution from $z\sim7$ to $z\sim5$ of $\sim+0.3$ dex in $\phi^{*}_{\rm \CII}$, and to an overall $\sim0.5$ dex increase in number density at $\log(L_{\rm \CII}/L_{\odot}) > 8.5$. While the "total" ALPINE fit of \citet{Yan20} likely overestimates the bright-end of the \CII LF, it nevertheless highlights a physically important component: the contribution of companion galaxies detected serendipitously in the ALPINE pointings, which were not part of the UV-selected target list and would largely
be missed by UV selection alone. These comparisons therefore suggest that the true \CII LF likely lies above the UV–based estimate, with dusty, massive, and companion sources providing a non-negligible fraction of the \CII luminosity density at early epochs. Our SMF-based determination is designed to recover this missing population, and we adopt it as our fiducial estimate of the intrinsic \CII LF at $z\sim7$.

The \CII LFs from all models are in qualitative agreement with our determinations. Interestingly, while no model sits at the same normalisation as our fiducial result, they are all broadly consistent with a steep faint-end slope. Most importantly, these other results do not suggest that we are missing a large number of extremely luminous \CII emitters at $z\sim7$ in the REBELS sample (except for the result from \citet{yue&ferrara19} and the SIDES simulation, see App. \ref{app:z6comp} at $z\sim6$). However, all models underestimate the number densities of \CII emitters overall and are in better agreement with or below our UV LF based determination. Our SMF based \CII LF suggests that a non-negligible fraction of $z\sim7$ \CII emitters are dusty, UV-faint galaxies, and that these sources account for the difference between the UV-based and SMF-based \CII LFs shown in Fig. \ref{fig:methodsplot}.

\begin{figure*}
\centering
\includegraphics[width=0.75\linewidth]{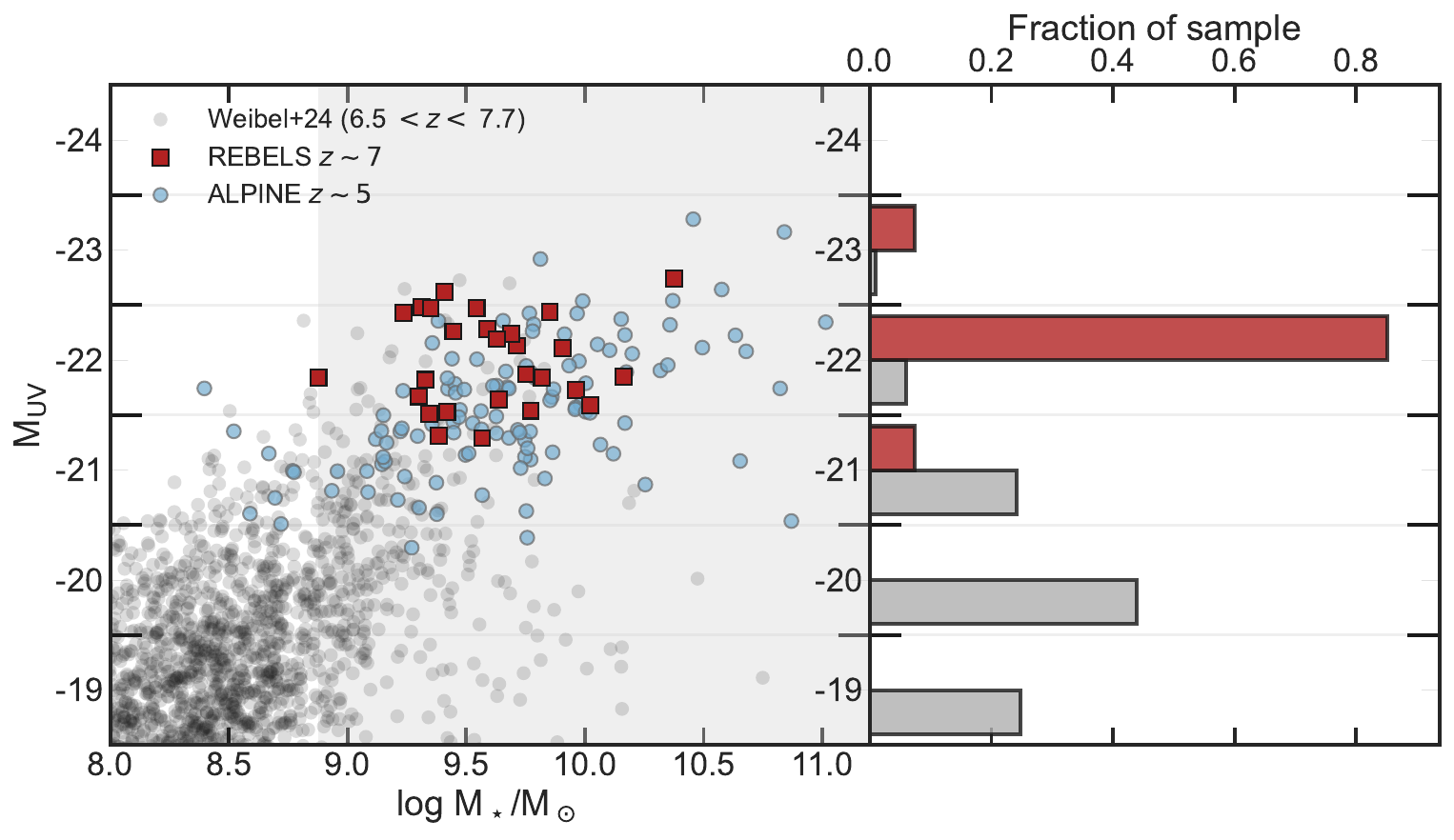}
\caption{Distribution of sources in the $M_{UV}-M_{\star}$ plane highlighting the differences in the samples used in this work. The SMF used for our empirical method is from \citet{Weibel24}, we plot the sources from their catalogue, within the redshift range of REBELS sources ($6.5<z<7.7$), as grey dots. REBELS sources are shown as red squares and ALPINE detections are also shown for reference as blue dots. This plot highlights the bias of both latter samples towards massive and UV-bright galaxies. The mass range probed by REBELS is shown the grey shaded area. Fractions of the total sample in each $M_{UV}$ bin in this mass range are shown on the right. REBELS sources and sources from \citet{Weibel24} are shown in red and grey, respectively, in $\Delta M_{UV} = 1$-mag bins of UV luminosity.}
\label{fig:Distrib}
\end{figure*}

Furthermore, as shown in Fig. \ref{fig:Distrib}, a large proportion of massive ($M_{\star}\gtrsim10^9M_\odot$) galaxies are still UV-faint ($M_{UV}>-21$), as per \citet{Weibel24}'s catalogue of >30,000 sources across the main JWST fields. Starting from the \citet{Weibel24}'s catalogue, we can also apply the $M_\mathrm{\star}-L_\mathrm{\CII}$ relation we derive in section \ref{sec:MethodSMF} and take the REBELS survey as a point of comparison. In REBELS redshift range of $6.5<z<7.7$, there are 159 galaxies with $\log{L_\mathrm{\CII}/L_\odot}>8.12$, i.e. the faintest \CII detected with REBELS. However, only 17 of those sources have $M_{UV}<-21.3$, i.e. the lower range of REBELS' absolute UV magnitudes. Under these assumptions, this implies that up to $\sim90\%$ of bright \CII sources could be missed by a UV-selected survey like REBELS. The differences in numbers of sources observed here are consistent with the 0.5-1.0 dex difference between our UV LF-based and SMF-based results across \CII luminosities above $\log(L_\mathrm{\CII}/L_\odot)>8.0$.

Recent JWST results have confirmed that a substantial fraction of the massive galaxy population at $z>6$ is heavily dust-obscured and therefore under-represented in UV-selected samples \citep[e.g.,][]{barrufet23b, labbe23b}, representing as much as $\sim45\%$ at $z\sim6-8$ of the integrated stellar mass density at $\log(M_\star/M_\odot) \gtrsim 9.25$ \citep{Gottumukkala24}. In addition, lower mass systems with high SFRs can also populate high $L_\mathrm{\CII}$ bins. Overall, these findings indicate that a large proportion of \CII emitters at high redshift likely remains undetected even in current deep ALMA follow-ups, highlighting the need for future wide-area, blind \CII surveys.

\section{Summary and Conclusions}
\label{sec:Conclusion}

In this paper, we used the REBELS ALMA survey and its pilot programs to derive, for the first time, the \CIImu line LF at $z\sim7$. We leverage the 27 galaxies with a detected ($>4\sigma$) \CII emission line in the redshift range $z=6.4-7.7$. Because the REBELS sample relies on a UV-selection to identify luminous \CII-emitting galaxies rather than making these identifications over a fixed volume, this complicates the determination of the \CII LF.

We employ three complementary approaches to derive the \CII LF using the UV LF and SMF at $z\sim7$ as proxies. 
The first method relies on the empirical $L_{UV} - L_\mathrm{\CII}$ relation derived here from the combined ALPINE and REBELS sample to transform the UV LF from \citet{Bouwens21a} into a \CII LF. The second method is similar but uses an 
empirical $M_\star-L_\mathrm{\CII}$ relation to transform the SMF from \citet{Weibel24} to a \CII LF.  The empirical $M_\star-L_\mathrm{\CII}$ relation exhibits a significantly tighter correlation (Spearman coefficient of 0.56 versus 0.28 for the $L_{UV} - L_\mathrm{\CII}$). The third method derives a step-wise LF result by following \citet{Yan20}, using completeness corrections based on the UV LF. 

At the bright end $(\log L_\mathrm{\CII}/L_\odot > 9)$, all three methods are in good agreement, yielding a characteristic luminosity of  $\log L_\mathrm{\CII}/L_\odot\simeq9.1$. At intermediate luminosities ($8.5 < \log L_\mathrm{\CII}/L_\odot < 9$), the SMF-based determination lies $\sim0.5$ dex above the UV LF-based methods. This difference grows close to 1 dex at the faint end, where the step-wise estimate also fails, becoming increasingly incomplete due to the initial UV selection of the target sample and the scatter in the $L_{UV}$-$L_{\CII}$ relation. 
We interpret this offset as representing a population of dust-obscured sources, systematically missed by UV-selected samples, that nonetheless have significant \CII luminosities. This population is not captured well by UV LF-based method, so we treat our UV LF based result as a lower limit on the true \CII LF at $z\sim7$. We adopt the SMF-based method as our fiducial determination because stellar mass exhibits a tighter correlation with \CII luminosity than UV luminosity according to our empirical relations, and because the SMF is less biased against dust-obscured systems.

Comparing our $z\sim7$ results with the ALPINE survey at $z\sim5$, we find a moderate evolution in the normalisation of $\sim0.3$ dex over $\sim400$ Myr. Importantly, when we apply the same SMF-based methodology to derive a $z\sim5$ \CII LF using the \citet{Weibel24} SMF, we obtain good agreement with the ALPINE targeted sample, validating our approach. This smooth evolution is consistent with the build-up of stellar mass and star formation rate density over this redshift range \citep[e.g.][]{Bouwens2015,Finkelstein2015,Bowler2020,Bouwens21a,Stefanon21}, and contrasts with the steeper evolution that would be inferred from the serendipitous ALPINE sample. Intriguingly, our $z\sim7$ determination is in excellent agreement with the $z\sim0$ \CII LF from \citet{Hemmati17}, which shows a remarkable consistency with the build-up of the cosmic star formation rate density over the same epoch.

Theoretical predictions are in qualitative agreement with our measurements.
The \texttt{Delphi} model is in very good agreement with our result at $z\sim7$ with either the \citet{Vallini15} or \citet{Romano22} relations. The SIDES simulation using the \citet{DeLooze14} SFR--$L_\mathrm{\CII}$ relation also shows a similar evolution from $z=6$ to $z=7$, and the existence of populations of similarly \CII bright sources at these redshifts.

However, all models systematically under-predict the number densities of \CII emitters relative to our fiducial SMF-based results, sometimes compensating by predicting the existence of populations of extremely bright \CII-emitters.
Furthermore, the existence of at least one UV-faint galaxy with $\log L_\mathrm{\CII}/L_\odot>9.5$ at $z\sim4-6$ in the ALPINE galaxy sample and the serendipitous detection of dusty sources at $z\sim7$ from REBELS \citep{Fudamoto21,vanleeuwen24} show that a UV-preselected galaxy sample almost certainly underestimates the \CII LF to some degree. Using the \citet{Weibel24} catalogue and our $M_\star - L_\mathrm{\CII}$ relation, we estimate that up to $\sim90\%$ of galaxies with log $L_\mathrm{\CII}/L_\odot > 8.12$ (the faintest REBELS detection) have $M_{UV} > -21.3$ and would thus be excluded from UV-bright samples like REBELS.

Assuming our \CII LF derived from the SMF-based method captures the totality of \CII-bright emitters, the difference between the UV-based and SMF-based methods suggests that 60\% to 70\% of sources with $\log(L_\mathrm{\CII}/L_\odot)  >8.5$ might be missed by focusing on most UV luminous galaxies and more than >80\% at lower luminosities. Thus large galaxy samples, using unbiased line scans over wider areas and also probing the immediate neighbourhoods around massive galaxies \citep[e.g][]{decarli2017,Fudamoto21,vanleeuwen24}, will be needed to estimate the required completeness corrections.

Furthermore, when deriving the SFRD implied by our \CII LF using the \citet{DeLooze14} calibration, we find that ${\sim}63\%$ of the cosmic SFRD at $z\sim7$ \citep{madaudickinson14} appears to originate from massive, \CII-bright galaxies.  Because \CII traces the total (obscured plus unobscured) star formation of the galaxies it samples, the $\sim0.6$ dex offset between our SMF- and UV-based SFRDs shows the impact that UV selection can have on the inferred properties of massive galaxies at these redshifts.  This population overlaps strongly with, but is not identical to, the dust-obscured population, and the derived offset here is our best estimate as to the likely lower limit on its contribution. The implied missed fraction sits at the upper end of current estimates underscoring the need to quantify the contribution of UV-faint, obscured galaxies to the SFRD at $z\sim7$.

Our results provide critical constraints for upcoming \CII intensity mapping experiments including TIME \citep{Sun21}, EoR-Spec \citep{Cothard20}, and CONCERTO \citep{Bethermin22}, which probe the aggregate \CII emission during reionization. Specifically, the predicted \CII power spectra expected for these surveys vary significantly depending on the shape of the \CII LF and its evolution at $z>6$. Additionally, bursty star formation is expected to boost the faint end of the \CII LF and distort the \CII power spectrum \citep{Liu24}, precisely in the regime below the reach of current targeted surveys. To fully characterize the \CII-emitting population at $z\sim7$—including UV-faint and companion sources—future surveys will require deeper wide-area blind \CII line scans. Such observations will be essential to determine whether the discrepancy between UV- and SMF-based estimates reflects a genuine population of dust-obscured \CII emitters or systematic uncertainties in various empirical scaling relations.

\section*{Acknowledgements}

The authors want to thank all the team members of ASPECS and ALPINE for many stimulating discussions regarding the \CII LF and \CII measurements of distant galaxies in general. Special thanks go to Bade Uzgil and Fabian Walter, as well as Lin Yan, Michele Ginolfi, Daniel Schaerer, and Matthieu Bethermin.  

This paper makes use of the following ALMA data: ADS/JAO.ALMA\#2019.1.01634.L. ALMA is a partnership of ESO (representing its member states), NSF (USA) and NINS (Japan), together with NRC (Canada), MOST and ASIAA (Taiwan), and KASI (Republic of Korea), in cooperation with the Republic of Chile. The Joint ALMA Observatory is operated by ESO, AUI/NRAO and NAOJ.
This research further made use of Astropy,\footnote{\url{http://www.astropy.org}} a community-developed core Python package for Astronomy \citep{astropy:2013, astropy:2018}, in addition to other freely available software\footnote{\url{https://github.com/jmeyers314/linmix}}.
This project made use of the pySIDES library\footnote{\url{https://cesamsi.lam.fr/instance/sides/home}} \citep[see][]{Bethermin22}, funded from the European Research Council (ERC) under the European Union’s Horizon 2020 research and innovation programme (grant agreement No 788212), and with support from the Excellence Initiative of Aix-Marseille University-AMidex, a French "Investissements d’Avenir.

P.A.O. acknowledges support from the Swiss National Science Foundation grants 190079 and 207349.
The Cosmic Dawn Center (DAWN) is funded by the Danish National Research Foundation under grant No. 140.
H.S.B.A. gratefully acknowledges support from Academia Sinica through grant AS-PD-1141-M01-2.
M.A. is supported by FONDECYT grant number 1252054, and gratefully acknowledges support from ANID Basal Project FB210003,  ANID MILENIO NCN2024\_112 and ANID + Vinculaci\'on Internacional + FOVI250261.
A.F. and A.P. acknowledge support from the ERC Advanced Grant INTERSTELLAR H2020/740120. Generous support from the Carl Friedrich von Siemens-Forschungspreis der Alexander von Humboldt-Stiftung Research Award is kindly acknowledged (A.F.). 
Y.F. acknowledges support from NAOJ ALMA Scientific Research Grant number 2020-16B.
MS acknowledges support from the European Research Commission Consolidator Grant 101088789 (SFEER), from the CIDEGENT/2021/059 grant by Generalitat Valenciana, and from project PID2023-149420NB-I00 funded by MICIU/AEI/10.13039/501100011033 and by ERDF/EU.
P.D. acknowledges support from the European Research Council's starting grant ERC StG-717001 (``DELPHI"), from the NWO grant 016.VIDI.189.162 (``ODIN") and the European Commission's and University of Groningen's CO-FUND Rosalind Franklin program.
H.I. acknowledges support from NAOJ ALMA Scientific Research Grant Code 2021-19A. H.I. acknowledges support from JSPS KAKENHI Grant Number JP19K23462.
R.S. acknowledges support from a STFC Ernest Rutherford Fellowship (ST/S004831/1).

\section*{Data Availability}

The data for this paper will be made available upon reasonable request to the first author. The underlying ALMA raw data are publicly available in the science archive and will be made available in reduced form by the REBELS collaboration.



\bibliographystyle{mnras}
\bibliography{MasterBiblio} 

@ARTICLE{vanleeuwen24,
       author = {{van Leeuwen}, I.~F. and {Bouwens}, R.~J. and {van der Werf}, P.~P. and {Hodge}, J.~A. and {Schouws}, S. and {Stefanon}, M. and {Algera}, H.~S.~B. and {Aravena}, M. and {Boogaard}, L.~A. and {Bowler}, R.~A.~A. and {da Cunha}, E. and {Dayal}, P. and {Decarli}, R. and {Gonzalez}, V. and {Inami}, H. and {de Looze}, I. and {Sommovigo}, L. and {Venemans}, B.~P. and {Walter}, F. and {Barrufet}, L. and {Ferrara}, A. and {Graziani}, L. and {Hygate}, A.~P.~S. and {Oesch}, P. and {Palla}, M. and {Rowland}, L. and {Schneider}, R.},
        title = "{Characterising the contribution of dust-obscured star formation at z {\ensuremath{\gtrsim}} 5 using 18 serendipitously identified [C II] emitters}",
      journal = {\mnras},
         year = 2024,
        month = nov,
       volume = {534},
       number = {3},
        pages = {2062-2085},
          doi = {10.1093/mnras/stae2171},
archivePrefix = {arXiv},
       eprint = {2409.11463},
 primaryClass = {astro-ph.GA},
       adsurl = {https://ui.adsabs.harvard.edu/abs/2024MNRAS.534.2062V}
}

@article{Bouwens2026,
  author  = {Rychard Bouwens},
  title   = {How JWST is Revolutionizing Our View of Early Galaxy Formation},
  journal = {Contemporary Physics},
  year    = {2026},
  note    = {Submitted},
}

@ARTICLE{Curti2023,
       author = {{Curti}, Mirko and {D'Eugenio}, Francesco and {Carniani}, Stefano and {Maiolino}, Roberto and {Sandles}, Lester and \
{Witstok}, Joris and {Baker}, William M. and {Bennett}, Jake S. and {Piotrowska}, Joanna M. and {Tacchella}, Sandro and {Charlot}, St\
ephane and {Nakajima}, Kimihiko and {Maheson}, Gabriel and {Mannucci}, Filippo and {Amiri}, Amirnezam and {Arribas}, Santiago and {Be\
lfiore}, Francesco and {Bonaventura}, Nina R. and {Bunker}, Andrew J. and {Chevallard}, Jacopo and {Cresci}, Giovanni and {Curtis-Lak\
e}, Emma and {Hayden-Pawson}, Connor and {Jones}, Gareth C. and {Kumari}, Nimisha and {Laseter}, Isaac and {Looser}, Tobias J. and {M\
arconi}, Alessandro and {Maseda}, Michael V. and {Scholtz}, Jan and {Smit}, Renske and {{\"U}bler}, Hannah and {Wallace}, Imaan E.~B.\
},
        title = "{The chemical enrichment in the early Universe as probed by JWST via direct metallicity measurements at z {\ensurema\
th{\sim}} 8}",
      journal = {\mnras},
         year = 2023,
        month = jan,
       volume = {518},
       number = {1},
        pages = {425-438},
          doi = {10.1093/mnras/stac2737},
archivePrefix = {arXiv},
       eprint = {2207.12375},
 primaryClass = {astro-ph.GA},
       adsurl = {https://ui.adsabs.harvard.edu/abs/2023MNRAS.518..425C}
}

@ARTICLE{Cameron2023,
       author = {{Cameron}, Alex J. and {Katz}, Harley and {Rey}, Martin P. and {Saxena}, Aayush},
        title = "{Nitrogen enhancements 440 Myr after the big bang: supersolar N/O, a tidal disruption event, or a dense stellar clus\
ter in GN-z11?}",
      journal = {\mnras},
         year = 2023,
        month = aug,
       volume = {523},
       number = {3},
        pages = {3516-3525},
          doi = {10.1093/mnras/stad1579},
archivePrefix = {arXiv},
       eprint = {2302.10142},
 primaryClass = {astro-ph.GA},
       adsurl = {https://ui.adsabs.harvard.edu/abs/2023MNRAS.523.3516C}
}

@ARTICLE{Heintz2024_dla,
       author = {{Heintz}, Kasper E. and {Watson}, Darach and {Brammer}, Gabriel and {Vejlgaard}, Simone and {Hutter}, Anne and {Stra\
it}, Victoria B. and {Matthee}, Jorryt and {Oesch}, Pascal A. and {Jakobsson}, P{\'a}ll and {Tanvir}, Nial R. and {Laursen}, Peter an\
d {Naidu}, Rohan P. and {Mason}, Charlotte A. and {Killi}, Meghana and {Jung}, Intae and {Hsiao}, Tiger Yu-Yang and {Abdurro'uf} and \
{Coe}, Dan and {Arrabal Haro}, Pablo and {Finkelstein}, Steven L. and {Toft}, Sune},
        title = "{Strong damped Lyman-{\ensuremath{\alpha}} absorption in young star-forming galaxies at redshifts 9 to 11}",
      journal = {Science},
         year = 2024,
        month = may,
       volume = {384},
       number = {6698},
        pages = {890-894},
          doi = {10.1126/science.adj0343},
archivePrefix = {arXiv},
       eprint = {2306.00647},
 primaryClass = {astro-ph.GA},
       adsurl = {https://ui.adsabs.harvard.edu/abs/2024Sci...384..890H}
}

@ARTICLE{Sanders2024,
       author = {{Sanders}, Ryan L. and {Shapley}, Alice E. and {Topping}, Michael W. and {Reddy}, Naveen A. and {Brammer}, Gabriel B\
.},
        title = "{Direct T $_{e}$-based Metallicities of z = 2─9 Galaxies with JWST/NIRSpec: Empirical Metallicity Calibrations Appli\
cable from Reionization to Cosmic Noon}",
      journal = {\apj},
         year = 2024,
        month = feb,
       volume = {962},
       number = {1},
          eid = {24},
        pages = {24},
          doi = {10.3847/1538-4357/ad15fc},
archivePrefix = {arXiv},
       eprint = {2303.08149},
 primaryClass = {astro-ph.GA},
       adsurl = {https://ui.adsabs.harvard.edu/abs/2024ApJ...962...24S}
}

@ARTICLE{Matthee2025,
       author = {{Matthee}, Jorryt},
        title = "{JWST provides a new view of cosmic dawn: latest developments in studies of early galaxies}",
      journal = {Contemporary Physics},
         year = 2025,
        month = oct,
       volume = {66},
       number = {1-4},
        pages = {116-151},
          doi = {10.1080/00107514.2025.2586370},
archivePrefix = {arXiv},
       eprint = {2511.04843},
 primaryClass = {astro-ph.GA},
       adsurl = {https://ui.adsabs.harvard.edu/abs/2025ConPh..66..116M}
}

@INPROCEEDINGS{Stark2026,
       author = {{Stark}, Daniel P. and {Topping}, Michael W. and {Endsley}, Ryan and {Tang}, Mengtao},
        title = "{Observations of the first galaxies in the Era of JWST}",
    booktitle = {Encyclopedia of Astrophysics, Volume 4},
         year = 2026,
       volume = {4},
        month = jan,
        pages = {453-499},
          doi = {10.1016/B978-0-443-21439-4.00128-0},
archivePrefix = {arXiv},
       eprint = {2501.17078},
 primaryClass = {astro-ph.GA},
       adsurl = {https://ui.adsabs.harvard.edu/abs/2026enap....4..453S}
}

@ARTICLE{decarli2017,
       author = {{Decarli}, R. and {Walter}, F. and {Venemans}, B.~P. and {Ba{\~n}ados}, E. and {Bertoldi}, F. and {Carilli}, C. and {Fan}, X. and {Farina}, E.~P. and {Mazzucchelli}, C. and {Riechers}, D. and {Rix}, H.-W. and {Strauss}, M.~A. and {Wang}, R. and {Yang}, Y.},
        title = "{Rapidly star-forming galaxies adjacent to quasars at redshifts exceeding 6}",
      journal = {\nat},
         year = 2017,
        month = may,
       volume = {545},
       number = {7655},
        pages = {457-461},
          doi = {10.1038/nature22358},
archivePrefix = {arXiv},
       eprint = {1705.08662},
 primaryClass = {astro-ph.GA},
       adsurl = {https://ui.adsabs.harvard.edu/abs/2017Natur.545..457D}
}

@ARTICLE{Yan20,
       author = {{Yan}, Lin and {Sajina}, A. and {Loiacono}, F. and {Lagache}, G. and {B{\'e}thermin}, M. and {Faisst}, A. and {Ginolfi}, M. and {F{\`e}vre}, O. Le and {Gruppioni}, C. and {Capak}, P.~L. and {Cassata}, P. and {Schaerer}, D. and {Silverman}, J.~D. and {Bardelli}, S. and {Dessauges-Zavadsky}, M. and {Cimatti}, A. and {Hathi}, N.~P. and {Lemaux}, B.~C. and {Ibar}, E. and {Jones}, G.~C. and {Koekemoer}, Anton M. and {Oesch}, P.~A. and {Talia}, M. and {Pozzi}, F. and {Riechers}, D.~A. and {Tasca}, L.~A.~M. and {Toft}, Sune and {Vallini}, L. and {Vergani}, D. and {Zamorani}, G. and {Zucca}, E.},
        title = "{The ALPINE-ALMA [C II] Survey: [C II] 158 {\ensuremath{\mu}}m Emission Line Luminosity Functions at z {\ensuremath{\sim}} 4-6}",
      journal = {\apj},
         year = 2020,
        month = dec,
       volume = {905},
       number = {2},
          eid = {147},
        pages = {147},
          doi = {10.3847/1538-4357/abc41c},
archivePrefix = {arXiv},
       eprint = {2006.04835},
 primaryClass = {astro-ph.GA},
       adsurl = {https://ui.adsabs.harvard.edu/abs/2020ApJ...905..147Y}
}

@article{astropy:2013,
Adsurl = {http://adsabs.harvard.edu/abs/2013A%26A...558A..33A},
Archiveprefix = {arXiv},
Author = {{Astropy Collaboration} and {Robitaille}, T.~P. and {Tollerud}, E.~J. and {Greenfield}, P. and {Droettboom}, M. and {Bray}, E. and {Aldcroft}, T. and {Davis}, M. and {Ginsburg}, A. and {Price-Whelan}, A.~M. and {Kerzendorf}, W.~E. and {Conley}, A. and {Crighton}, N. and {Barbary}, K. and {Muna}, D. and {Ferguson}, H. and {Grollier}, F. and {Parikh}, M.~M. and {Nair}, P.~H. and {Unther}, H.~M. and {Deil}, C. and {Woillez}, J. and {Conseil}, S. and {Kramer}, R. and {Turner}, J.~E.~H. and {Singer}, L. and {Fox}, R. and {Weaver}, B.~A. and {Zabalza}, V. and {Edwards}, Z.~I. and {Azalee Bostroem}, K. and {Burke}, D.~J. and {Casey}, A.~R. and {Crawford}, S.~M. and {Dencheva}, N. and {Ely}, J. and {Jenness}, T. and {Labrie}, K. and {Lim}, P.~L. and {Pierfederici}, F. and {Pontzen}, A. and {Ptak}, A. and {Refsdal}, B. and {Servillat}, M. and {Streicher}, O.},
Doi = {10.1051/0004-6361/201322068},
Eid = {A33},
Eprint = {1307.6212},
Journal = {\aap},
Month = oct,
Pages = {A33},
Primaryclass = {astro-ph.IM},
Title = {{Astropy: A community Python package for astronomy}},
Volume = 558,
Year = 2013}

@ARTICLE{astropy:2018,
       author = {{Astropy Collaboration} and {Price-Whelan}, A.~M. and
         {Sip{\H{o}}cz}, B.~M. and {G{\"u}nther}, H.~M. and {Lim}, P.~L. and
         {Crawford}, S.~M. and {Conseil}, S. and {Shupe}, D.~L. and
         {Craig}, M.~W. and {Dencheva}, N. and {Ginsburg}, A. and {Vand
        erPlas}, J.~T. and {Bradley}, L.~D. and {P{\'e}rez-Su{\'a}rez}, D. and
         {de Val-Borro}, M. and {Aldcroft}, T.~L. and {Cruz}, K.~L. and
         {Robitaille}, T.~P. and {Tollerud}, E.~J. and {Ardelean}, C. and
         {Babej}, T. and {Bach}, Y.~P. and {Bachetti}, M. and {Bakanov}, A.~V. and
         {Bamford}, S.~P. and {Barentsen}, G. and {Barmby}, P. and
         {Baumbach}, A. and {Berry}, K.~L. and {Biscani}, F. and {Boquien}, M. and
         {Bostroem}, K.~A. and {Bouma}, L.~G. and {Brammer}, G.~B. and
         {Bray}, E.~M. and {Breytenbach}, H. and {Buddelmeijer}, H. and
         {Burke}, D.~J. and {Calderone}, G. and {Cano Rodr{\'\i}guez}, J.~L. and
         {Cara}, M. and {Cardoso}, J.~V.~M. and {Cheedella}, S. and {Copin}, Y. and
         {Corrales}, L. and {Crichton}, D. and {D'Avella}, D. and {Deil}, C. and
         {Depagne}, {\'E}. and {Dietrich}, J.~P. and {Donath}, A. and
         {Droettboom}, M. and {Earl}, N. and {Erben}, T. and {Fabbro}, S. and
         {Ferreira}, L.~A. and {Finethy}, T. and {Fox}, R.~T. and
         {Garrison}, L.~H. and {Gibbons}, S.~L.~J. and {Goldstein}, D.~A. and
         {Gommers}, R. and {Greco}, J.~P. and {Greenfield}, P. and
         {Groener}, A.~M. and {Grollier}, F. and {Hagen}, A. and {Hirst}, P. and
         {Homeier}, D. and {Horton}, A.~J. and {Hosseinzadeh}, G. and {Hu}, L. and
         {Hunkeler}, J.~S. and {Ivezi{\'c}}, {\v{Z}}. and {Jain}, A. and
         {Jenness}, T. and {Kanarek}, G. and {Kendrew}, S. and {Kern}, N.~S. and
         {Kerzendorf}, W.~E. and {Khvalko}, A. and {King}, J. and {Kirkby}, D. and
         {Kulkarni}, A.~M. and {Kumar}, A. and {Lee}, A. and {Lenz}, D. and
         {Littlefair}, S.~P. and {Ma}, Z. and {Macleod}, D.~M. and
         {Mastropietro}, M. and {McCully}, C. and {Montagnac}, S. and
         {Morris}, B.~M. and {Mueller}, M. and {Mumford}, S.~J. and {Muna}, D. and
         {Murphy}, N.~A. and {Nelson}, S. and {Nguyen}, G.~H. and
         {Ninan}, J.~P. and {N{\"o}the}, M. and {Ogaz}, S. and {Oh}, S. and
         {Parejko}, J.~K. and {Parley}, N. and {Pascual}, S. and {Patil}, R. and
         {Patil}, A.~A. and {Plunkett}, A.~L. and {Prochaska}, J.~X. and
         {Rastogi}, T. and {Reddy Janga}, V. and {Sabater}, J. and
         {Sakurikar}, P. and {Seifert}, M. and {Sherbert}, L.~E. and
         {Sherwood-Taylor}, H. and {Shih}, A.~Y. and {Sick}, J. and
         {Silbiger}, M.~T. and {Singanamalla}, S. and {Singer}, L.~P. and
         {Sladen}, P.~H. and {Sooley}, K.~A. and {Sornarajah}, S. and
         {Streicher}, O. and {Teuben}, P. and {Thomas}, S.~W. and
         {Tremblay}, G.~R. and {Turner}, J.~E.~H. and {Terr{\'o}n}, V. and
         {van Kerkwijk}, M.~H. and {de la Vega}, A. and {Watkins}, L.~L. and
         {Weaver}, B.~A. and {Whitmore}, J.~B. and {Woillez}, J. and
         {Zabalza}, V. and {Astropy Contributors}},
        title = "{The Astropy Project: Building an Open-science Project and Status of the v2.0 Core Package}",
      journal = {\aj},
         year = 2018,
        month = sep,
       volume = {156},
       number = {3},
          eid = {123},
        pages = {123},
          doi = {10.3847/1538-3881/aabc4f},
archivePrefix = {arXiv},
       eprint = {1801.02634},
 primaryClass = {astro-ph.IM},
       adsurl = {https://ui.adsabs.harvard.edu/abs/2018AJ....156..123A}
}

@ARTICLE{Kelly07,
       author = {{Kelly}, Brandon C.},
        title = "{Some Aspects of Measurement Error in Linear Regression of Astronomical Data}",
      journal = {\apj},
         year = 2007,
        month = aug,
       volume = {665},
       number = {2},
        pages = {1489-1506},
          doi = {10.1086/519947},
archivePrefix = {arXiv},
       eprint = {0705.2774},
 primaryClass = {astro-ph},
       adsurl = {https://ui.adsabs.harvard.edu/abs/2007ApJ...665.1489K}
}

@ARTICLE{Smit18,
       author = {{Smit}, Renske and {Bouwens}, Rychard J. and {Carniani}, Stefano and {Oesch}, Pascal A. and {Labb{\'e}}, Ivo and {Illingworth}, Garth D. and {van der Werf}, Paul and {Bradley}, Larry D. and {Gonzalez}, Valentino and {Hodge}, Jacqueline A. and {Holwerda}, Benne W. and {Maiolino}, Roberto and {Zheng}, Wei},
        title = "{Rotation in [C II]-emitting gas in two galaxies at a redshift of 6.8}",
      journal = {\nat},
         year = 2018,
        month = jan,
       volume = {553},
       number = {7687},
        pages = {178-181},
          doi = {10.1038/nature24631},
archivePrefix = {arXiv},
       eprint = {1706.04614},
 primaryClass = {astro-ph.GA},
       adsurl = {https://ui.adsabs.harvard.edu/abs/2018Natur.553..178S}
}

@ARTICLE{Schouws22a,
       author = {{Schouws}, Sander and {Stefanon}, Mauro and {Bouwens}, Rychard and {Smit}, Renske and {Hodge}, Jacqueline and {Labb{\'e}}, Ivo and {Algera}, Hiddo and {Boogaard}, Leindert and {Carniani}, Stefano and {Fudamoto}, Yoshinobu and {Holwerda}, Benne W. and {Illingworth}, Garth D. and {Maiolino}, Roberto and {Maseda}, Michael and {Oesch}, Pascal and {van der Werf}, Paul},
        title = "{Significant Dust-obscured Star Formation in Luminous Lyman-break Galaxies at z   7-8}",
      journal = {\apj},
         year = 2022,
        month = mar,
       volume = {928},
       number = {1},
          eid = {31},
        pages = {31},
          doi = {10.3847/1538-4357/ac4605},
archivePrefix = {arXiv},
       eprint = {2105.12133},
 primaryClass = {astro-ph.GA},
       adsurl = {https://ui.adsabs.harvard.edu/abs/2022ApJ...928...31S}
}

@ARTICLE{Schouws22b,
       author = {{Schouws}, Sander and {Bouwens}, Rychard and {Smit}, Renske and {Hodge}, Jacqueline and {Stefanon}, Mauro and {Witstok}, Joris and {Hilhorst}, Juli{\"e}tte and {Labb{\'e}}, Ivo and {Algera}, Hiddo and {Boogaard}, Leindert and {Maseda}, Michael and {Oesch}, Pascal and {R{\"o}ttgering}, Huub and {van der Werf}, Paul},
        title = "{ALMA as a Redshift Machine: Using [C II] to Efficiently Confirm Galaxies in the Epoch of Reionization}",
      journal = {\apj},
         year = 2023,
        month = sep,
       volume = {954},
       number = {1},
          eid = {103},
        pages = {103},
          doi = {10.3847/1538-4357/ace10c},
archivePrefix = {arXiv},
       eprint = {2202.04080},
 primaryClass = {astro-ph.GA},
       adsurl = {https://ui.adsabs.harvard.edu/abs/2023ApJ...954..103S}
}

@ARTICLE{Bethermin20,
       author = {{B{\'e}thermin}, M. and {Fudamoto}, Y. and {Ginolfi}, M. and {Loiacono}, F. and {Khusanova}, Y. and {Capak}, P.~L. and {Cassata}, P. and {Faisst}, A. and {Le F{\`e}vre}, O. and {Schaerer}, D. and {Silverman}, J.~D. and {Yan}, L. and {Amorin}, R. and {Bardelli}, S. and {Boquien}, M. and {Cimatti}, A. and {Davidzon}, I. and {Dessauges-Zavadsky}, M. and {Fujimoto}, S. and {Gruppioni}, C. and {Hathi}, N.~P. and {Ibar}, E. and {Jones}, G.~C. and {Koekemoer}, A.~M. and {Lagache}, G. and {Lemaux}, B.~C. and {Moreau}, C. and {Oesch}, P.~A. and {Pozzi}, F. and {Riechers}, D.~A. and {Talia}, M. and {Toft}, S. and {Vallini}, L. and {Vergani}, D. and {Zamorani}, G. and {Zucca}, E.},
        title = "{The ALPINE-ALMA [CII] survey: Data processing, catalogs, and statistical source properties}",
      journal = {\aap},
         year = 2020,
        month = nov,
       volume = {643},
          eid = {A2},
        pages = {A2},
          doi = {10.1051/0004-6361/202037649},
archivePrefix = {arXiv},
       eprint = {2002.00962},
 primaryClass = {astro-ph.GA},
       adsurl = {https://ui.adsabs.harvard.edu/abs/2020A&A...643A...2B}
}

@ARTICLE{LeFevre20,
       author = {{Le F{\`e}vre}, O. and {B{\'e}thermin}, M. and {Faisst}, A. and {Jones}, G.~C. and {Capak}, P. and {Cassata}, P. and {Silverman}, J.~D. and {Schaerer}, D. and {Yan}, L. and {Amorin}, R. and {Bardelli}, S. and {Boquien}, M. and {Cimatti}, A. and {Dessauges-Zavadsky}, M. and {Giavalisco}, M. and {Hathi}, N.~P. and {Fudamoto}, Y. and {Fujimoto}, S. and {Ginolfi}, M. and {Gruppioni}, C. and {Hemmati}, S. and {Ibar}, E. and {Koekemoer}, A. and {Khusanova}, Y. and {Lagache}, G. and {Lemaux}, B.~C. and {Loiacono}, F. and {Maiolino}, R. and {Mancini}, C. and {Narayanan}, D. and {Morselli}, L. and {M{\'e}ndez-Hern{\`a}ndez}, Hugo and {Oesch}, P.~A. and {Pozzi}, F. and {Romano}, M. and {Riechers}, D. and {Scoville}, N. and {Talia}, M. and {Tasca}, L.~A.~M. and {Thomas}, R. and {Toft}, S. and {Vallini}, L. and {Vergani}, D. and {Walter}, F. and {Zamorani}, G. and {Zucca}, E.},
        title = "{The ALPINE-ALMA [CII] survey. Survey strategy, observations, and sample properties of 118 star-forming galaxies at 4 < z < 6}",
      journal = {\aap},
         year = 2020,
        month = nov,
       volume = {643},
          eid = {A1},
        pages = {A1},
          doi = {10.1051/0004-6361/201936965},
archivePrefix = {arXiv},
       eprint = {1910.09517},
 primaryClass = {astro-ph.CO},
       adsurl = {https://ui.adsabs.harvard.edu/abs/2020A&A...643A...1L}
}

@ARTICLE{Uzgil21,
       author = {{Uzgil}, Bade D. and {Oesch}, Pascal A. and {Walter}, Fabian and {Aravena}, Manuel and {Boogaard}, Leindert and {Carilli}, Chris and {Decarli}, Roberto and {D{\'\i}az-Santos}, Tanio and {Fudamoto}, Yoshi and {Inami}, Hanae and {Bouwens}, Rychard and {Cortes}, Paulo C. and {Cox}, Pierre and {Daddi}, Emmanuele and {Gonz{\'a}lez-L{\'o}pez}, Jorge and {Labbe}, Ivo and {Popping}, Gerg{\"o} and {Riechers}, Dominik and {Stefanon}, Mauro and {Van der Werf}, Paul and {Weiss}, Axel},
        title = "{The ALMA Spectroscopic Survey in the HUDF: A Search for [C II] Emitters at 6 {\ensuremath{\leq}} z {\ensuremath{\leq}} 8}",
      journal = {\apj},
         year = 2021,
        month = may,
       volume = {912},
       number = {1},
          eid = {67},
        pages = {67},
          doi = {10.3847/1538-4357/abe86b},
archivePrefix = {arXiv},
       eprint = {2102.10706},
 primaryClass = {astro-ph.GA},
       adsurl = {https://ui.adsabs.harvard.edu/abs/2021ApJ...912...67U}
}

@ARTICLE{Stacey91,
       author = {{Stacey}, G.~J. and {Geis}, N. and {Genzel}, R. and {Lugten}, J.~B. and {Poglitsch}, A. and {Sternberg}, A. and {Townes}, C.~H.},
        title = "{The 158 Micron [C ii] Line: A Measure of Global Star Formation Activity in Galaxies}",
      journal = {\apj},
         year = 1991,
        month = jun,
       volume = {373},
        pages = {423},
          doi = {10.1086/170062},
       adsurl = {https://ui.adsabs.harvard.edu/abs/1991ApJ...373..423S}
}

@ARTICLE{Romano22,
       author = {{Romano}, M. and {Morselli}, L. and {Cassata}, P. and {Ginolfi}, M. and {Schaerer}, D. and {B{\'e}thermin}, M. and {Capak}, P. and {Faisst}, A. and {Le F{\`e}vre}, O. and {Silverman}, J.~D. and {Yan}, L. and {Bardelli}, S. and {Boquien}, M. and {Dessauges-Zavadsky}, M. and {Fujimoto}, S. and {Hathi}, N.~P. and {Jones}, G.~C. and {Koekemoer}, A.~M. and {Lemaux}, B.~C. and {M{\'e}ndez-Hern{\'a}ndez}, H. and {Narayanan}, D. and {Talia}, M. and {Vergani}, D. and {Zamorani}, G. and {Zucca}, E.},
        title = "{The ALPINE-ALMA [CII] survey: The population of [CII]-undetected galaxies and their role in the L$_{[CII]}$-SFR relation}",
      journal = {\aap},
         year = 2022,
        month = apr,
       volume = {660},
          eid = {A14},
        pages = {A14},
          doi = {10.1051/0004-6361/202142265},
archivePrefix = {arXiv},
       eprint = {2202.02334},
 primaryClass = {astro-ph.GA},
       adsurl = {https://ui.adsabs.harvard.edu/abs/2022A&A...660A..14R}
}

@ARTICLE{Schaerer20,
       author = {{Schaerer}, D. and {Ginolfi}, M. and {B{\'e}thermin}, M. and {Fudamoto}, Y. and {Oesch}, P.~A. and {Le F{\`e}vre}, O. and {Faisst}, A. and {Capak}, P. and {Cassata}, P. and {Silverman}, J.~D. and {Yan}, Lin and {Jones}, G.~C. and {Amorin}, R. and {Bardelli}, S. and {Boquien}, M. and {Cimatti}, A. and {Dessauges-Zavadsky}, M. and {Giavalisco}, M. and {Hathi}, N.~P. and {Fujimoto}, S. and {Ibar}, E. and {Koekemoer}, A. and {Lagache}, G. and {Lemaux}, B.~C. and {Loiacono}, F. and {Maiolino}, R. and {Narayanan}, D. and {Morselli}, L. and {M{\'e}ndez-Hern{\`a}ndez}, H. and {Pozzi}, F. and {Riechers}, D. and {Talia}, M. and {Toft}, S. and {Vallini}, L. and {Vergani}, D. and {Zamorani}, G. and {Zucca}, E.},
        title = "{The ALPINE-ALMA [C II] survey. Little to no evolution in the [C II]-SFR relation over the last 13 Gyr}",
      journal = {\aap},
         year = 2020,
        month = nov,
       volume = {643},
          eid = {A3},
        pages = {A3},
          doi = {10.1051/0004-6361/202037617},
archivePrefix = {arXiv},
       eprint = {2002.00979},
 primaryClass = {astro-ph.GA},
       adsurl = {https://ui.adsabs.harvard.edu/abs/2020A&A...643A...3S}
}

@ARTICLE{Hollenbach99,
       author = {{Hollenbach}, D.~J. and {Tielens}, A.~G.~G.~M.},
        title = "{Photodissociation regions in the interstellar medium of galaxies}",
      journal = {Reviews of Modern Physics},
         year = 1999,
        month = jan,
       volume = {71},
       number = {1},
        pages = {173-230},
          doi = {10.1103/RevModPhys.71.173},
       adsurl = {https://ui.adsabs.harvard.edu/abs/1999RvMP...71..173H}
}

@ARTICLE{Wolfire95,
       author = {{Wolfire}, M.~G. and {Hollenbach}, D. and {McKee}, C.~F. and {Tielens}, A.~G.~G.~M. and {Bakes}, E.~L.~O.},
        title = "{The Neutral Atomic Phases of the Interstellar Medium}",
      journal = {\apj},
         year = 1995,
        month = apr,
       volume = {443},
        pages = {152},
          doi = {10.1086/175510},
       adsurl = {https://ui.adsabs.harvard.edu/abs/1995ApJ...443..152W}
}

@ARTICLE{Sommovigo22,
       author = {{Sommovigo}, L. and {Ferrara}, A. and {Pallottini}, A. and {Dayal}, P. and {Bouwens}, R.~J. and {Smit}, R. and {da Cunha}, E. and {De Looze}, I. and {Bowler}, R.~A.~A. and {Hodge}, J. and {Inami}, H. and {Oesch}, P. and {Endsley}, R. and {Gonzalez}, V. and {Schouws}, S. and {Stark}, D. and {Stefanon}, M. and {Aravena}, M. and {Graziani}, L. and {Riechers}, D. and {Schneider}, R. and {van der Werf}, P. and {Algera}, H. and {Barrufet}, L. and {Fudamoto}, Y. and {Hygate}, A.~P.~S. and {Labb{\'e}}, I. and {Li}, Y. and {Nanayakkara}, T. and {Topping}, M.},
        title = "{The ALMA REBELS Survey: cosmic dust temperature evolution out to z   7}",
      journal = {\mnras},
         year = 2022,
        month = jul,
       volume = {513},
       number = {3},
        pages = {3122-3135},
          doi = {10.1093/mnras/stac302},
archivePrefix = {arXiv},
       eprint = {2202.01227},
 primaryClass = {astro-ph.GA},
       adsurl = {https://ui.adsabs.harvard.edu/abs/2022MNRAS.513.3122S}
}

@ARTICLE{Dayal22,
       author = {{Dayal}, P. and {Ferrara}, A. and {Sommovigo}, L. and {Bouwens}, R. and {Oesch}, P.~A. and {Smit}, R. and {Gonzalez}, V. and {Schouws}, S. and {Stefanon}, M. and {Kobayashi}, C. and {Bremer}, J. and {Algera}, H.~S.~B. and {Aravena}, M. and {Bowler}, R.~A.~A. and {da Cunha}, E. and {Fudamoto}, Y. and {Graziani}, L. and {Hodge}, J. and {Inami}, H. and {De Looze}, I. and {Pallottini}, A. and {Riechers}, D. and {Schneider}, R. and {Stark}, D. and {Endsley}, R.},
        title = "{The ALMA REBELS survey: the dust content of z   7 Lyman break galaxies}",
      journal = {\mnras},
         year = 2022,
        month = may,
       volume = {512},
       number = {1},
        pages = {989-1002},
          doi = {10.1093/mnras/stac537},
archivePrefix = {arXiv},
       eprint = {2202.11118},
 primaryClass = {astro-ph.GA},
       adsurl = {https://ui.adsabs.harvard.edu/abs/2022MNRAS.512..989D}
}

@ARTICLE{Topping22,
       author = {{Topping}, Michael W. and {Stark}, Daniel P. and {Endsley}, Ryan and {Bouwens}, Rychard J. and {Schouws}, Sander and {Smit}, Renske and {Stefanon}, Mauro and {Inami}, Hanae and {Bowler}, Rebecca A.~A. and {Oesch}, Pascal and {Gonzalez}, Valentino and {Dayal}, Pratika and {da Cunha}, Elisabete and {Algera}, Hiddo and {van der Werf}, Paul and {Pallottini}, Andrea and {Barrufet}, Laia and {Schneider}, Raffaella and {De Looze}, Ilse and {Sommovigo}, Laura and {Whitler}, Lily and {Graziani}, Luca and {Fudamoto}, Yoshinobu and {Ferrara}, Andrea},
        title = "{The ALMA REBELS Survey: specific star formation rates in the reionization era}",
      journal = {\mnras},
         year = 2022,
        month = oct,
       volume = {516},
       number = {1},
        pages = {975-991},
          doi = {10.1093/mnras/stac2291},
archivePrefix = {arXiv},
       eprint = {2203.07392},
 primaryClass = {astro-ph.GA},
       adsurl = {https://ui.adsabs.harvard.edu/abs/2022MNRAS.516..975T}
}

@ARTICLE{Topping21,
       author = {{Topping}, Michael W. and {Shapley}, Alice E. and {Stark}, Daniel P. and {Endsley}, Ryan and {Robertson}, Brant and {Greene}, Jenny E. and {Furlanetto}, Steven R. and {Tang}, Mengtao},
        title = "{The Detection of Ionized Carbon Emission at z   8}",
      journal = {\apjl},
         year = 2021,
        month = aug,
       volume = {917},
       number = {2},
          eid = {L36},
        pages = {L36},
          doi = {10.3847/2041-8213/ac1a79},
archivePrefix = {arXiv},
       eprint = {2107.06295},
 primaryClass = {astro-ph.GA},
       adsurl = {https://ui.adsabs.harvard.edu/abs/2021ApJ...917L..36T}
}

@ARTICLE{RamosPadilla22,
       author = {{Ramos Padilla}, A.~F. and {Wang}, L. and {van der Tak}, F.~F.~S. and {Trager}, S.~C.},
        title = "{Diagnosing the interstellar medium of galaxies with far-infrared emission lines. II. [C II], [O I], [O III], [N II], and [N III] up to z = 6}",
      journal = {\aap},
         year = 2023,
        month = nov,
       volume = {679},
          eid = {A131},
        pages = {A131},
          doi = {10.1051/0004-6361/202243358},
archivePrefix = {arXiv},
       eprint = {2205.11955},
 primaryClass = {astro-ph.GA},
       adsurl = {https://ui.adsabs.harvard.edu/abs/2023A&A...679A.131R}
}

@ARTICLE{Bouwens21a,
       author = {{Bouwens}, R.~J. and {Oesch}, P.~A. and {Stefanon}, M. and {Illingworth}, G. and {Labb{\'e}}, I. and {Reddy}, N. and {Atek}, H. and {Montes}, M. and {Naidu}, R. and {Nanayakkara}, T. and {Nelson}, E. and {Wilkins}, S.},
        title = "{New Determinations of the UV Luminosity Functions from z   9 to 2 Show a Remarkable Consistency with Halo Growth and a Constant Star Formation Efficiency}",
      journal = {\aj},
         year = 2021,
        month = aug,
       volume = {162},
       number = {2},
          eid = {47},
        pages = {47},
          doi = {10.3847/1538-3881/abf83e},
archivePrefix = {arXiv},
       eprint = {2102.07775},
 primaryClass = {astro-ph.GA},
       adsurl = {https://ui.adsabs.harvard.edu/abs/2021AJ....162...47B}
}

@ARTICLE{Robertson21,
       author = {{Robertson}, Brant E.},
        title = "{Galaxy Formation and Reionization: Key Unknowns and Expected Breakthroughs by the James Webb Space Telescope}",
      journal = {\araa},
         year = 2022,
        month = aug,
       volume = {60},
        pages = {121-158},
          doi = {10.1146/annurev-astro-120221-044656},
archivePrefix = {arXiv},
       eprint = {2110.13160},
 primaryClass = {astro-ph.CO},
       adsurl = {https://ui.adsabs.harvard.edu/abs/2022ARA&A..60..121R}
}

@ARTICLE{Bowler2020,
       author = {{Bowler}, R.~A.~A. and {Jarvis}, M.~J. and {Dunlop}, J.~S. and {McLure}, R.~J. and {McLeod}, D.~J. and {Adams}, N.~J. and {Milvang-Jensen}, B. and {McCracken}, H.~J.},
        title = "{A lack of evolution in the very bright end of the galaxy luminosity function from z ≃ 8 to 10}",
      journal = {\mnras},
         year = 2020,
        month = apr,
       volume = {493},
       number = {2},
        pages = {2059-2084},
          doi = {10.1093/mnras/staa313},
archivePrefix = {arXiv},
       eprint = {1911.12832},
 primaryClass = {astro-ph.GA},
       adsurl = {https://ui.adsabs.harvard.edu/abs/2020MNRAS.493.2059B}
}

@ARTICLE{Schmidt68,
       author = {{Schmidt}, Maarten},
        title = "{Space Distribution and Luminosity Functions of Quasi-Stellar Radio Sources}",
      journal = {\apj},
         year = 1968,
        month = feb,
       volume = {151},
        pages = {393},
          doi = {10.1086/149446},
       adsurl = {https://ui.adsabs.harvard.edu/abs/1968ApJ...151..393S}
}

@ARTICLE{Hodge20,
       author = {{Hodge}, J.~A. and {da Cunha}, E.},
        title = "{High-redshift star formation in the Atacama large millimetre/submillimetre array era}",
      journal = {Royal Society Open Science},
         year = 2020,
        month = dec,
       volume = {7},
       number = {12},
          eid = {200556},
        pages = {200556},
          doi = {10.1098/rsos.200556},
archivePrefix = {arXiv},
       eprint = {2004.00934},
 primaryClass = {astro-ph.GA},
       adsurl = {https://ui.adsabs.harvard.edu/abs/2020RSOS....700556H}
}

@ARTICLE{Fudamoto22,
       author = {{Fudamoto}, Y. and {Smit}, R. and {Bowler}, R.~A.~A. and {Oesch}, P.~A. and {Bouwens}, R. and {Stefanon}, M. and {Inami}, H. and {Endsley}, R. and {Gonzalez}, V. and {Schouws}, S. and {Stark}, D. and {Algera}, H.~S.~B. and {Aravena}, M. and {Barrufet}, L. and {da Cunha}, E. and {Dayal}, P. and {Ferrara}, A. and {Graziani}, L. and {Hodge}, J.~A. and {Hygate}, A.~P.~S. and {Inoue}, A.~K. and {Nanayakkara}, T. and {Pallottini}, A. and {Pizzati}, E. and {Schneider}, R. and {Sommovigo}, L. and {Sugahara}, Y. and {Topping}, M. and {van der Werf}, P. and {Bethermin}, M. and {Cassata}, P. and {Dessauges-Zavadsky}, M. and {Ibar}, E. and {Faisst}, A.~L. and {Fujimoto}, S. and {Ginolfi}, M. and {Hathi}, N. and {Jones}, G.~C. and {Pozzi}, F. and {Schaerer}, D.},
        title = "{The ALMA REBELS Survey: Average [C II] 158 {\ensuremath{\mu}}m Sizes of Star-forming Galaxies from z   7 to z   4}",
      journal = {\apj},
         year = 2022,
        month = aug,
       volume = {934},
       number = {2},
          eid = {144},
        pages = {144},
          doi = {10.3847/1538-4357/ac7a47},
archivePrefix = {arXiv},
       eprint = {2206.01886},
 primaryClass = {astro-ph.GA},
       adsurl = {https://ui.adsabs.harvard.edu/abs/2022ApJ...934..144F}
}

@ARTICLE{Fujimoto19,
       author = {{Fujimoto}, Seiji and {Ouchi}, Masami and {Ferrara}, Andrea and {Pallottini}, Andrea and {Ivison}, R.~J. and {Behrens}, Christoph and {Gallerani}, Simona and {Arata}, Shohei and {Yajima}, Hidenobu and {Nagamine}, Kentaro},
        title = "{First Identification of 10 kpc [C II] 158 {\ensuremath{\mu}}m Halos around Star-forming Galaxies at z = 5-7}",
      journal = {\apj},
         year = 2019,
        month = dec,
       volume = {887},
       number = {2},
          eid = {107},
        pages = {107},
          doi = {10.3847/1538-4357/ab480f},
archivePrefix = {arXiv},
       eprint = {1902.06760},
 primaryClass = {astro-ph.GA},
       adsurl = {https://ui.adsabs.harvard.edu/abs/2019ApJ...887..107F}
}

@ARTICLE{Heintz21,
       author = {{Heintz}, Kasper E. and {Watson}, Darach and {Oesch}, Pascal A. and {Narayanan}, Desika and {Madden}, Suzanne C.},
        title = "{Measuring the H I Content of Individual Galaxies Out to the Epoch of Reionization with [C II]}",
      journal = {\apj},
         year = 2021,
        month = dec,
       volume = {922},
       number = {2},
          eid = {147},
        pages = {147},
          doi = {10.3847/1538-4357/ac2231},
archivePrefix = {arXiv},
       eprint = {2108.13442},
 primaryClass = {astro-ph.GA},
       adsurl = {https://ui.adsabs.harvard.edu/abs/2021ApJ...922..147H}
}

@ARTICLE{Gallerani18,
       author = {{Gallerani}, S. and {Pallottini}, A. and {Feruglio}, C. and {Ferrara}, A. and {Maiolino}, R. and {Vallini}, L. and {Riechers}, D.~A. and {Pavesi}, R.},
        title = "{ALMA suggests outflows in z {\ensuremath{\sim}} 5.5 galaxies}",
      journal = {\mnras},
         year = 2018,
        month = jan,
       volume = {473},
       number = {2},
        pages = {1909-1917},
          doi = {10.1093/mnras/stx2458},
archivePrefix = {arXiv},
       eprint = {1604.05714},
 primaryClass = {astro-ph.GA},
       adsurl = {https://ui.adsabs.harvard.edu/abs/2018MNRAS.473.1909G}
}

@ARTICLE{Ginolfi20,
       author = {{Ginolfi}, M. and {Jones}, G.~C. and {B{\'e}thermin}, M. and {Fudamoto}, Y. and {Loiacono}, F. and {Fujimoto}, S. and {Le F{\'e}vre}, O. and {Faisst}, A. and {Schaerer}, D. and {Cassata}, P. and {Silverman}, J.~D. and {Yan}, L. and {Capak}, P. and {Bardelli}, S. and {Boquien}, M. and {Carraro}, R. and {Dessauges-Zavadsky}, M. and {Giavalisco}, M. and {Gruppioni}, C. and {Ibar}, E. and {Khusanova}, Y. and {Lemaux}, B.~C. and {Maiolino}, R. and {Narayanan}, D. and {Oesch}, P. and {Pozzi}, F. and {Rodighiero}, G. and {Talia}, M. and {Toft}, S. and {Vallini}, L. and {Vergani}, D. and {Zamorani}, G.},
        title = "{The ALPINE-ALMA [C II] survey: Star-formation-driven outflows and circumgalactic enrichment in the early Universe}",
      journal = {\aap},
         year = 2020,
        month = jan,
       volume = {633},
          eid = {A90},
        pages = {A90},
          doi = {10.1051/0004-6361/201936872},
archivePrefix = {arXiv},
       eprint = {1910.04770},
 primaryClass = {astro-ph.GA},
       adsurl = {https://ui.adsabs.harvard.edu/abs/2020A&A...633A..90G}
}

@ARTICLE{Zanella18,
       author = {{Zanella}, A. and {Daddi}, E. and {Magdis}, G. and {Diaz Santos}, T. and {Cormier}, D. and {Liu}, D. and {Cibinel}, A. and {Gobat}, R. and {Dickinson}, M. and {Sargent}, M. and {Popping}, G. and {Madden}, S.~C. and {Bethermin}, M. and {Hughes}, T.~M. and {Valentino}, F. and {Rujopakarn}, W. and {Pannella}, M. and {Bournaud}, F. and {Walter}, F. and {Wang}, T. and {Elbaz}, D. and {Coogan}, R.~T.},
        title = "{The [C II] emission as a molecular gas mass tracer in galaxies at low and high redshifts}",
      journal = {\mnras},
         year = 2018,
        month = dec,
       volume = {481},
       number = {2},
        pages = {1976-1999},
          doi = {10.1093/mnras/sty2394},
archivePrefix = {arXiv},
       eprint = {1808.10331},
 primaryClass = {astro-ph.GA},
       adsurl = {https://ui.adsabs.harvard.edu/abs/2018MNRAS.481.1976Z}
}

@ARTICLE{Dessauges20,
       author = {{Dessauges-Zavadsky}, M. and {Ginolfi}, M. and {Pozzi}, F. and {B{\'e}thermin}, M. and {Le F{\`e}vre}, O. and {Fujimoto}, S. and {Silverman}, J.~D. and {Jones}, G.~C. and {Vallini}, L. and {Schaerer}, D. and {Faisst}, A.~L. and {Khusanova}, Y. and {Fudamoto}, Y. and {Cassata}, P. and {Loiacono}, F. and {Capak}, P.~L. and {Yan}, L. and {Amorin}, R. and {Bardelli}, S. and {Boquien}, M. and {Cimatti}, A. and {Gruppioni}, C. and {Hathi}, N.~P. and {Ibar}, E. and {Koekemoer}, A.~M. and {Lemaux}, B.~C. and {Narayanan}, D. and {Oesch}, P.~A. and {Rodighiero}, G. and {Romano}, M. and {Talia}, M. and {Toft}, S. and {Vergani}, D. and {Zamorani}, G. and {Zucca}, E.},
        title = "{The ALPINE-ALMA [C II] survey. Molecular gas budget in the early Universe as traced by [C II]}",
      journal = {\aap},
         year = 2020,
        month = nov,
       volume = {643},
          eid = {A5},
        pages = {A5},
          doi = {10.1051/0004-6361/202038231},
archivePrefix = {arXiv},
       eprint = {2004.10771},
 primaryClass = {astro-ph.GA},
       adsurl = {https://ui.adsabs.harvard.edu/abs/2020A&A...643A...5D}
}

@ARTICLE{Dayal2018,
       author = {{Dayal}, Pratika and {Ferrara}, Andrea},
        title = "{Early galaxy formation and its large-scale effects}",
      journal = {\physrep},
         year = 2018,
        month = dec,
       volume = {780},
        pages = {1-64},
          doi = {10.1016/j.physrep.2018.10.002},
archivePrefix = {arXiv},
       eprint = {1809.09136},
 primaryClass = {astro-ph.GA},
       adsurl = {https://ui.adsabs.harvard.edu/abs/2018PhR...780....1D}
}

@ARTICLE{Popping19,
       author = {{Popping}, Gerg{\"o} and {Narayanan}, Desika and
         {Somerville}, Rachel S. and {Faisst}, Andreas L. and
         {Krumholz}, Mark R.},
        title = "{The art of modelling CO, [C I], and [C II] in cosmological galaxy formation models}",
      journal = {\mnras},
         year = "2019",
        month = "Feb",
       volume = {482},
       number = {4},
        pages = {4906-4932},
          doi = {10.1093/mnras/sty2969},
archivePrefix = {arXiv},
       eprint = {1805.11093},
 primaryClass = {astro-ph.GA},
       adsurl = {https://ui.adsabs.harvard.edu/abs/2019MNRAS.482.4906P}
}

@ARTICLE{Loiacono20,
       author = {{Loiacono}, Federica and {Decarli}, Roberto and {Gruppioni}, Carlotta and {Talia}, Margherita and {Cimatti}, Andrea and {Zamorani}, Gianni and {Pozzi}, Francesca and {Yan}, Lin and {Lemaux}, Brian C. and {Riechers}, Dominik A. and {Le F{\`e}vre}, Olivier and {B{\`e}thermin}, Matthieu and {Capak}, Peter and {Cassata}, Paolo and {Faisst}, Andreas and {Schaerer}, Daniel and {Silverman}, John D. and {Bardelli}, Sandro and {Boquien}, M{\'e}d{\'e}ric and {Burkutean}, Sandra and {Dessauges-Zavadsky}, Miroslava and {Fudamoto}, Yoshinobu and {Fujimoto}, Seiji and {Ginolfi}, Michele and {Hathi}, Nimish P. and {Jones}, Gareth C. and {Khusanova}, Yana and {Koekemoer}, Anton M. and {Lagache}, Guilaine and {Lubin}, Lori M. and {Massardi}, Marcella and {Oesch}, Pascal and {Romano}, Michael and {Vallini}, Livia and {Vergani}, Daniela and {Zucca}, Elena},
        title = "{The ALPINE-ALMA [C II] survey. Luminosity function of serendipitous [C II] line emitters at z {\ensuremath{\sim}} 5}",
      journal = {\aap},
         year = 2021,
        month = feb,
       volume = {646},
          eid = {A76},
        pages = {A76},
          doi = {10.1051/0004-6361/202038607},
archivePrefix = {arXiv},
       eprint = {2006.04837},
 primaryClass = {astro-ph.GA},
       adsurl = {https://ui.adsabs.harvard.edu/abs/2021A&A...646A..76L}
}

@ARTICLE{Hemmati17,
       author = {{Hemmati}, Shoubaneh and {Yan}, Lin and {Diaz-Santos}, Tanio and
         {Armus}, Lee and {Capak}, Peter and {Faisst}, Andreas and
         {Masters}, Daniel},
        title = "{The Local [C II] 158 {\ensuremath{\mu}}m Emission Line Luminosity Function}",
      journal = {\apj},
         year = "2017",
        month = "Jan",
       volume = {834},
       number = {1},
          eid = {36},
        pages = {36},
          doi = {10.3847/1538-4357/834/1/36},
archivePrefix = {arXiv},
       eprint = {1611.03092},
 primaryClass = {astro-ph.GA},
       adsurl = {https://ui.adsabs.harvard.edu/abs/2017ApJ...834...36H}
}

@ARTICLE{Yamaguchi17,
       author = {{Yamaguchi}, Yuki and {Kohno}, Kotaro and {Tamura}, Yoichi and
         {Oguri}, Masamune and {Ezawa}, Hajime and {Hayatsu}, Natsuki H. and
         {Kitayama}, Tetsu and {Matsuda}, Yuichi and {Matsuo}, Hiroshi and
         {Oshima}, Tai and {Ota}, Naomi and {Izumi}, Takuma and
         {Umehata}, Hideki},
        title = "{Blind Millimeter Line Emitter Search using ALMA Data Toward Gravitational Lensing Clusters}",
      journal = {\apj},
         year = "2017",
        month = "Aug",
       volume = {845},
       number = {2},
          eid = {108},
        pages = {108},
          doi = {10.3847/1538-4357/aa80e0},
archivePrefix = {arXiv},
       eprint = {1707.05547},
 primaryClass = {astro-ph.GA},
       adsurl = {https://ui.adsabs.harvard.edu/abs/2017ApJ...845..108Y}
}

@ARTICLE{Carilli13,
       author = {{Carilli}, C.~L. and {Walter}, F.},
        title = "{Cool Gas in High-Redshift Galaxies}",
      journal = {\araa},
         year = 2013,
        month = aug,
       volume = {51},
       number = {1},
        pages = {105-161},
          doi = {10.1146/annurev-astro-082812-140953},
archivePrefix = {arXiv},
       eprint = {1301.0371},
 primaryClass = {astro-ph.CO},
       adsurl = {https://ui.adsabs.harvard.edu/abs/2013ARA&A..51..105C}
}

@ARTICLE{Finkelstein2015,
       author = {{Finkelstein}, Steven L. and {Ryan}, Russell E., Jr. and {Papovich}, Casey and {Dickinson}, Mark and {Song}, Mimi and {Somerville}, Rachel S. and {Ferguson}, Henry C. and {Salmon}, Brett and {Giavalisco}, Mauro and {Koekemoer}, Anton M. and {Ashby}, Matthew L.~N. and {Behroozi}, Peter and {Castellano}, Marco and {Dunlop}, James S. and {Faber}, Sandy M. and {Fazio}, Giovanni G. and {Fontana}, Adriano and {Grogin}, Norman A. and {Hathi}, Nimish and {Jaacks}, Jason and {Kocevski}, Dale D. and {Livermore}, Rachael and {McLure}, Ross J. and {Merlin}, Emiliano and {Mobasher}, Bahram and {Newman}, Jeffrey A. and {Rafelski}, Marc and {Tilvi}, Vithal and {Willner}, S.~P.},
        title = "{The Evolution of the Galaxy Rest-frame Ultraviolet Luminosity Function over the First Two Billion Years}",
      journal = {\apj},
         year = 2015,
        month = sep,
       volume = {810},
       number = {1},
          eid = {71},
        pages = {71},
          doi = {10.1088/0004-637X/810/1/71},
archivePrefix = {arXiv},
       eprint = {1410.5439},
 primaryClass = {astro-ph.GA},
       adsurl = {https://ui.adsabs.harvard.edu/abs/2015ApJ...810...71F}
}

@ARTICLE{Bouwens2015,
       author = {{Bouwens}, R.~J. and {Illingworth}, G.~D. and {Oesch}, P.~A. and {Trenti}, M. and {Labb{\'e}}, I. and {Bradley}, L. and {Carollo}, M. and {van Dokkum}, P.~G. and {Gonzalez}, V. and {Holwerda}, B. and {Franx}, M. and {Spitler}, L. and {Smit}, R. and {Magee}, D.},
        title = "{UV Luminosity Functions at Redshifts z {\ensuremath{\sim}} 4 to z {\ensuremath{\sim}} 10: 10,000 Galaxies from HST Legacy Fields}",
      journal = {\apj},
         year = 2015,
        month = apr,
       volume = {803},
       number = {1},
          eid = {34},
        pages = {34},
          doi = {10.1088/0004-637X/803/1/34},
archivePrefix = {arXiv},
       eprint = {1403.4295},
 primaryClass = {astro-ph.CO},
       adsurl = {https://ui.adsabs.harvard.edu/abs/2015ApJ...803...34B}
}

@ARTICLE{Schechter1976,
       author = {{Schechter}, P.},
        title = "{An analytic expression for the luminosity function for galaxies.}",
      journal = {\apj},
         year = 1976,
        month = jan,
       volume = {203},
        pages = {297-306},
          doi = {10.1086/154079},
       adsurl = {https://ui.adsabs.harvard.edu/abs/1976ApJ...203..297S}
}

@ARTICLE{Stefanon21,
       author = {{Stefanon}, Mauro and {Bouwens}, Rychard J. and {Labb{\'e}}, Ivo and {Illingworth}, Garth D. and {Gonzalez}, Valentino and {Oesch}, Pascal A.},
        title = "{Galaxy Stellar Mass Functions from z   10 to z   6 using the Deepest Spitzer/Infrared Array Camera Data: No Significant Evolution in the Stellar-to-halo Mass Ratio of Galaxies in the First Gigayear of Cosmic Time}",
      journal = {\apj},
         year = 2021,
        month = nov,
       volume = {922},
       number = {1},
          eid = {29},
        pages = {29},
          doi = {10.3847/1538-4357/ac1bb6},
archivePrefix = {arXiv},
       eprint = {2103.16571},
 primaryClass = {astro-ph.GA},
       adsurl = {https://ui.adsabs.harvard.edu/abs/2021ApJ...922...29S}
}

@ARTICLE{DeLooze14,
       author = {{De Looze}, Ilse and {Cormier}, Diane and {Lebouteiller}, Vianney and {Madden}, Suzanne and {Baes}, Maarten and {Bendo}, George J. and {Boquien}, M{\'e}d{\'e}ric and {Boselli}, Alessandro and {Clements}, David L. and {Cortese}, Luca and {Cooray}, Asantha and {Galametz}, Maud and {Galliano}, Fr{\'e}d{\'e}ric and {Graci{\'a}-Carpio}, Javier and {Isaak}, Kate and {Karczewski}, Oskar {\L}. and {Parkin}, Tara J. and {Pellegrini}, Eric W. and {R{\'e}my-Ruyer}, Aur{\'e}lie and {Spinoglio}, Luigi and {Smith}, Matthew W.~L. and {Sturm}, Eckhard},
        title = "{The applicability of far-infrared fine-structure lines as star formation rate tracers over wide ranges of metallicities and galaxy types}",
      journal = {\aap},
         year = 2014,
        month = aug,
       volume = {568},
          eid = {A62},
        pages = {A62},
          doi = {10.1051/0004-6361/201322489},
archivePrefix = {arXiv},
       eprint = {1402.4075},
 primaryClass = {astro-ph.GA},
       adsurl = {https://ui.adsabs.harvard.edu/abs/2014A&A...568A..62D}
}

@ARTICLE{Bouwens21REBELS,
       author = {{Bouwens}, R.~J. and {Smit}, R. and {Schouws}, S. and {Stefanon}, M. and {Bowler}, R. and {Endsley}, R. and {Gonzalez}, V. and {Inami}, H. and {Stark}, D. and {Oesch}, P. and {Hodge}, J. and {Aravena}, M. and {da Cunha}, E. and {Dayal}, P. and {de Looze}, I. and {Ferrara}, A. and {Fudamoto}, Y. and {Graziani}, L. and {Li}, C. and {Nanayakkara}, T. and {Pallottini}, A. and {Schneider}, R. and {Sommovigo}, L. and {Topping}, M. and {van der Werf}, P. and {Algera}, H. and {Barrufet}, L. and {Hygate}, A. and {Labb{\'e}}, I. and {Riechers}, D. and {Witstok}, J.},
        title = "{Reionization Era Bright Emission Line Survey: Selection and Characterization of Luminous Interstellar Medium Reservoirs in the z > 6.5 Universe}",
      journal = {\apj},
         year = 2022,
        month = jun,
       volume = {931},
       number = {2},
          eid = {160},
        pages = {160},
          doi = {10.3847/1538-4357/ac5a4a},
archivePrefix = {arXiv},
       eprint = {2106.13719},
 primaryClass = {astro-ph.GA},
       adsurl = {https://ui.adsabs.harvard.edu/abs/2022ApJ...931..160B}
}

@ARTICLE{Bakx20,
       author = {{Bakx}, Tom J.~L.~C. and {Tamura}, Yoichi and {Hashimoto}, Takuya and {Inoue}, Akio K. and {Lee}, Minju M. and {Mawatari}, Ken and {Ota}, Kazuaki and {Umehata}, Hideki and {Zackrisson}, Erik and {Hatsukade}, Bunyo and {Kohno}, Kotaro and {Matsuda}, Yuichi and {Matsuo}, Hiroshi and {Okamoto}, Takashi and {Shibuya}, Takatoshi and {Shimizu}, Ikkoh and {Taniguchi}, Yoshiaki and {Yoshida}, Naoki},
        title = "{ALMA uncovers the [C II] emission and warm dust continuum in a z = 8.31 Lyman break galaxy}",
      journal = {\mnras},
         year = 2020,
        month = apr,
       volume = {493},
       number = {3},
        pages = {4294-4307},
          doi = {10.1093/mnras/staa509},
archivePrefix = {arXiv},
       eprint = {2001.02812},
 primaryClass = {astro-ph.GA},
       adsurl = {https://ui.adsabs.harvard.edu/abs/2020MNRAS.493.4294B}
}

@ARTICLE{Capak15,
       author = {{Capak}, P.~L. and {Carilli}, C. and {Jones}, G. and {Casey}, C.~M. and {Riechers}, D. and {Sheth}, K. and {Carollo}, C.~M. and {Ilbert}, O. and {Karim}, A. and {Lefevre}, O. and {Lilly}, S. and {Scoville}, N. and {Smolcic}, V. and {Yan}, L.},
        title = "{Galaxies at redshifts 5 to 6 with systematically low dust content and high [C II] emission}",
      journal = {\nat},
         year = 2015,
        month = jun,
       volume = {522},
       number = {7557},
        pages = {455-458},
          doi = {10.1038/nature14500},
archivePrefix = {arXiv},
       eprint = {1503.07596},
 primaryClass = {astro-ph.GA},
       adsurl = {https://ui.adsabs.harvard.edu/abs/2015Natur.522..455C}
}

@ARTICLE{Carniani18,
       author = {{Carniani}, S. and {Maiolino}, R. and {Smit}, R. and {Amor{\'\i}n}, R.},
        title = "{ALMA Detection of Extended [C II] Emission in Himiko at z = 6.6}",
      journal = {\apjl},
         year = 2018,
        month = feb,
       volume = {854},
       number = {1},
          eid = {L7},
        pages = {L7},
          doi = {10.3847/2041-8213/aaab45},
archivePrefix = {arXiv},
       eprint = {1712.01890},
 primaryClass = {astro-ph.GA},
       adsurl = {https://ui.adsabs.harvard.edu/abs/2018ApJ...854L...7C}
}

@ARTICLE{Marrone18,
       author = {{Marrone}, D.~P. and {Spilker}, J.~S. and {Hayward}, C.~C. and {Vieira}, J.~D. and {Aravena}, M. and {Ashby}, M.~L.~N. and {Bayliss}, M.~B. and {B{\'e}thermin}, M. and {Brodwin}, M. and {Bothwell}, M.~S. and {Carlstrom}, J.~E. and {Chapman}, S.~C. and {Chen}, Chian-Chou and {Crawford}, T.~M. and {Cunningham}, D.~J.~M. and {De Breuck}, C. and {Fassnacht}, C.~D. and {Gonzalez}, A.~H. and {Greve}, T.~R. and {Hezaveh}, Y.~D. and {Lacaille}, K. and {Litke}, K.~C. and {Lower}, S. and {Ma}, J. and {Malkan}, M. and {Miller}, T.~B. and {Morningstar}, W.~R. and {Murphy}, E.~J. and {Narayanan}, D. and {Phadke}, K.~A. and {Rotermund}, K.~M. and {Sreevani}, J. and {Stalder}, B. and {Stark}, A.~A. and {Strandet}, M.~L. and {Tang}, M. and {Wei{\ss}}, A.},
        title = "{Galaxy growth in a massive halo in the first billion years of cosmic history}",
      journal = {\nat},
         year = 2018,
        month = jan,
       volume = {553},
       number = {7686},
        pages = {51-54},
          doi = {10.1038/nature24629},
archivePrefix = {arXiv},
       eprint = {1712.03020},
 primaryClass = {astro-ph.GA},
       adsurl = {https://ui.adsabs.harvard.edu/abs/2018Natur.553...51M}
}

@ARTICLE{Bethermin22,
       author = {{B{\'e}thermin}, M. and {Gkogkou}, A. and {Van Cuyck}, M. and {Lagache}, G. and {Beelen}, A. and {Aravena}, M. and {Benoit}, A. and {Bounmy}, J. and {Calvo}, M. and {Catalano}, A. and {de Batz de Trenquelleon}, B. and {De Breuck}, C. and {Fasano}, A. and {Ferrara}, A. and {Goupy}, J. and {Hoarau}, C. and {Horellou}, C. and {Hu}, W. and {Julia}, A. and {Knudsen}, K. and {Lambert}, J.-C. and {Macias-Perez}, J. and {Marpaud}, J. and {Monfardini}, A. and {Pallottini}, A. and {Ponthieu}, N. and {Roehlly}, Y. and {Vallini}, L. and {Walter}, F. and {Weiss}, A.},
        title = "{CONCERTO: High-fidelity simulation of millimeter line emissions of galaxies and [CII] intensity mapping}",
      journal = {\aap},
         year = 2022,
        month = nov,
       volume = {667},
          eid = {A156},
        pages = {A156},
          doi = {10.1051/0004-6361/202243888},
archivePrefix = {arXiv},
       eprint = {2204.12827},
 primaryClass = {astro-ph.GA},
       adsurl = {https://ui.adsabs.harvard.edu/abs/2022A&A...667A.156B}
}

@ARTICLE{Cothard20,
       author = {{Cothard}, N.~F. and {Choi}, S.~K. and {Duell}, C.~J. and {Herter}, T. and {Hubmayr}, J. and {McMahon}, J. and {Niemack}, M.~D. and {Nikola}, T. and {Sierra}, C. and {Stacey}, G.~J. and {Vavagiakis}, E.~M. and {Wollack}, E.~J. and {Zou}, B.},
        title = "{The Design of the CCAT-prime Epoch of Reionization Spectrometer Instrument}",
      journal = {Journal of Low Temperature Physics},
         year = 2020,
        month = jan,
       volume = {199},
       number = {3-4},
        pages = {898-907},
          doi = {10.1007/s10909-019-02297-1},
archivePrefix = {arXiv},
       eprint = {1911.11687},
 primaryClass = {astro-ph.IM},
       adsurl = {https://ui.adsabs.harvard.edu/abs/2020JLTP..199..898C}
}

@ARTICLE{Sun21,
       author = {{Sun}, G. and {Chang}, T. -C. and {Uzgil}, B.~D. and {Bock}, J.~J. and {Bradford}, C.~M. and {Butler}, V. and {Caze-Cortes}, T. and {Cheng}, Y. -T. and {Cooray}, A. and {Crites}, A.~T. and {Hailey-Dunsheath}, S. and {Emerson}, N. and {Frez}, C. and {Hoscheit}, B.~L. and {Hunacek}, J. and {Keenan}, R.~P. and {Li}, C.~T. and {Madonia}, P. and {Marrone}, D.~P. and {Moncelsi}, L. and {Shiu}, C. and {Trumper}, I. and {Turner}, A. and {Weber}, A. and {Wei}, T.~S. and {Zemcov}, M.},
        title = "{Probing Cosmic Reionization and Molecular Gas Growth with TIME}",
      journal = {\apj},
         year = 2021,
        month = jul,
       volume = {915},
       number = {1},
          eid = {33},
        pages = {33},
          doi = {10.3847/1538-4357/abfe62},
archivePrefix = {arXiv},
       eprint = {2012.09160},
 primaryClass = {astro-ph.GA},
       adsurl = {https://ui.adsabs.harvard.edu/abs/2021ApJ...915...33S}
}

@ARTICLE{Decarli20,
       author = {{Decarli}, Roberto and {Aravena}, Manuel and {Boogaard}, Leindert and {Carilli}, Chris and {Gonz{\'a}lez-L{\'o}pez}, Jorge and {Walter}, Fabian and {Cortes}, Paulo C. and {Cox}, Pierre and {da Cunha}, Elisabete and {Daddi}, Emanuele and {D{\'\i}az-Santos}, Tanio and {Hodge}, Jacqueline A. and {Inami}, Hanae and {Neeleman}, Marcel and {Novak}, Mladen and {Oesch}, Pascal and {Popping}, Gerg{\"o} and {Riechers}, Dominik and {Smail}, Ian and {Uzgil}, Bade and {van der Werf}, Paul and {Wagg}, Jeff and {Weiss}, Axel},
        title = "{The ALMA Spectroscopic Survey in the Hubble Ultra Deep Field: Multiband Constraints on Line-luminosity Functions and the Cosmic Density of Molecular Gas}",
      journal = {\apj},
         year = 2020,
        month = oct,
       volume = {902},
       number = {2},
          eid = {110},
        pages = {110},
          doi = {10.3847/1538-4357/abaa3b},
archivePrefix = {arXiv},
       eprint = {2009.10744},
 primaryClass = {astro-ph.GA},
       adsurl = {https://ui.adsabs.harvard.edu/abs/2020ApJ...902..110D}
}

@ARTICLE{Faisst20,
       author = {{Faisst}, A.~L. and {Schaerer}, D. and {Lemaux}, B.~C. and {Oesch}, P.~A. and {Fudamoto}, Y. and {Cassata}, P. and {B{\'e}thermin}, M. and {Capak}, P.~L. and {Le F{\`e}vre}, O. and {Silverman}, J.~D. and {Yan}, L. and {Ginolfi}, M. and {Koekemoer}, A.~M. and {Morselli}, L. and {Amor{\'\i}n}, R. and {Bardelli}, S. and {Boquien}, M. and {Brammer}, G. and {Cimatti}, A. and {Dessauges-Zavadsky}, M. and {Fujimoto}, S. and {Gruppioni}, C. and {Hathi}, N.~P. and {Hemmati}, S. and {Ibar}, E. and {Jones}, G.~C. and {Khusanova}, Y. and {Loiacono}, F. and {Pozzi}, F. and {Talia}, M. and {Tasca}, L.~A.~M. and {Riechers}, D.~A. and {Rodighiero}, G. and {Romano}, M. and {Scoville}, N. and {Toft}, S. and {Vallini}, L. and {Vergani}, D. and {Zamorani}, G. and {Zucca}, E.},
        title = "{The ALPINE-ALMA [C II] Survey: Multiwavelength Ancillary Data and Basic Physical Measurements}",
      journal = {\apjs},
         year = 2020,
        month = apr,
       volume = {247},
       number = {2},
          eid = {61},
        pages = {61},
          doi = {10.3847/1538-4365/ab7ccd},
archivePrefix = {arXiv},
       eprint = {1912.01621},
 primaryClass = {astro-ph.GA},
       adsurl = {https://ui.adsabs.harvard.edu/abs/2020ApJS..247...61F}
}

@ARTICLE{Katz19,
       author = {{Katz}, Harley and {Galligan}, Thomas P. and {Kimm}, Taysun and {Rosdahl}, Joakim and {Haehnelt}, Martin G. and {Blaizot}, Jeremy and {Devriendt}, Julien and {Slyz}, Adrianne and {Laporte}, Nicolas and {Ellis}, Richard},
        title = "{Probing cosmic dawn with emission lines: predicting infrared and nebular line emission for ALMA and JWST}",
      journal = {\mnras},
         year = 2019,
        month = aug,
       volume = {487},
       number = {4},
        pages = {5902-5921},
          doi = {10.1093/mnras/stz1672},
archivePrefix = {arXiv},
       eprint = {1901.01272},
 primaryClass = {astro-ph.GA},
       adsurl = {https://ui.adsabs.harvard.edu/abs/2019MNRAS.487.5902K}
}

@ARTICLE{Ferrara19,
       author = {{Ferrara}, A. and {Vallini}, L. and {Pallottini}, A. and {Gallerani}, S. and {Carniani}, S. and {Kohandel}, M. and {Decataldo}, D. and {Behrens}, C.},
        title = "{A physical model for [C II] line emission from galaxies}",
      journal = {\mnras},
         year = 2019,
        month = oct,
       volume = {489},
       number = {1},
        pages = {1-12},
          doi = {10.1093/mnras/stz2031},
archivePrefix = {arXiv},
       eprint = {1908.07536},
 primaryClass = {astro-ph.GA},
       adsurl = {https://ui.adsabs.harvard.edu/abs/2019MNRAS.489....1F}
}

@ARTICLE{Fudamoto21,
       author = {{Fudamoto}, Y. and {Oesch}, P.~A. and {Schouws}, S. and {Stefanon}, M. and {Smit}, R. and {Bouwens}, R.~J. and {Bowler}, R.~A.~A. and {Endsley}, R. and {Gonzalez}, V. and {Inami}, H. and {Labbe}, I. and {Stark}, D. and {Aravena}, M. and {Barrufet}, L. and {da Cunha}, E. and {Dayal}, P. and {Ferrara}, A. and {Graziani}, L. and {Hodge}, J. and {Hutter}, A. and {Li}, Y. and {De Looze}, I. and {Nanayakkara}, T. and {Pallottini}, A. and {Riechers}, D. and {Schneider}, R. and {Ucci}, G. and {van der Werf}, P. and {White}, C.},
        title = "{Normal, dust-obscured galaxies in the epoch of reionization}",
      journal = {\nat},
         year = 2021,
        month = sep,
       volume = {597},
       number = {7877},
        pages = {489-492},
          doi = {10.1038/s41586-021-03846-z},
archivePrefix = {arXiv},
       eprint = {2109.10378},
 primaryClass = {astro-ph.GA},
       adsurl = {https://ui.adsabs.harvard.edu/abs/2021Natur.597..489F}
}

@ARTICLE{Gruppioni20,
       author = {{Gruppioni}, C. and {B{\'e}thermin}, M. and {Loiacono}, F. and {Le F{\`e}vre}, O. and {Capak}, P. and {Cassata}, P. and {Faisst}, A.~L. and {Schaerer}, D. and {Silverman}, J. and {Yan}, L. and {Bardelli}, S. and {Boquien}, M. and {Carraro}, R. and {Cimatti}, A. and {Dessauges-Zavadsky}, M. and {Ginolfi}, M. and {Fujimoto}, S. and {Hathi}, N.~P. and {Jones}, G.~C. and {Khusanova}, Y. and {Koekemoer}, A.~M. and {Lagache}, G. and {Lemaux}, B.~C. and {Oesch}, P.~A. and {Pozzi}, F. and {Riechers}, D.~A. and {Rodighiero}, G. and {Romano}, M. and {Talia}, M. and {Vallini}, L. and {Vergani}, D. and {Zamorani}, G. and {Zucca}, E.},
        title = "{The ALPINE-ALMA [CII] survey. The nature, luminosity function, and star formation history of dusty galaxies up to z ≃ 6}",
      journal = {\aap},
         year = 2020,
        month = nov,
       volume = {643},
          eid = {A8},
        pages = {A8},
          doi = {10.1051/0004-6361/202038487},
archivePrefix = {arXiv},
       eprint = {2006.04974},
 primaryClass = {astro-ph.GA},
       adsurl = {https://ui.adsabs.harvard.edu/abs/2020A&A...643A...8G}
}

@ARTICLE{Hashimoto19,
       author = {{Hashimoto}, Takuya and {Inoue}, Akio K. and {Mawatari}, Ken and {Tamura}, Yoichi and {Matsuo}, Hiroshi and {Furusawa}, Hisanori and {Harikane}, Yuichi and {Shibuya}, Takatoshi and {Knudsen}, Kirsten K. and {Kohno}, Kotaro and {Ono}, Yoshiaki and {Zackrisson}, Erik and {Okamoto}, Takashi and {Kashikawa}, Nobunari and {Oesch}, Pascal A. and {Ouchi}, Masami and {Ota}, Kazuaki and {Shimizu}, Ikkoh and {Taniguchi}, Yoshiaki and {Umehata}, Hideki and {Watson}, Darach},
        title = "{Big Three Dragons: A z = 7.15 Lyman-break galaxy detected in [O III] 88 {\ensuremath{\mu}}m, [C II] 158 {\ensuremath{\mu}}m, and dust continuum with ALMA}",
      journal = {\pasj},
         year = 2019,
        month = aug,
       volume = {71},
       number = {4},
          eid = {71},
        pages = {71},
          doi = {10.1093/pasj/psz049},
archivePrefix = {arXiv},
       eprint = {1806.00486},
 primaryClass = {astro-ph.GA},
       adsurl = {https://ui.adsabs.harvard.edu/abs/2019PASJ...71...71H}
}

@ARTICLE{Lagache18,
       author = {{Lagache}, G. and {Cousin}, M. and {Chatzikos}, M.},
        title = "{The [CII] 158 {\ensuremath{\mu}}m line emission in high-redshift galaxies}",
      journal = {\aap},
         year = 2018,
        month = jan,
       volume = {609},
          eid = {A130},
        pages = {A130},
          doi = {10.1051/0004-6361/201732019},
archivePrefix = {arXiv},
       eprint = {1711.00798},
 primaryClass = {astro-ph.GA},
       adsurl = {https://ui.adsabs.harvard.edu/abs/2018A&A...609A.130L}
}

@ARTICLE{Matthee19,
       author = {{Matthee}, J. and {Sobral}, D. and {Boogaard}, L.~A. and {R{\"o}ttgering}, H. and {Vallini}, L. and {Ferrara}, A. and {Paulino-Afonso}, A. and {Boone}, F. and {Schaerer}, D. and {Mobasher}, B.},
        title = "{Resolved UV and [C II] Structures of Luminous Galaxies within the Epoch of Reionization}",
      journal = {\apj},
         year = 2019,
        month = aug,
       volume = {881},
       number = {2},
          eid = {124},
        pages = {124},
          doi = {10.3847/1538-4357/ab2f81},
archivePrefix = {arXiv},
       eprint = {1903.08171},
 primaryClass = {astro-ph.GA},
       adsurl = {https://ui.adsabs.harvard.edu/abs/2019ApJ...881..124M}
}

@ARTICLE{Maiolino15,
       author = {{Maiolino}, R. and {Carniani}, S. and {Fontana}, A. and {Vallini}, L. and {Pentericci}, L. and {Ferrara}, A. and {Vanzella}, E. and {Grazian}, A. and {Gallerani}, S. and {Castellano}, M. and {Cristiani}, S. and {Brammer}, G. and {Santini}, P. and {Wagg}, J. and {Williams}, R.},
        title = "{The assembly of `normal' galaxies at z {\ensuremath{\sim}} 7 probed by ALMA}",
      journal = {\mnras},
         year = 2015,
        month = sep,
       volume = {452},
       number = {1},
        pages = {54-68},
          doi = {10.1093/mnras/stv1194},
archivePrefix = {arXiv},
       eprint = {1502.06634},
 primaryClass = {astro-ph.GA},
       adsurl = {https://ui.adsabs.harvard.edu/abs/2015MNRAS.452...54M}
}

@ARTICLE{HerreraCamus15,
       author = {{Herrera-Camus}, R. and {Bolatto}, A.~D. and {Wolfire}, M.~G. and {Smith}, J.~D. and {Croxall}, K.~V. and {Kennicutt}, R.~C. and {Calzetti}, D. and {Helou}, G. and {Walter}, F. and {Leroy}, A.~K. and {Draine}, B. and {Brandl}, B.~R. and {Armus}, L. and {Sandstrom}, K.~M. and {Dale}, D.~A. and {Aniano}, G. and {Meidt}, S.~E. and {Boquien}, M. and {Hunt}, L.~K. and {Galametz}, M. and {Tabatabaei}, F.~S. and {Murphy}, E.~J. and {Appleton}, P. and {Roussel}, H. and {Engelbracht}, C. and {Beirao}, P.},
        title = "{[C II] 158 {\ensuremath{\mu}}m Emission as a Star Formation Tracer}",
      journal = {\apj},
         year = 2015,
        month = feb,
       volume = {800},
       number = {1},
          eid = {1},
        pages = {1},
          doi = {10.1088/0004-637X/800/1/1},
archivePrefix = {arXiv},
       eprint = {1409.7123},
 primaryClass = {astro-ph.GA},
       adsurl = {https://ui.adsabs.harvard.edu/abs/2015ApJ...800....1H}
}

@ARTICLE{Inami22,
       author = {{Inami}, Hanae and {Algera}, Hiddo S.~B. and {Schouws}, Sander and {Sommovigo}, Laura and {Bouwens}, Rychard and {Smit}, Renske and {Stefanon}, Mauro and {Bowler}, Rebecca A.~A. and {Endsley}, Ryan and {Ferrara}, Andrea and {Oesch}, Pascal and {Stark}, Daniel and {Aravena}, Manuel and {Barrufet}, Laia and {da Cunha}, Elisabete and {Dayal}, Pratika and {De Looze}, Ilse and {Fudamoto}, Yoshinobu and {Gonzalez}, Valentino and {Graziani}, Luca and {Hodge}, Jacqueline A. and {Hygate}, Alexander P.~S. and {Nanayakkara}, Themiya and {Pallottini}, Andrea and {Riechers}, Dominik A. and {Schneider}, Raffaella and {Topping}, Michael and {van der Werf}, Paul},
        title = "{The ALMA REBELS Survey: dust continuum detections at z > 6.5}",
      journal = {\mnras},
         year = 2022,
        month = sep,
       volume = {515},
       number = {3},
        pages = {3126-3143},
          doi = {10.1093/mnras/stac1779},
archivePrefix = {arXiv},
       eprint = {2203.15136},
 primaryClass = {astro-ph.GA},
       adsurl = {https://ui.adsabs.harvard.edu/abs/2022MNRAS.515.3126I}
}

@ARTICLE{Riechers17,
       author = {{Riechers}, Dominik A. and {Leung}, T.~K. Daisy and {Ivison}, Rob J. and {P{\'e}rez-Fournon}, Ismael and {Lewis}, Alexander J.~R. and {Marques-Chaves}, Rui and {Oteo}, Iv{\'a}n and {Clements}, Dave L. and {Cooray}, Asantha and {Greenslade}, Josh and {Mart{\'\i}nez-Navajas}, Paloma and {Oliver}, Seb and {Rigopoulou}, Dimitra and {Scott}, Douglas and {Weiss}, Axel},
        title = "{Rise of the Titans: A Dusty, Hyper-luminous {\textquotedblleft}870 {\ensuremath{\mu}}m Riser{\textquotedblright} Galaxy at z {\ensuremath{\sim}} 6}",
      journal = {\apj},
         year = 2017,
        month = nov,
       volume = {850},
       number = {1},
          eid = {1},
        pages = {1},
          doi = {10.3847/1538-4357/aa8ccf},
archivePrefix = {arXiv},
       eprint = {1705.09660},
 primaryClass = {astro-ph.GA},
       adsurl = {https://ui.adsabs.harvard.edu/abs/2017ApJ...850....1R}
}

@MISC{Schouws22InPrep,
       author = {{Schouws}, S. and {Bouwens}, R. J. and others},
        title = {The REBELS ALMA Survey: [CII] line measurements},
         note = {in prep.},
         year = {2026}
}

@ARTICLE{Barrufet23,
       author = {{Barrufet}, L. and {Oesch}, P.~A. and {Bouwens}, R. and {Inami}, H. and {Sommovigo}, L. and {Algera}, H. and {da Cunha}, E. and {Aravena}, M. and {Dayal}, P. and {Ferrara}, A. and {Fudamoto}, Y. and {Gonzalez}, V. and {Graziani}, L. and {Hygate}, A.~P.~S. and {de Looze}, I. and {Nanayakkara}, T. and {Pallottini}, A. and {Schneider}, R. and {Stefanon}, M. and {Topping}, M. and {van der Werf}, P.},
        title = "{The ALMA REBELS Survey: the first infrared luminosity function measurement at z {\ensuremath{\sim}} 7}",
      journal = {\mnras},
         year = 2023,
        month = jul,
       volume = {522},
       number = {3},
        pages = {3926-3934},
          doi = {10.1093/mnras/stad1259},
archivePrefix = {arXiv},
       eprint = {2303.11321},
 primaryClass = {astro-ph.GA},
       adsurl = {https://ui.adsabs.harvard.edu/abs/2023MNRAS.522.3926B}}

@ARTICLE{Weibel24,
       author = {{Weibel}, Andrea and {Oesch}, Pascal A. and {Barrufet}, Laia and {Gottumukkala}, Rashmi and {Ellis}, Richard S. and {Santini}, Paola and {Weaver}, John R. and {Allen}, Natalie and {Bouwens}, Rychard and {Bowler}, Rebecca A.~A. and {Brammer}, Gabe and {Carnall}, Adam C. and {Cullen}, Fergus and {Dayal}, Pratika and {Dickinson}, Mark and {Donnan}, Callum T. and {Dunlop}, James S. and {Giavalisco}, Mauro and {Grogin}, Norman A. and {Illingworth}, Garth D. and {Koekemoer}, Anton M. and {Labbe}, Ivo and {Marchesini}, Danilo and {McLeod}, Derek J. and {McLure}, Ross J. and {Naidu}, Rohan P. and {P{\'e}rez-Gonz{\'a}lez}, Pablo G. and {Shuntov}, Marko and {Stefanon}, Mauro and {Toft}, Sune and {Xiao}, Mengyuan},
        title = "{Galaxy build-up in the first 1.5 Gyr of cosmic history: insights from the stellar mass function at z   4-9 from JWST NIRCam observations}",
      journal = {\mnras},
         year = 2024,
        month = sep,
       volume = {533},
       number = {2},
        pages = {1808-1838},
          doi = {10.1093/mnras/stae1891},
archivePrefix = {arXiv},
       eprint = {2403.08872},
 primaryClass = {astro-ph.GA},
       adsurl = {https://ui.adsabs.harvard.edu/abs/2024MNRAS.533.1808W}
}

@ARTICLE{dayal2014a,
   author = {{Dayal}, P. and {Ferrara}, A. and {Dunlop}, J.~S. and {Pacucci}, F.
	},
    title = "{Essential physics of early galaxy formation}",
  journal = {\mnras},
archivePrefix = "arXiv",
   eprint = {1405.4862},
     year = 2014,
    month = dec,
   volume = 445,
    pages = {2545-2557},
      doi = {10.1093/mnras/stu1848},
   adsurl = {http://adsabs.harvard.edu/abs/2014MNRAS.445.2545D}
}

@ARTICLE{Vallini15,
       author = {{Vallini}, L. and {Gallerani}, S. and {Ferrara}, A. and {Pallottini}, A. and {Yue}, B.},
        title = "{On the [CII]-SFR Relation in High Redshift Galaxies}",
      journal = {\apj},
         year = 2015,
        month = nov,
       volume = {813},
       number = {1},
          eid = {36},
        pages = {36},
          doi = {10.1088/0004-637X/813/1/36},
archivePrefix = {arXiv},
       eprint = {1507.00340},
 primaryClass = {astro-ph.GA},
       adsurl = {https://ui.adsabs.harvard.edu/abs/2015ApJ...813...36V}
}

@ARTICLE{kathri25,
       author = {{Khatri}, Prachi and {Romano-D{\'\i}az}, Emilio and {Porciani}, Cristiano},
        title = "{The [C II] line emission as an interstellar medium probe in the MARIGOLD galaxies}",
      journal = {\aap},
         year = 2025,
        month = may,
       volume = {697},
          eid = {A174},
        pages = {A174},
          doi = {10.1051/0004-6361/202453048},
archivePrefix = {arXiv},
       eprint = {2411.09755},
 primaryClass = {astro-ph.GA},
       adsurl = {https://ui.adsabs.harvard.edu/abs/2025A&A...697A.174K}
}

@ARTICLE{Barrufet23b,
       author = {{Barrufet}, L. and {Oesch}, P.~A. and {Weibel}, A. and {Brammer}, G. and {Bezanson}, R. and {Bouwens}, R. and {Fudamoto}, Y. and {Gonzalez}, V. and {Gottumukkala}, R. and {Illingworth}, G. and {Heintz}, K.~E. and {Holden}, B. and {Labbe}, I. and {Magee}, D. and {Naidu}, R.~P. and {Nelson}, E. and {Stefanon}, M. and {Smit}, R. and {van Dokkum}, P. and {Weaver}, J.~R. and {Williams}, C.~C.},
        title = "{Unveiling the nature of infrared bright, optically dark galaxies with early JWST data}",
      journal = {\mnras},
         year = 2023,
        month = jun,
       volume = {522},
       number = {1},
        pages = {449-456},
          doi = {10.1093/mnras/stad947},
archivePrefix = {arXiv},
       eprint = {2207.14733},
 primaryClass = {astro-ph.GA},
       adsurl = {https://ui.adsabs.harvard.edu/abs/2023MNRAS.522..449B}
}

@ARTICLE{labbe23b,
       author = {{Labb{\'e}}, Ivo and {van Dokkum}, Pieter and {Nelson}, Erica and {Bezanson}, Rachel and {Suess}, Katherine A. and {Leja}, Joel and {Brammer}, Gabriel and {Whitaker}, Katherine and {Mathews}, Elijah and {Stefanon}, Mauro and {Wang}, Bingjie},
        title = "{A population of red candidate massive galaxies  600 Myr after the Big Bang}",
      journal = {\nat},
         year = 2023,
        month = apr,
       volume = {616},
       number = {7956},
        pages = {266-269},
          doi = {10.1038/s41586-023-05786-2},
archivePrefix = {arXiv},
       eprint = {2207.12446},
 primaryClass = {astro-ph.GA},
       adsurl = {https://ui.adsabs.harvard.edu/abs/2023Natur.616..266L}
}

@ARTICLE{Gottumukkala24,
       author = {{Gottumukkala}, R. and {Barrufet}, L. and {Oesch}, P.~A. and {Weibel}, A. and {Allen}, N. and {Alcalde Pampliega}, B. and {Nelson}, E.~J. and {Williams}, C.~C. and {Brammer}, G. and {Fudamoto}, Y. and {Gonz{\'a}lez}, V. and {Heintz}, K.~E. and {Illingworth}, G. and {Magee}, D. and {Naidu}, R.~P. and {Shuntov}, M. and {Stefanon}, M. and {Toft}, S. and {Valentino}, F. and {Xiao}, M.},
        title = "{Unveiling the hidden Universe with JWST: the contribution of dust-obscured galaxies to the stellar mass function at z   3 - 8}",
      journal = {\mnras},
         year = 2024,
        month = may,
       volume = {530},
       number = {1},
        pages = {966-983},
          doi = {10.1093/mnras/stae754},
archivePrefix = {arXiv},
       eprint = {2310.03787},
 primaryClass = {astro-ph.GA},
       adsurl = {https://ui.adsabs.harvard.edu/abs/2024MNRAS.530..966G}
}

@ARTICLE{herard-demanche25,
       author = {{Herard-Demanche}, Thomas and {Bouwens}, Rychard J. and {Oesch}, Pascal A. and {Naidu}, Rohan P. and {Decarli}, Roberto and {Nelson}, Erica J. and {Brammer}, Gabriel and {Weibel}, Andrea and {Xiao}, Mengyuan and {Stefanon}, Mauro and {Walter}, Fabian and {Matthee}, Jorryt and {Meyer}, Romain A. and {Wuyts}, Stijn and {Reddy}, Naveen and {Rowland}, Lucie and {van Leeuwen}, Ivana and {Haro}, Pablo Arrabal and {Dannerbauer}, Helmut and {Shapley}, Alice E. and {Chisholm}, John and {van Dokkum}, Pieter and {Labbe}, Ivo and {Illingworth}, Garth and {Schaerer}, Daniel and {Shivaei}, Irene},
        title = "{Mapping dusty galaxy growth at z > 5 with FRESCO: detection of H{\ensuremath{\alpha}} in submm galaxy HDF850.1 and the surrounding overdense structures}",
      journal = {\mnras},
         year = 2025,
        month = feb,
       volume = {537},
       number = {2},
        pages = {788-808},
          doi = {10.1093/mnras/staf030},
archivePrefix = {arXiv},
       eprint = {2309.04525},
 primaryClass = {astro-ph.GA},
       adsurl = {https://ui.adsabs.harvard.edu/abs/2025MNRAS.537..788H}
}

@ARTICLE{Heintz23,
       author = {{Heintz}, K.~E. and {Shapley}, A.~E. and {Sanders}, R.~L. and {Killi}, M. and {Watson}, D. and {Magdis}, G. and {Valentino}, F. and {Ginolfi}, M. and {Narayanan}, D. and {Greve}, T.~R. and {Fynbo}, J.~P.~U. and {Vizgan}, D. and {Wilson}, S.~N.},
        title = "{Gauging the mass of metals in the gas phase of galaxies from the Local Universe to the Epoch of Reionization}",
      journal = {\aap},
         year = 2023,
        month = oct,
       volume = {678},
          eid = {A30},
        pages = {A30},
          doi = {10.1051/0004-6361/202346573},
archivePrefix = {arXiv},
       eprint = {2308.14813},
 primaryClass = {astro-ph.GA},
       adsurl = {https://ui.adsabs.harvard.edu/abs/2023A&A...678A..30H}
}

@ARTICLE{Harvey24,
       author = {{Harvey}, Thomas and {Conselice}, Christopher J. and {Adams}, Nathan J. and {Austin}, Duncan and {Juod{\v{z}}balis}, Ignas and {Trussler}, James and {Li}, Qiong and {Ormerod}, Katherine and {Ferreira}, Leonardo and {Lovell}, Christopher C. and {Duan}, Qiao and {Westcott}, Lewi and {Harris}, Honor and {Bhatawdekar}, Rachana and {Coe}, Dan and {Cohen}, Seth H. and {Caruana}, Joseph and {Cheng}, Cheng and {Driver}, Simon P. and {Frye}, Brenda and {Furtak}, Lukas J. and {Grogin}, Norman A. and {Hathi}, Nimish P. and {Holwerda}, Benne W. and {Jansen}, Rolf A. and {Koekemoer}, Anton M. and {Marshall}, Madeline A. and {Nonino}, Mario and {Vijayan}, Aswin P. and {Wilkins}, Stephen M. and {Windhorst}, Rogier and {Willmer}, Christopher N.~A. and {Yan}, Haojing and {Zitrin}, Adi},
        title = "{EPOCHS. IV. SED Modeling Assumptions and Their Impact on the Stellar Mass Function at 6.5 {\ensuremath{\leq}} z {\ensuremath{\leq}} 13.5 Using PEARLS and Public JWST Observations}",
      journal = {\apj},
         year = 2025,
        month = jan,
       volume = {978},
       number = {1},
          eid = {89},
        pages = {89},
          doi = {10.3847/1538-4357/ad8c29},
archivePrefix = {arXiv},
       eprint = {2403.03908},
 primaryClass = {astro-ph.GA},
       adsurl = {https://ui.adsabs.harvard.edu/abs/2025ApJ...978...89H}
}

@ARTICLE{Weaver23,
       author = {{Weaver}, J.~R. and {Davidzon}, I. and {Toft}, S. and {Ilbert}, O. and {McCracken}, H.~J. and {Gould}, K.~M.~L. and {Jespersen}, C.~K. and {Steinhardt}, C. and {Lagos}, C.~D.~P. and {Capak}, P.~L. and {Casey}, C.~M. and {Chartab}, N. and {Faisst}, A.~L. and {Hayward}, C.~C. and {Kartaltepe}, J.~S. and {Kauffmann}, O.~B. and {Koekemoer}, A.~M. and {Kokorev}, V. and {Laigle}, C. and {Liu}, D. and {Long}, A. and {Magdis}, G.~E. and {McPartland}, C.~J.~R. and {Milvang-Jensen}, B. and {Mobasher}, B. and {Moneti}, A. and {Peng}, Y. and {Sanders}, D.~B. and {Shuntov}, M. and {Sneppen}, A. and {Valentino}, F. and {Zalesky}, L. and {Zamorani}, G.},
        title = "{COSMOS2020: The galaxy stellar mass function. The assembly and star formation cessation of galaxies at 0.2< z {\ensuremath{\leq}} 7.5}",
      journal = {\aap},
         year = 2023,
        month = sep,
       volume = {677},
          eid = {A184},
        pages = {A184},
          doi = {10.1051/0004-6361/202245581},
archivePrefix = {arXiv},
       eprint = {2212.02512},
 primaryClass = {astro-ph.GA},
       adsurl = {https://ui.adsabs.harvard.edu/abs/2023A&A...677A.184W}
}

@ARTICLE{Aravena24,
       author = {{Aravena}, M. and {Heintz}, K. and {Dessauges-Zavadsky}, M. and {Oesch}, P. and {Algera}, H. and {Bouwens}, R. and {da Cunha}, E. and {Dayal}, P. and {De Looze}, I. and {Ferrara}, A. and {Fudamoto}, Y. and {Gonzalez}, V. and {Graziani}, L. and {Hygate}, A.~P.~S. and {Inami}, H. and {Pallottini}, A. and {Schneider}, R. and {Schouws}, S. and {Sommovigo}, L. and {Topping}, M. and {van der Werf}, P. and {Palla}, M.},
        title = "{The ALMA Reionization Era Bright Emission Line Survey: The molecular gas content of galaxies at z 7}",
      journal = {\aap},
         year = 2024,
        month = feb,
       volume = {682},
          eid = {A24},
        pages = {A24},
          doi = {10.1051/0004-6361/202347281},
archivePrefix = {arXiv},
       eprint = {2309.15948},
 primaryClass = {astro-ph.GA},
       adsurl = {https://ui.adsabs.harvard.edu/abs/2024A&A...682A..24A}
}

@ARTICLE{Rowland24,
       author = {{Rowland}, Lucie E. and {Hodge}, Jacqueline and {Bouwens}, Rychard and {Mancera Pi{\~n}a}, Pavel E. and {Hygate}, Alexander and {Algera}, Hiddo and {Aravena}, Manuel and {Bowler}, Rebecca and {da Cunha}, Elisabete and {Dayal}, Pratika and {Ferrara}, Andrea and {Herard-Demanche}, Thomas and {Inami}, Hanae and {van Leeuwen}, Ivana and {de Looze}, Ilse and {Oesch}, Pascal and {Pallottini}, Andrea and {Phillips}, Si{\^a}n and {Rybak}, Matus and {Schouws}, Sander and {Smit}, Renske and {Sommovigo}, Laura and {Stefanon}, Mauro and {van der Werf}, Paul},
        title = "{REBELS-25: discovery of a dynamically cold disc galaxy at z = 7.31}",
      journal = {\mnras},
         year = 2024,
        month = dec,
       volume = {535},
       number = {3},
        pages = {2068-2091},
          doi = {10.1093/mnras/stae2217},
archivePrefix = {arXiv},
       eprint = {2405.06025},
 primaryClass = {astro-ph.GA},
       adsurl = {https://ui.adsabs.harvard.edu/abs/2024MNRAS.535.2068R}
}

@ARTICLE{Algera26,
       author = {{Algera}, Hiddo S.~B. and {Rowland}, Lucie and {Stefanon}, Mauro and {Palla}, Marco and {Sommovigo}, Laura and {Inami}, Hanae and {Bouwens}, Rychard and {Aravena}, Manuel and {Bowler}, Rebecca A.~A. and {Dayal}, Pratika and {De Looze}, Ilse and {Ferrara}, Andrea and {Fisher}, Rebecca and {Graziani}, Luca and {Gulis}, Cindy and {Heintz}, Kasper and {Hodge}, Jacqueline and {Laza-Ramos}, Andr{\'e}s and {van Leeuwen}, Ivana and {Pallottini}, Andrea and {Phillips}, Si{\^a}n and {Schouws}, Sander and {Smit}, Renske and {Stark}, Daniel P. and {van der Werf}, Paul},
        title = "{REBELS-IFU: dust build-up in massive galaxies at redshift 7}",
      journal = {\mnras},
         year = 2026,
        month = jan,
       volume = {545},
       number = {2},
          eid = {staf1897},
        pages = {staf1897},
          doi = {10.1093/mnras/staf1897},
archivePrefix = {arXiv},
       eprint = {2501.10508},
 primaryClass = {astro-ph.GA},
       adsurl = {https://ui.adsabs.harvard.edu/abs/2026MNRAS.545f1897A}
}

@ARTICLE{Algera23,
       author = {{Algera}, Hiddo S.~B. and {Inami}, Hanae and {Oesch}, Pascal A. and {Sommovigo}, Laura and {Bouwens}, Rychard J. and {Topping}, Michael W. and {Schouws}, Sander and {Stefanon}, Mauro and {Stark}, Daniel P. and {Aravena}, Manuel and {Barrufet}, Laia and {da Cunha}, Elisabete and {Dayal}, Pratika and {Endsley}, Ryan and {Ferrara}, Andrea and {Fudamoto}, Yoshinobu and {Gonzalez}, Valentino and {Graziani}, Luca and {Hodge}, Jacqueline A. and {Hygate}, Alexander P.~S. and {de Looze}, Ilse and {Nanayakkara}, Themiya and {Schneider}, Raffaella and {van der Werf}, Paul P.},
        title = "{The ALMA REBELS survey: the dust-obscured cosmic star formation rate density at redshift 7}",
      journal = {\mnras},
         year = 2023,
        month = feb,
       volume = {518},
       number = {4},
        pages = {6142-6157},
          doi = {10.1093/mnras/stac3195},
archivePrefix = {arXiv},
       eprint = {2208.08243},
 primaryClass = {astro-ph.GA},
       adsurl = {https://ui.adsabs.harvard.edu/abs/2023MNRAS.518.6142A}
}

@ARTICLE{Bowler24,
       author = {{Bowler}, R.~A.~A. and {Inami}, H. and {Sommovigo}, L. and {Smit}, R. and {Algera}, H.~S.~B. and {Aravena}, M. and {Barrufet}, L. and {Bouwens}, R. and {da Cunha}, E. and {Cullen}, F. and {Dayal}, P. and {De Looze}, I. and {Dunlop}, J.~S. and {Fudamoto}, Y. and {Mauerhofer}, V. and {McLure}, R.~J. and {Stefanon}, M. and {Schneider}, R. and {Ferrara}, A. and {Graziani}, L. and {Hodge}, J.~A. and {Nanayakkara}, T. and {Palla}, M. and {Schouws}, S. and {Stark}, D.~P. and {van der Werf}, P.~P.},
        title = "{The ALMA REBELS survey: obscured star formation in massive Lyman-break galaxies at z= 4-8 revealed by the IRX-{\ensuremath{\beta}} and M$_{{\ensuremath{\star}}}$ relations}",
      journal = {\mnras},
         year = 2024,
        month = jan,
       volume = {527},
       number = {3},
        pages = {5808-5828},
          doi = {10.1093/mnras/stad3578},
archivePrefix = {arXiv},
       eprint = {2309.17386},
 primaryClass = {astro-ph.GA},
       adsurl = {https://ui.adsabs.harvard.edu/abs/2024MNRAS.527.5808B}
}

@ARTICLE{rowland26a,
       author = {{Rowland}, Lucie E. and {Stefanon}, Mauro and {Bouwens}, Rychard and {Hodge}, Jacqueline and {Algera}, Hiddo and {Fisher}, Rebecca and {Dayal}, Pratika and {Pallottini}, Andrea and {Stark}, Daniel P. and {Heintz}, Kasper E. and {Aravena}, Manuel and {Bowler}, Rebecca A.~A. and {Cescon}, Karin and {Endsley}, Ryan and {Ferrara}, Andrea and {Fudamoto}, Yoshinobu and {Gonzalez}, Valentino and {Graziani}, Luca and {Gulis}, Cindy and {Herard-Demanche}, Thomas and {Inami}, Hanae and {Laza-Ramos}, Andr{\`e}s and {van Leeuwen}, Ivana and {de Looze}, Ilse and {Nanayakkara}, Themiya and {Oesch}, Pascal and {Ormerod}, Katherine and {Palla}, Marco and {Sartorio}, Nina S. and {Schouws}, Sander and {Smit}, Renske and {Sommovigo}, Laura and {Toft}, Sune and {Weaver}, John R. and {van der Werf}, Paul},
        title = "{REBELS-IFU: evidence for metal-rich massive galaxies at z {\ensuremath{\sim}} 6{\ensuremath{-}}8}",
      journal = {\mnras},
         year = 2026,
        month = feb,
       volume = {546},
       number = {2},
          eid = {staf2023},
        pages = {staf2023},
          doi = {10.1093/mnras/staf2023},
archivePrefix = {arXiv},
       eprint = {2501.10559},
 primaryClass = {astro-ph.GA},
       adsurl = {https://ui.adsabs.harvard.edu/abs/2026MNRAS.546f2023R}
}

@ARTICLE{Fisher25,
       author = {{Fisher}, R. and {Bowler}, R.~A.~A. and {Stefanon}, M. and {Rowland}, L.~E. and {Algera}, H.~S.~B. and {Aravena}, M. and {Bouwens}, R. and {Dayal}, P. and {Ferrara}, A. and {Fudamoto}, Y. and {Gulis}, C. and {Hodge}, J.~A. and {Inami}, H. and {Ormerod}, K. and {Pallottini}, A. and {Phillips}, S.~G. and {Sartorio}, N.~S. and {Schouws}, S. and {Smit}, R. and {Sommovigo}, L. and {Stark}, D.~P. and {van der Werf}, P.~P.},
        title = "{REBELS-IFU: dust attenuation curves of 12 massive galaxies at z ≃ 7}",
      journal = {\mnras},
         year = 2025,
        month = may,
       volume = {539},
       number = {1},
        pages = {109-126},
          doi = {10.1093/mnras/staf485},
archivePrefix = {arXiv},
       eprint = {2501.10541},
 primaryClass = {astro-ph.GA},
       adsurl = {https://ui.adsabs.harvard.edu/abs/2025MNRAS.539..109F}
}

@ARTICLE{madaudickinson14,
       author = {{Madau}, Piero and {Dickinson}, Mark},
        title = "{Cosmic Star-Formation History}",
      journal = {\araa},
         year = 2014,
        month = aug,
       volume = {52},
        pages = {415-486},
          doi = {10.1146/annurev-astro-081811-125615},
archivePrefix = {arXiv},
       eprint = {1403.0007},
 primaryClass = {astro-ph.CO},
       adsurl = {https://ui.adsabs.harvard.edu/abs/2014ARA&A..52..415M}
}



\appendix

\section{Additional tests and comparisons}

\subsection{Other SMFs}
We adopted the SMF from \citet{Weibel24} for our fiducial \CII LF estimation. We also show how the same derivation would have performed with the SMFs from \citet{Stefanon21}, \citet{Weaver23} and \citet{Harvey24} in Fig.~\ref{app:othersmfs} and Tab. \ref{app:otherLFpars}.
We favour \citet{Weibel24} for the following reason. It is measured over four independent, contiguous JWST/NIRCam fields (CEERS-EGS, PRIMER-COSMOS, PRIMER-UDS and JADES-GOODS-S) spanning $\sim500\,\mathrm{arcmin}^2$ and $>30{,}000$ rest-optically selected galaxies reduced through a single homogeneous pipeline. This allows us to better control for field-to-field "cosmic variance" while also incorporating fainter and redder objects missed by pre-JWST selections. Although the individual Schechter parameters differ owing to strong degeneracies between $\phi^*$, $L^*$ and $\alpha$, the resulting LFs agree within uncertainties over the luminosity range constrained by our data. Reassuringly, the fiducial \CII LF we derive from \citet{Weibel24} lies roughly near the middle of the \CII LF results presented in Fig.~\ref{app:othersmfs}.
\begin{table}
\centering
\caption{\CII LF parameters derived from Schechter function fits at $z\sim7$
using different SMFs from the literature.}
\label{app:otherLFpars}
\begin{tabular}{lcccc}
\hline
SMF & Redshift & $\log\phi^*_{\mathrm{\CII}}$ &
$\log L^*_{\mathrm{\CII}}$ & $\alpha_{\mathrm{\CII}}$ \\
\hline
S21 & $z\sim7$ & $-4.73^{+0.27}_{-0.44}$ & $9.17^{+0.15}_{-0.14}$ & $-2.19^{+0.16}_{-0.24}$ \\
W23 & $z\sim7$ & $-2.92^{+0.29}_{-0.48}$ & $9.12^{+0.26}_{-0.20}$ & $-1.77^{+0.07}_{-0.11}$ \\
H25 & $z\sim7$ & $-5.98^{+0.37}_{-0.56}$ & $10.17^{+0.51}_{-0.44}$ & $-2.50^{+0.23}_{-0.31}$ \\
\hline
\end{tabular}
\par\medskip
\begin{minipage}{\columnwidth}
\small \textit{Notes.} $\phi^*_{\mathrm{\CII}}$ is in
$\mathrm{Mpc^{-3}\,dex^{-1}}$ and $L^*_{\mathrm{\CII}}$ in $L_\odot$. S21, W23 and H25 respectively correspond to the SMFs presented in \citet{Stefanon21},
\citet{Weaver23} and \citet{Harvey24}.
\end{minipage}
\end{table}

\begin{figure}
    \centering
    \includegraphics[width=\linewidth]{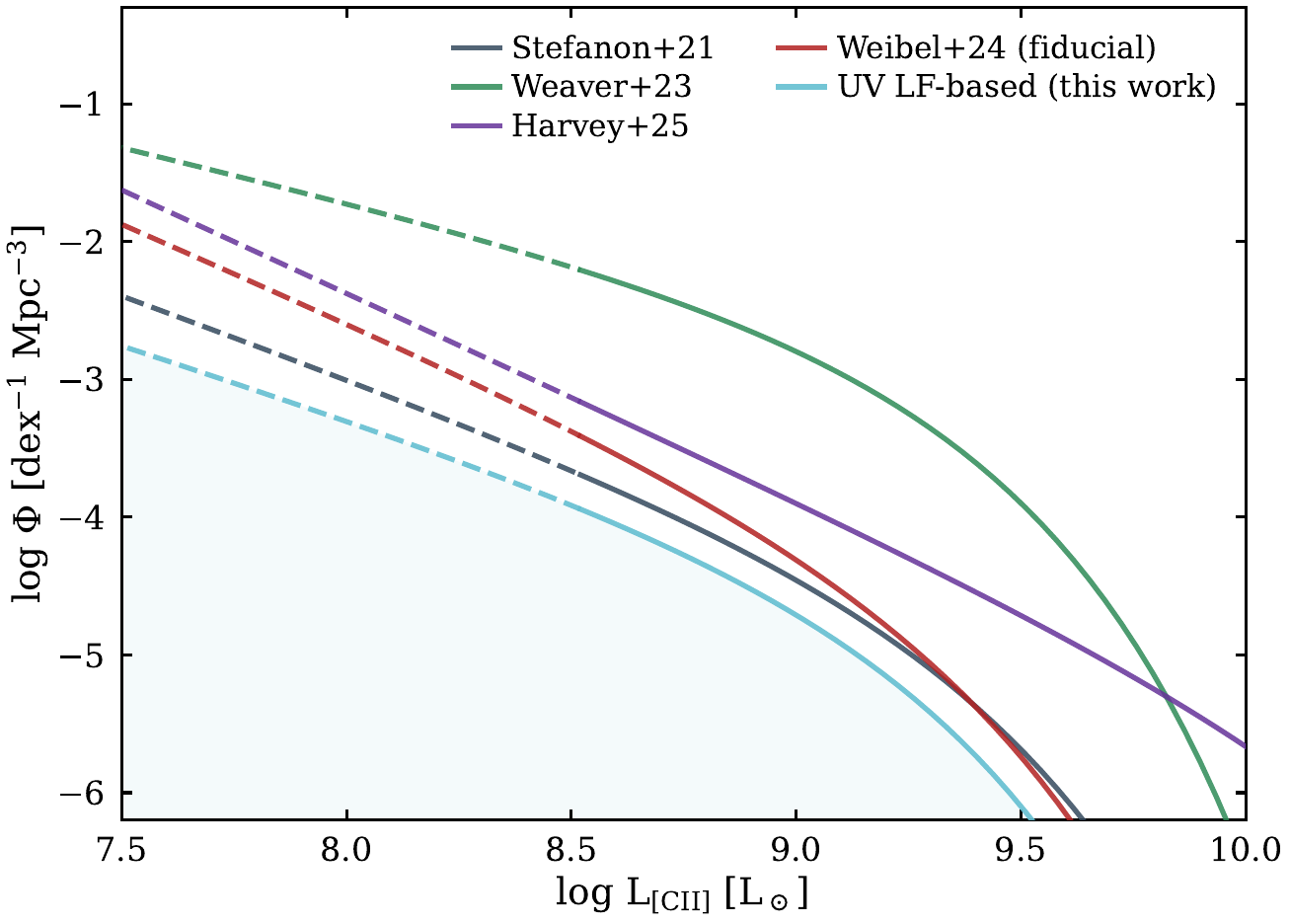}
    \caption{Comparison of the \CII luminosity function at $z\sim7$ derived using the empirical method and different SMFs from the literature. These estimates use \citet[dark blue]{Stefanon21}, \citet[green]{Weaver23}, \citet[purple]{Harvey24}, and \citet[red]{Weibel24}, with the latter adopted as our fiducial choice. The UV LF-based estimate from this work is shown in blue. Solid lines indicate the luminosity range constrained by our data, while dashed lines show the extrapolation of our fit below the completeness threshold of our sample.}
    \label{app:othersmfs}
\end{figure}

\subsection{Comparison with theoretical works at \texorpdfstring{$z\sim6$}{z~6}}
\label{app:z6comp}
To allow a fair comparison with other works, we also compute the \CII LF using the SMF from \citet{Weibel24} at $z\sim6$. Several other theoretical predictions have been made at this redshift rather than at $z\sim7$. Our empirical method, which is based on a sample of galaxies from $z\sim 4-8$, should apply at $z\sim6$. 
We can first compare with \citet{Popping19}, which combines a semi-analytical model of galaxy formation with chemical equilibrium networks and numerical radiative transfer models to derive the \CII emission of galaxies at $z\sim6$. Their results are consistent with our SMF-based method at the faint-end, but this is primarily below the completeness limit of our empirical relations.  At the bright end, however, the \citet{Popping19} results significantly underpredict the number of galaxies with $L_\mathrm{\CII}>10^9\,$L$_\odot$, as their predicted LF drops well below our measurements at those luminosities.

\citet{Liu24} at $z=6$ shows a good agreement with our determinations, for which we show only the $\epsilon_{\rm ff}=0.015$ fiducial for clarity. Their model appears to be consistent with our SMF-based \CII LF at the bright-end while also showing a similarly steep faint-end slope. Their variant at $\epsilon_{\rm ff}=0.045$ (not shown) is also consistent across the range and within our uncertainties. However, we note that their low burstiness variant at $\epsilon_{\rm ff}=0.005$ predicts more \CII-bright emitters than we infer and a shallower faint-end slope.

Both the predictions from \citet{yue&ferrara19} and \citet{Bethermin22} predict the existence of extremely \CII-bright galaxies at $z\sim6$.
\citet{yue&ferrara19} derive their $z\sim6$ \CII LF semi-empirically, convolving
the dust-corrected $z\sim6$ UV LF of a log-normal relation between \CII and SFR and is thus closer in approach to our UV LF based method. Their bright-end excess is consistent with both our result and the \citet{Bethermin22}. We note, however, that this excess can also be explained by a scatter upward due to a broad $L_\mathrm{\CII}$--SFR relation, rather than a real population of extremely bright \CII-emitters.

\begin{figure}
    \centering
    \includegraphics[width=\linewidth]{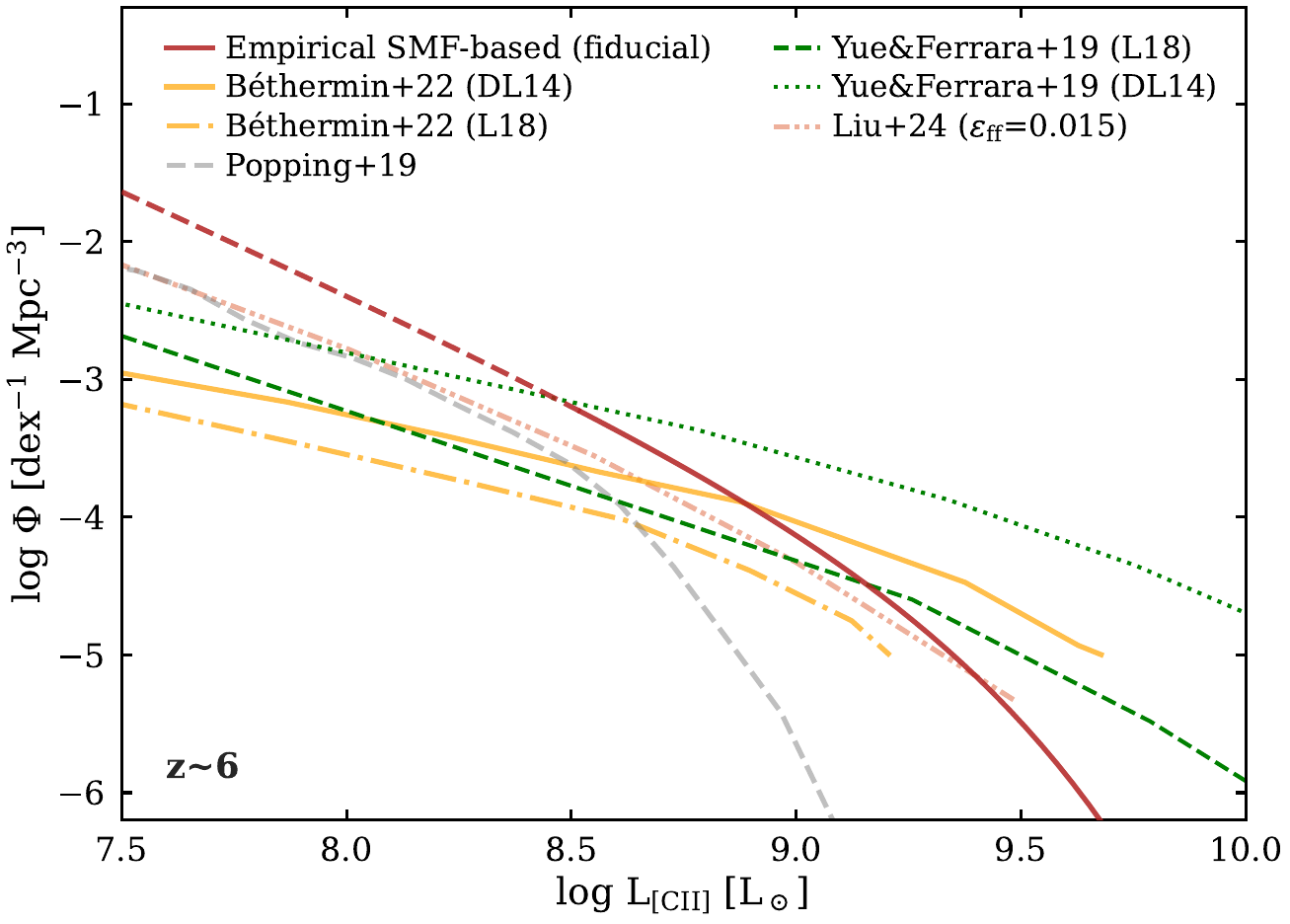}
    \caption{Theoretical predictions of the \CII LF at $z\sim6$. We show our SMF-based \CII LF computed at $z\sim6$ in red, that of \citet{Popping19} in grey, \citet{yue&ferrara19} in green, the fiducial bursty model from \citet{Liu24} in pink, and the SIDES simulation in orange \citep{Bethermin22}, using two different $L_\mathrm{\CII}$--SFR relations.}
    \label{fig:placeholder}
\end{figure}


\bsp	

\label{lastpage}
\end{document}